\documentclass[12pt]{jpp}

\usepackage{subfig}
\usepackage{graphicx}
\usepackage{float}
\usepackage{cite}
\usepackage[labelfont=bf, textfont=it]{caption}
\usepackage{hyperref}
\usepackage{xcolor}
\usepackage{booktabs}
\usepackage{amsmath}

\definecolor{myred}{rgb}{0.0, 0.0, 0.0}

\begin{document}

\title{A quasi-linear model of electromagnetic turbulent transport and its application to flux-driven transport predictions for STEP}

\author{M. Giacomin\aff{1,2}
  \corresp{\email{maurizio.giacomin@unipd.it}},
  D. Dickinson\aff{2}, W. Dorland\aff{3,4}, N. R. Mandell\aff{5}, A. Bokshi\aff{2}, F. J. Casson\aff{6}, H. G. Dudding\aff{6}, D. Kennedy\aff{6}, B. S. Patel\aff{6}, and  C. M. Roach\aff{6}}

\affiliation{\aff{1} Dipartimento di Fisica ``G. Galilei'', Università degli Studi di Padova, Padua, Italy
\aff{2} York Plasma Institute, University of York, York, YO10 5DD, United Kingdom
\aff{3} Department of Physics, University of Maryland, College Park, MD 20740, USA
\aff{4} Institute for Research in Electronics and Applied Physics, University of Maryland, College Park, MD 20742, USA
\aff{5} Princeton Plasma Physics Laboratory, Princeton, 08543, NJ, USA
\aff{6} UKAEA (United Kingdom Atomic Energy Authority), Culham Campus, Abingdon, Oxfordshire, OX14 3DB, United Kingdom}

\maketitle

\begin{abstract}
A quasi-linear reduced transport model is developed from a database of high-$\beta$ electromagnetic nonlinear gyrokinetic simulations performed with Spherical Tokamak for Energy Production (STEP) relevant parameters. 
The quasi-linear model is fully electromagnetic and accounts for the effect of equilibrium flow shear using a novel approach. Its flux predictions are shown to agree quantitatively with predictions from local nonlinear gyrokinetic simulations across a broad range of STEP-relevant local equilibria. 
This reduced transport model is implemented in the T3D transport solver that is used to perform the first flux-driven simulations for STEP to account for transport from hybrid-KBM turbulence, which dominates over a wide region of the core plasma.
Nonlinear gyrokinetic simulations of the final transport steady state from T3D return turbulent fluxes that are consistent with the reduced model, indicating that the quasi-linear model may also be appropriate for describing the transport steady state.  Within the assumption considered here, our simulations support the existence of a transport steady state in STEP with a fusion power comparable to that in the burning flat-top of the conceptual design, but do not demonstrate how this state can be accessed. 
   
\end{abstract}

\section{Introduction}

The design of future tokamaks requires accurate evaluations of the core turbulent transport, which is most reliably modelled using high-fidelity gyrokinetic simulations that have been validated extensively against experiments. However, these simulations are computationally very expensive and,  for reasons of computational tractability, reduced turbulent transport models are more routinely used in the design of plasma scenarios.
Transport steady state fusion plasma solutions can be calculated in reasonable times with relatively modest compute resource, using reduced physics-based transport models, such as TGLF~\citep{staebler2005,staebler2007} and QuaLiKiz~\citep{bourdelle2015,citrin2017}.  These models combine a fast and simplified approach to find the linearly dominant eigenmodes (by either solving gyrofluid equations, or using a simplified gyrokinetic solver) together with a saturation rule \citep{staebler2016,casati2009,stephens2021,dudding2022}.   
We note, however, that these reduced models of turbulent transport have mainly been validated in electrostatic turbulence regimes~\citep{mariani2018,rodriguez2019,luda2021,marin2021,casson2020,ho2023,angioni2022} and their application to high-$\beta$ spherical tokamaks, where electromagnetic turbulence is expected to be important~\citep{kennedy2023a,giacomin2024}, needs to be carefully assessed~\citep{patelthesis}.

Various flat-top operating points of the Spherical Tokamak for Energy Production (STEP)~\citep{wilson2020,meyer2022} have been developed using the integrated modelling suite JINTRAC~\citep{romanelli2014} and the JETTO transport module with a simple Bohm Gyro-Bohm transport modelling tuned to a confinement scaling~\citep{meyer2022,tholerus2024}.
A gyrokinetic linear analysis of one of the most promising STEP flat-top operating points, STEP-EC-HD \cite{tholerus2024}, is reported in \citet{kennedy2023a}.  This reveals dominant microinstabilities at the ion gyroscale, that share many properties of the Kinetic Ballooning Mode (KBM). This dominant mode, labelled hybrid-KBM in \citet{kennedy2023a}, is unstable over a wide radial range of the plasma, and connects to Ion Temperature Gradient (ITG) and Trapped Electron Mode (TEM) instabilities at low $\beta$.
A subsequent nonlinear gyrokinetic investigation, detailed in  \citet{giacomin2024}, shows that, if equilibrium flow shear is neglected, turbulent transport from hybrid-KBMs significantly exceeds the available sources at mid-radius in STEP-EC-HD, thus highlighting the importance of accounting for hybrid-KBM turbulence in developing consistent flat-top operating points for STEP.
The saturation level of hybrid-KBM turbulence far above threshold is, however, very sensitive to equilibrium flow shear (the sensitivity to flow shear is decreased near threshold, as shown in this work) and, at moderate shearing rates, transport fluxes reduce significantly towards values consistent with the assumed sources in STEP.
The work carried out by \citet{giacomin2024} also shows that the heat flux depends strongly on the parameter $\beta'=\partial \beta/\partial \rho$ (where $\beta = 2\mu_0 p/B^2$, $p$ is the total pressure, $B$ is the magnetic field and $\rho$ is the radial coordinate).
As $\beta'$ is increased above its nominal value, hybrid-KBMs get stabilised and more slowly growing Microtearing Modes (MTMs), which are more resilient, take over as the dominant modes.  On the STEP surface analysed in \citet{giacomin2024}, MTMs drive a significantly lower level of turbulent transport that is more compatible with the assumed sources.  A reduced transport model for electromagnetic hybrid-KBM turbulence is clearly required to design flat-top operating points in STEP and to simulate access to them~\citep{mitchell2023}. 

High-fidelity tools to evolve plasma equilibrium profiles on the transport time scale have been developed in recent years by coupling transport solvers to local or global gyrokinetic codes. These codes, such as \textcolor{myred}{TANGO-GENE}~\citep{disiena2022}, \textcolor{myred}{PORTALS}-CGYRO~\citep{rodriguez2022} and TRINITY~\citep{barnes2010}, exploit the enormous separation between the timescales associated with turbulence and equilibrium transport, to evolve plasma profiles self-consistently with the turbulence.
\textcolor{myred}{For instance, PORTALS-CGYRO has been recently applied to perform the first flux-matched nonlinear gyrokinetic predictions of SPARC profiles, where Gaussian process regression techniques have been applied to optimize the transport solver~\citep{rodriguez2022}.}   
Naturally high-fidelity models require considerably more computational resources than using reduced transport models\footnote{This is true even when Gaussian process regression techniques are used to reduce the number of solver iterations~\citep{rodriguez2022}.}; and their application to STEP is computationally prohibitively expensive, because each of the many nonlinear local simulations required by a transport solver costs more than $10^5$ core-hours. Furthermore, starting from an initial state that strongly violates power balance increases significantly the number of nonlinear gyrokinetic simulations required for profile convergence (power balance is very poorly satisfied at several radial locations in STEP-EC-HD as assumed in scenario design with JINTRAC, which provides the initial condition in our flux-driven transport calculations).
\textcolor{myred}{There is therefore an increasing interest in developing low-cost and high-fidelity transport solvers that can be exploited in integrated modeling, such as the fast and accurate turbulent transport model recently developed by \citet{citrin2023}, where the saturation rule, calibrated to a set of nonlinear GENE simulations, is combined with a neural network regression to perform flux-matched predictions of the ITER baseline scenario.}

In this work, we develop a quasi-linear reduced transport model from a database of nonlinear local gyrokinetic simulations of hybrid-KBM turbulence for a range of STEP local equilibria. The reduced transport model is based on a quasi-linear metric that combines a saturation rule and quasi-linear flux weights evaluated from linear gyrokinetic calculations. The functional dependence of the heat flux on this quasi-linear metric is developed from a fit to the database of nonlinear simulations. 
This reduced transport model is implemented in T3D (or Trinity3D)~\citep{t3d}, a new Python version of TRINITY \citep{barnes2010} which is used in this paper to perform the first flux-driven simulations to model transport from hybrid-KBM turbulence in STEP.

The paper is organized as follows. Sec.~\ref{sec:overview} presents a brief overview of linear and nonlinear simulations from various radial locations in a STEP flat-top operating point. The reduced turbulent transport model, obtained from a fit to the nonlinear gyrokinetic database, is presented in Sec.~\ref{sec:ql_metric}. Sec.~\ref{sec:t3d} describes the implementation of the reduced model in T3D and its exploitation for the first predictive transport simulations to model hybrid-KBM turbulence in STEP. The robustness of the reduced model is assessed in Sec.~\ref{sec:analysis}, by comparing the reduced model transport fluxes against a new set of nonlinear gyrokinetic simulations for different surfaces of the new T3D transport steady state.  Finally the main conclusions are presented in Sec.~\ref{sec:conclusion}. 

 \section{Overview of STEP local gyrokinetic simulations}
\label{sec:overview}

This section gives a brief overview of the linear and nonlinear gyrokinetic analysis of the STEP-EC-HD flat-top operating point reported in \cite{kennedy2023a} and \cite{giacomin2024}, and motivates the need for a new reduced model of transport from electromagnetic turbulence in STEP-like regimes. 

Table~\ref{tab:equilibrium} gives local equilibrium parameters from six radial surfaces (with low order rational values of the safety factor $q$)  in STEP-EC-HD~\citep{meyer2022,tholerus2024} \footnote{This STEP-EC-HD equilibrium is at SimDB UUID:2bb77572-d832-11ec-b2e3-679f5f37cafe, alias: smars/jetto/step/88888/apr2922/seq-1.  Local quantities were obtained from the global STEP-EC-HD equilibrium using the \texttt{pyrokinetics} Python library~\citep{pyrokinetics}.}: $\Psi_n$ represents the normalized poloidal magnetic flux; $q$ the safety factor; $\hat{s}$ the magnetic shear; $\rho$ is a normalised flux label (half the flux-surface mid-plane diameter normalized to the minor radius $a$); $\kappa$ and $\kappa'$ the elongation and its radial derivative; $\delta$ and $\delta'$ the triangularity and its radial derivative; and $\Delta'$ the radial derivative of the Shafranov shift; local value of $\beta_e = 2\mu_0 p_e/B_0^2$, and its radial derivative $\beta'$; $\rho_* = \rho_s/a$  where $\rho_s = c_s/\Omega_D$ is the local ion sound Larmor radius, with $c_s = \sqrt{T_e/m_D}$, $\Omega_D = e B_0/m_D$, $T_e$ the local electron temperature, $m_D$ the deuterium mass and $B_0$ the toroidal magnetic field at the central major radius on the surface; the normalised logarithmic radial derivatives of the electron and deuterium density, $a/L_{n_s} \equiv (1/n_s)\partial n_s/\partial \rho$, and temperature, $a/L_{T_s}\equiv (1/T_s)\partial T_s/\partial \rho$; $n_e$ the value of the electron density (the deuterium and tritium density are equal, i.e., $n_D=n_T=n_e/2$); $T_e$ and $T_D$ the electron and deuterium temperature (the tritium and deuterium temperatures are equal); and the target values of the heat and particle flux computed in JETTO from the expected heat and particle sources of STEP-EC-HD.

\begin{table}
    \centering
    \begin{tabular}{lcccccc}
    \toprule
      \multicolumn{7}{c}{\textbf{Local parameters of STEP-EC-HD}}\\
    \midrule
    $\Psi_n$    & 0.15  & 0.36 & 0.49 & 0.58 & 0.71 & 0.80 \\
    $q$         & 2.5  & 3.0   & 3.5  & 4.0  &  5.0 & 6.0\\
    $\hat{s}$   & 0.3 & 0.6 &   1.2  & 1.6  &  2.2 & 3.2\\
    $\rho$    & 0.34 & 0.54  & 0.64 & 0.70 & 0.80 & 0.85\\
    $\kappa'$   &-0.09 & -0.09 & 0.06 & 0.19 & 0.49 & 0.63\\
    $\delta$    & 0.16 & 0.23  & 0.29 & 0.32 & 0.40 & 0.45\\ 
    $\delta'$   & 0.44 & 0.36  & 0.46 & 0.54 & 0.70 & 1.05 \\
    $\Delta'$   &-0.23 & -0.34 &-0.40 &-0.44 &-0.49 &-0.53\\
    $\beta_e$     & 0.17 & 0.12& 0.09 & 0.07 & 0.05 & 0.03\\
    $\beta'$    &-0.38 & -0.45 & -0.48&-0.47 &-0.44 & -0.40\\
    $\rho_*$   &0.0033 &0.0028 &0.0026& 0.0024&0.0022& 0.0020\\ 
    $a/L_{n_e}$ & 0.09 & 0.45  & 1.06 & 1.54 & 2.58 & 3.43\\
    $a/L_{T_e}$ & 0.92 & 1.32  & 1.58 & 1.77 & 2.15 & 2.58\\
    $a/L_{n_D}$ & 0.09 & 0.45  & 1.06 & 1.54 & 2.58 & 3.43\\
    $a/L_{T_D}$ & 1.19 & 1.67  & 1.82 & 1.96 & 2.41 & 3.28\\
    $n_e$ [10$^{20}$ m$^{-3}$] & 2.0 & 1.9   & 1.8 & 1.7 & 1.4 & 1.2\\
    $T_e$ [keV]                & 15.6 & 11.8 & 10.3& 9.2 & 7.7 & 6.8\\
    $T_D$ [keV]                & 17.8 & 12.5 & 10.6& 9.4 & 7.7 & 6.8\\
    Area [m$^2$] & 217 & 319   & 362  & 389 &  435 & 456 \\
    $Q_\mathrm{JETTO}$ [MW/m$^2$] & 0.8 & 0.9 & 0.8 & 0.7 & 0.6 & 0.5 \\
    $\Gamma_\mathrm{JETTO}$ [$10^{20}$m$^{-2}$s$^{-1}$] & 0.01 & 0.06 & 0.11 & 0.16 & 0.21 & 0.22 \\
    \bottomrule
    \end{tabular}
    \caption{Local parameters on low order rational surfaces of STEP-EC-HD. The deuterium and tritium density are equal to half the electron density. The deuterium and tritium temperature are equal. All parameters are defined in the main text. }
    \label{tab:equilibrium}
\end{table}

\subsection{Linear simulation results}
\label{sec:linear}

Faithfully capturing the linear physics will be critical to the quasi-linear model, so here we briefly revisit the linear analysis of STEP-EC-HD, detailed in \cite{kennedy2023a}. 
Linear local gyrokinetic simulations were carried out on six radial surfaces using the GS2 code~\citep{dorland2000,gs2}. These calculations included three plasma species (electron, deuterium and tritium) and used the following numerical resolutions: extended parallel domain in ballooning angle spanning $-7\pi <\theta< 7 \pi$ ($n_\mathrm{period}=4$), 65 parallel grid points in each [$-\pi$, $\pi$] interval ($n_\theta=65$), 40 pitch-angles ($n_\lambda = 40$), and 12 energy grid points ($n_\epsilon = 12$). The extended ballooning domain is increased to $n_\mathrm{period}=32$ on the innermost surface at $\Psi_n=0.15$ in order to  resolve more extended MTMs found on this surface. 
All calculations are fully electromagnetic, including $\delta\phi$, $\delta A_\parallel$ and $\delta B_\parallel$ \footnote{See \cite{kennedy2024} for a detailed discussion on the importance of $\delta B_\parallel$ in STEP.}, and were carried at $\theta_0=0$ unless otherwise specified.

\begin{figure}
    \centering
    \subfloat[]{\includegraphics[height=0.24\textheight]{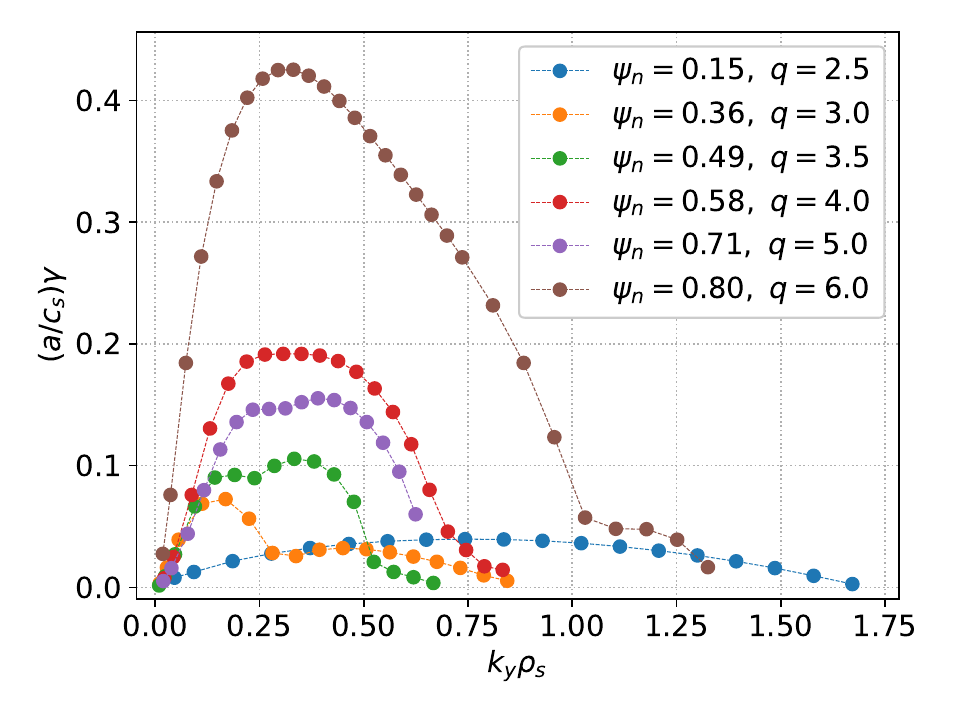}}
    \subfloat[]{\includegraphics[height=0.24\textheight]{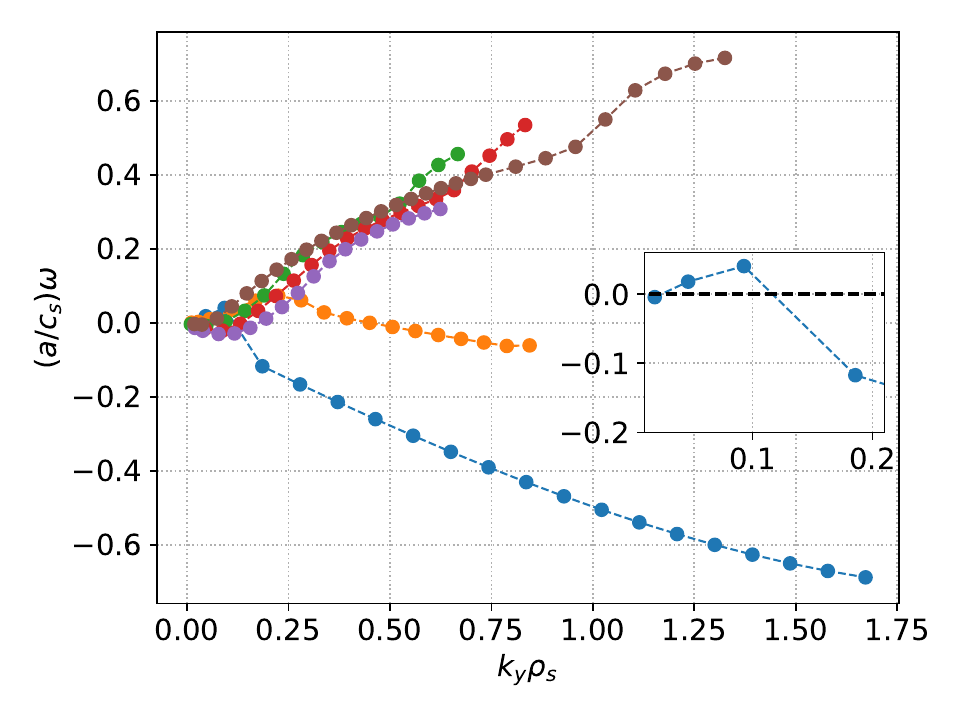}}
    \caption{Growth rate (a) and mode frequency (b) as functions of the binormal wave vector $k_y$ at various radial surfaces of STEP-EC-HD. \textcolor{myred}{ The inset in (b) shows the transition to positive frequency values at low $k_y$ of $\Psi_n = 0.15$.} }
    \label{fig:linear}
\end{figure}

Fig.~\ref{fig:linear} shows the growth rate and the mode frequency as functions of the binormal wave vector $k_y\rho_s$ on the six surfaces. No unstable modes are found at $k_y\rho_s>1.8$ on any of the surfaces considered here. The maximum normalised local growth rate generally increases from the innermost to the outermost radial surface, and a sign change in the dominant mode frequency from mostly negative across most of the spectrum at $\Psi_n=0.15$ to mostly positive at $\Psi_n=0.36$, clearly indicates an instability transition going outwards from the core. 

\begin{figure}
    \centering
    \subfloat[]{\includegraphics[height=0.23\textheight]{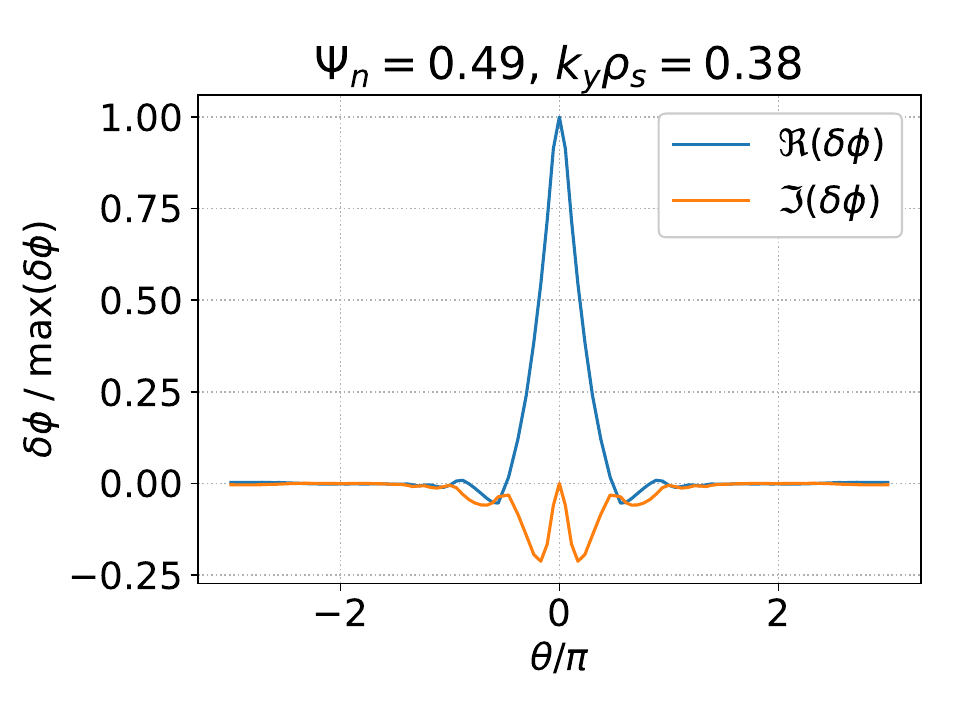}}\quad
    \subfloat[]{\includegraphics[height=0.23\textheight]{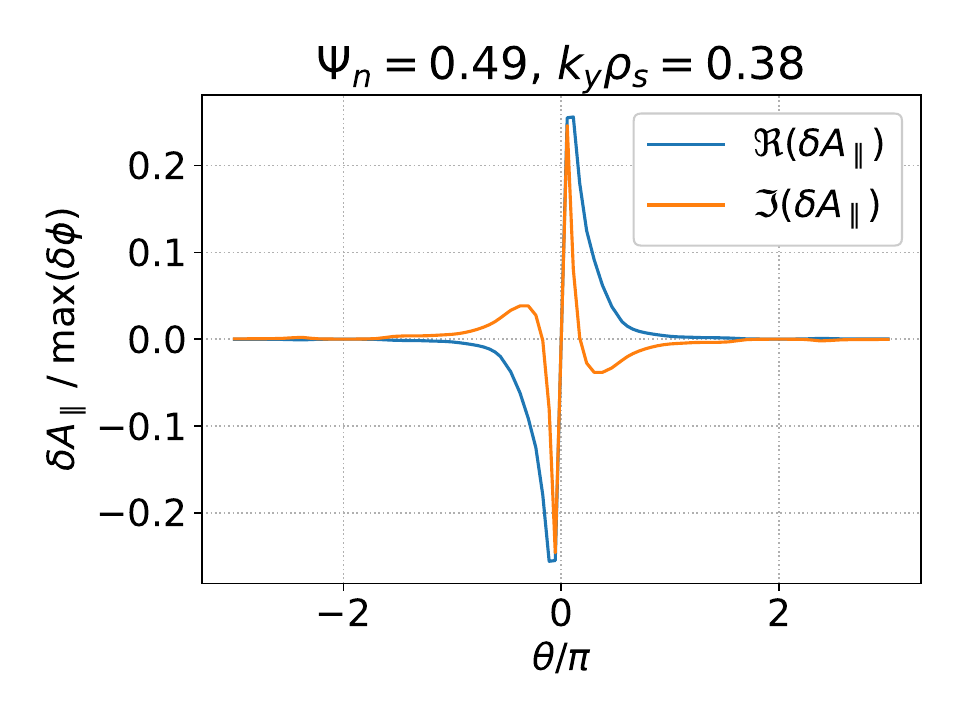}}
    \caption{Parallel mode structure of $\delta\phi$ (a) and $\delta A_\parallel$ (b) both normalized to $\max(\delta\phi)$ for the most unstable mode with $k_y\rho_s \simeq 0.4$. Figure adapted from \cite{kennedy2023a}.}
    \label{fig:psin049_mode}
\end{figure}

The dominant mode at $\Psi_n=0.49$ has been labelled as a hybrid-KBM because it shares many properties with KBM and exhibits ITG and TEM contributions to the linear drive \cite{kennedy2023a}. Fig.~\ref{fig:psin049_mode} shows the parallel mode structure of $\delta \phi$ and $\delta A_\parallel$ of the dominant hybrid-KBM at $k_y\rho_s\simeq 0.4$ on $\Psi_n=0.49$. 

\begin{figure}
    \centering
    \subfloat[]{\includegraphics[height=0.23\textheight]{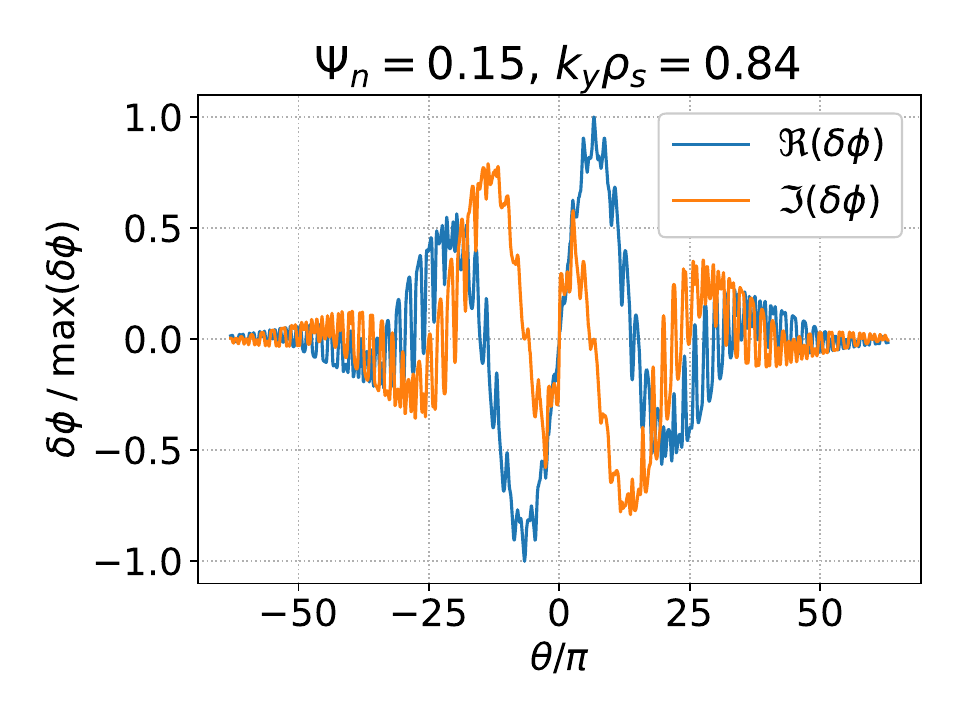}}\quad
    \subfloat[]{\includegraphics[height=0.23\textheight]{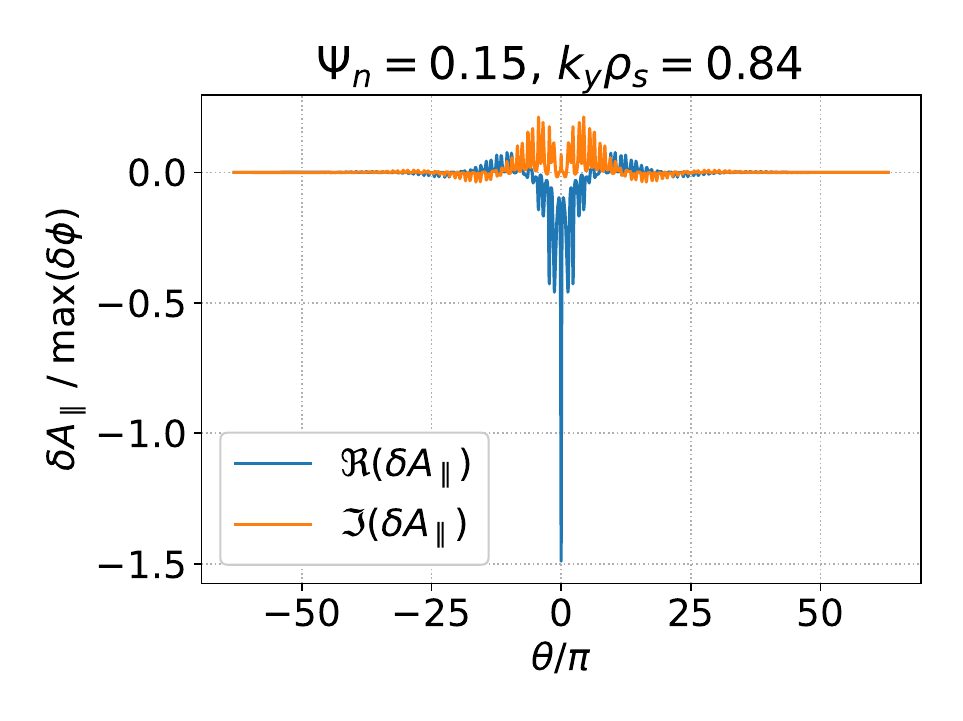}}\\
    \subfloat[]{\includegraphics[height=0.23\textheight]{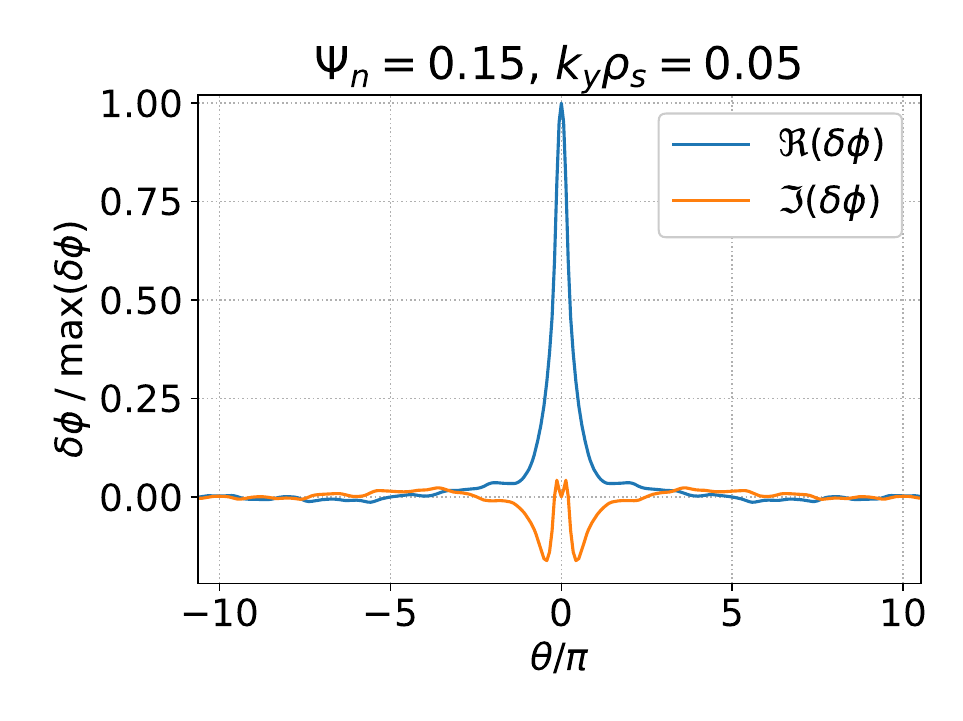}}\quad
    \subfloat[]{\includegraphics[height=0.23\textheight]{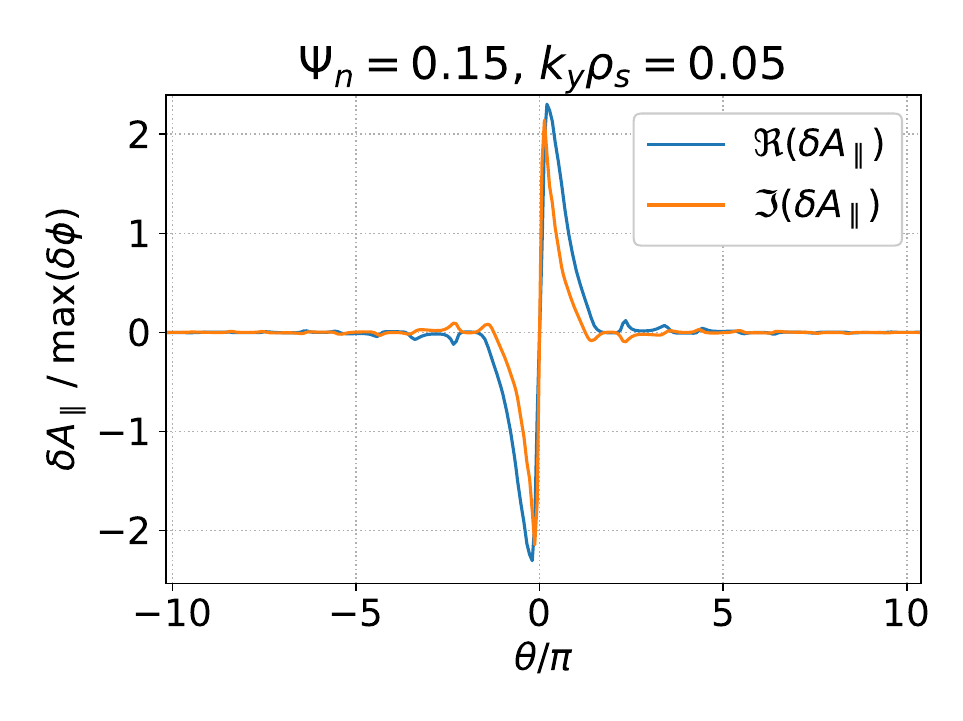}}
    \caption{Parallel mode structure of $\delta\phi$ [(a) and (c)] and $\delta A_\parallel$ [(b) and (d)] both normalised to $\max(\delta\phi)$ for the unstable mode with $k_y\rho_s \simeq 0.84$ (top row) and $k_y\rho_s \simeq 0.05$ (bottom row) at $\Psi_n=0.15$.}
    \label{fig:psin015_mode}
\end{figure}

On the innermost surface at $\Psi_n=0.15$, however, the dominant modes have negative frequency and very different parallel mode structures.  These are illustrated in Fig.~\ref{fig:psin015_mode}~[(a) and (b)] for the fastest growing mode at $k_y\rho_s \simeq 0.84$: $\delta \phi$ has odd parity and is highly extended in $\theta$ whilst $\delta A_\parallel$ is very localised at $\theta=0$ with even parity. Amplitudes of $e\delta \phi/T_e$ and $\delta A_\parallel/(\rho_s B_0)$ are comparable, indicating that the mode is strongly electromagnetic. These negative frequency modes propagate in the electron diamagnetic drift direction, and have the typical properties expected of MTMs~\citep{applegate2004}. On the same surface at the lowest $k_y$, on the other hand, the dominant mode switches to a positive frequency \textcolor{myred}{(see the inset in Fig.~\ref{fig:linear}~(b))} and very different parallel mode structures in $\delta\phi$ and $\delta A_\parallel$.  The parallel mode structure of a low $k_y$ mode with positive frequency is shown in  Fig.~\ref{fig:psin015_mode}~[(c) and (d)]: the parallel structure closely resemble Fig.~\ref{fig:psin049_mode}, indicating that the hybrid-KBM is linearly dominant at low $k_y$. Although these low $k_y$ hybrid-KBMs are only weakly unstable, they nevertheless contribute significantly to the turbulent transport, as will be  discussed in Sec.~\ref{sec:nonlinear}.

\begin{figure}
    \centering
    \subfloat[$\Psi_n=0.15$]{\includegraphics[height=0.23\textheight]{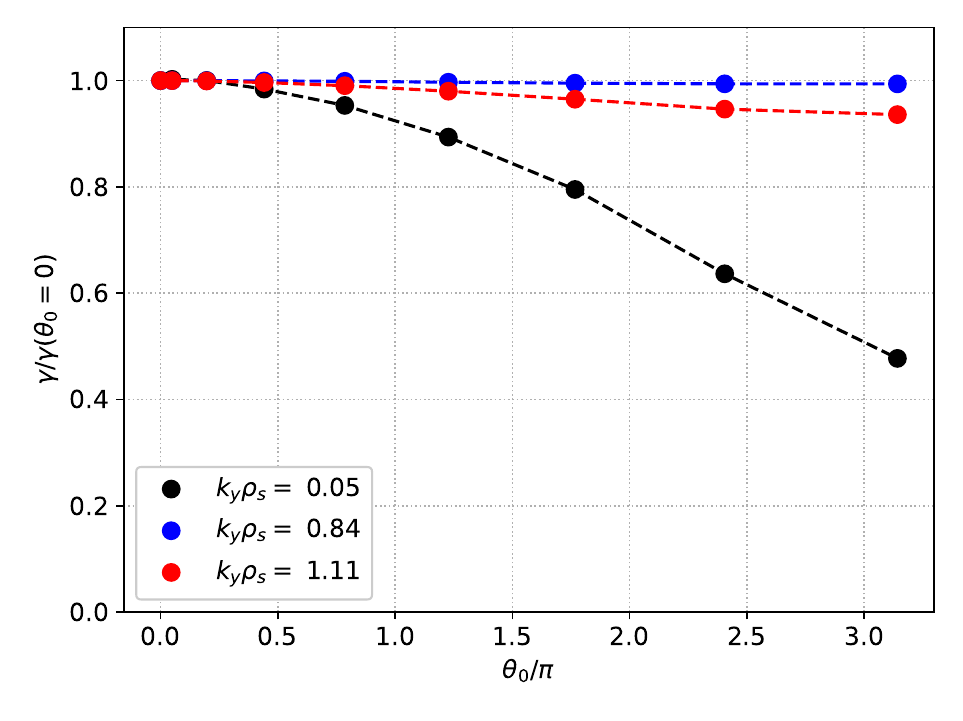}}\quad
    \subfloat[$\Psi_n=0.8$]{\includegraphics[height=0.23\textheight]{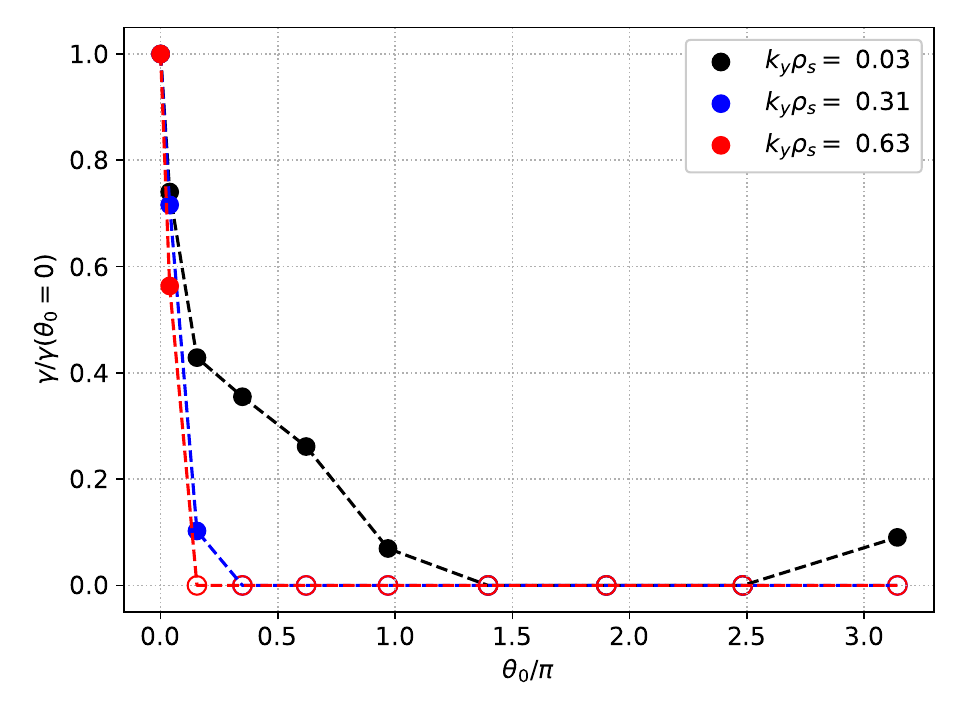}}
    \caption{Dependence of the growth rate on $\theta_0$ at three different values of $k_y\rho_s$ on the surfaces at $\Psi_n=0.15$ (a) and $\Psi_n = 0.8$ (b). Stable modes are shown with $\gamma = 0$ and open markers.}
    \label{fig:theta0}
\end{figure}

The analysis of \cite{kennedy2023a} showed that hybrid-KBM is highly ballooned with growth rates that are extremely sensitive to the ballooning parameter, $\theta_0$.  This suggests that hybrid-KBM turbulence should be sensitive to equilibrium flow shear, as was subsequently confirmed in \cite{giacomin2024}. Fig.~\ref{fig:theta0}~(b) shows the dependence on $\theta_0$ of growth rates on the outermost surface at $\Psi_n=0.8$, where hybrid-KBMs similar to the modes at $\Psi_n=0.49$ dominate: the hybrid-KBM growth rates are strongly suppressed as $\theta_0$ increases from zero (the slowly growing unstable mode on this surface at $\theta_0=\pi$ and $k_y\rho_s \simeq 0.03$ is a MTM).  
Fig.~\ref{fig:theta0}~(a), however, shows that the situation is quite different on the innermost surface at $\Psi_n=0.15$:  here the MTM growth rates at $k_y\rho_s \simeq 0.8$ and $k_y\rho_s \simeq 1.3$ are insensitive to $\theta_0$ \footnote{See \cite{hardman2023} and \cite{patel2023a} for a detailed discussion of the relationship between $\gamma^{\rm MTM}(\theta_0)$ and the sensitivity of MTM turbulence to equilibrium flow shear.}, though 
in contrast the dominant hybrid-KBM at \textcolor{myred}{$k_y\rho_s\simeq 0.05$} has a growth rate that decreases with $\theta_0$ albeit remaining unstable across the entire $\theta_0$ domain.

\subsection{Nonlinear simulations results}
\label{sec:nonlinear}

This section presents a brief overview of local nonlinear gyrokinetic simulations carried out on three radial surfaces, $\Psi_n\in\{0.15,\;0.36,\; 0.8\}$, of STEP-EC-HD. This analysis extends the investigation carried out at $\Psi_n=0.49$ by \cite{giacomin2024} providing further nonlinear simulations to be used in the development of a reduced turbulent transport model to account for hybrid-KBMs. The nonlinear gyrokinetic simulations are carried out with the GENE code~\citep{gene} using the advanced Sugama collision operator~\citep{sugama2009}. The numerical resolutions used in these simulations is reported in table~\ref{tab:nl_res}. We highlight the higher radial resolution required by MTMs at $\Psi_n=0.15$. Simulations are performed without and with equilibrium flow shear. All the simulations are performed over a time interval that covers at least a growth time of the slowest growing low $k_y$ mode of each surface, corresponding to several growth times of the most unstable mode.

\begin{table}
    \centering
    \begin{tabular}{cccccccc}
    \toprule
    $\boldsymbol{\Psi_n}$ & $\boldsymbol{\Delta k_y\rho_s}$ & $\boldsymbol{k_{y,\mathrm{max}}\rho_s}$ & $\boldsymbol{\Delta k_x\rho_s}$ & $\boldsymbol{k_{x,\mathrm{max}}\rho_s}$ &  $\boldsymbol{n_\theta}$ & $\boldsymbol{n_\mu}$ & $\boldsymbol{n_v}$ \\
    \midrule
    0.15 & 0.02 & 1.28 & 0.02 & 2.56 & 32 & 16 & 32\\
    0.36 & 0.016 & 1.02 & 0.03 & 0.96 & 32 & 16 & 32\\
    0.8 & 0.02 & 1.28 & 0.025 & 1.6 & 32 & 16 & 32\\
    \bottomrule
    \end{tabular}
    \caption{Numerical resolution of the nonlinear GENE simulations at the different radial locations. In the table, $n_\mu$ is number of grid points in the $\mu=v_{\perp}^2/(2B)$ direction and $n_v$ is  number of velocity grid points.}
    \label{tab:nl_res}
\end{table}

\begin{figure}
    \centering
    \subfloat[]{\includegraphics[width=0.49\textwidth]{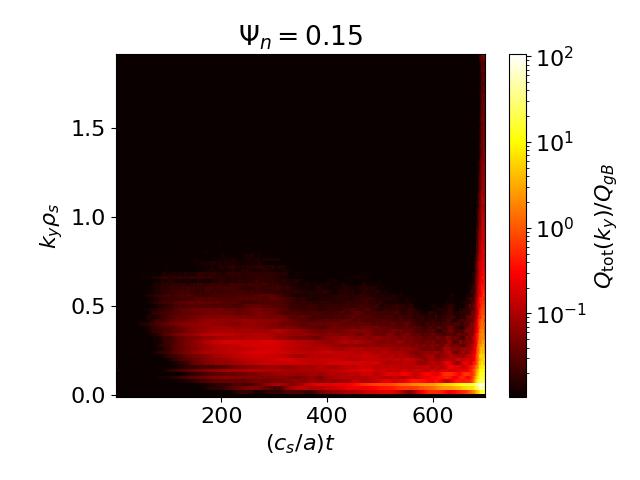}}
    \subfloat[]{\includegraphics[width=0.49\textwidth]{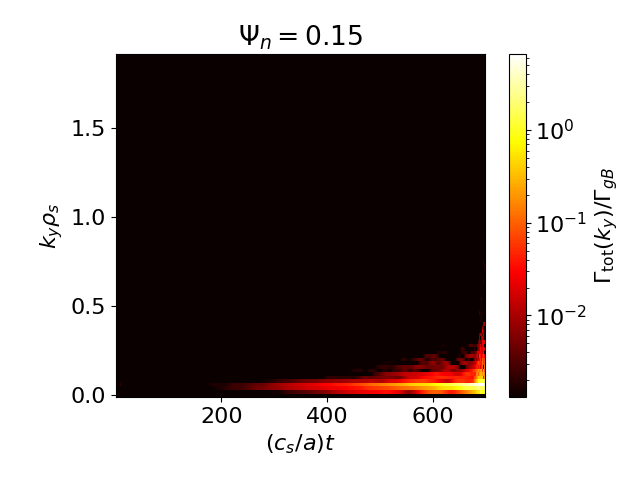}}
    \caption{Total heat (a) and particle (b) flux as functions of  $k_y$ and time from the nonlinear simulation at $\Psi_n=0.15$. The equilibrium flow shear is not included.}
    \label{fig:psin015_nl}
\end{figure}

\vspace*{2mm}
\noindent
{\em Innermost surface dominated linearly by MTMs ($\Psi_n=0.15$):}\\
Fig.~\ref{fig:psin015_nl} shows the time trace of the total heat and particle flux $k_y$ spectrum, normalized to $Q_\mathrm{gB} = \rho_*^2 n_e T_e c_s$ and  $\Gamma_\mathrm{gB} = \rho_*^2 n_e c_s$, from a nonlinear simulation at $\Psi_n=0.15$ without equilibrium flow shear. 
The heat flux contribution from $k_y\rho_s \sim 0.25$ increases until $t\simeq 250\;a/c_s$ where it reaches values of the order of $Q_\mathrm{gB}$, and then remains approximately constant until $t\simeq 600\;a/c_s$.  
The particle flux contributions from $k_y\rho_s > 0.2$ are negligible, as expected from MTM driven turbulent transport in this $k_y$ range on this surface (see, e.g., \citet{doerk2011,guttenfelder2011};\citet{hatch2016} and \citet{giacomin2023a} for details on MTM turbulence).
On the other hand, the heat and particle flux contributions from $k_y\rho_s < 0.2$ steadily grow (the total heat flux grows to exceed 100~$Q_\mathrm{gB}$) and show no signs of saturation within the simulated time interval. 
Fig.~\ref{fig:psin015_nl} suggests that the large turbulent fluxes at low $k_y$ are driven by the hybrid-KBMs that are linearly unstable at the lowest binormal wavenumbers in Fig.~\ref{fig:linear}.

\begin{figure}
    \centering
    \subfloat[]{\includegraphics[width=0.48\textwidth]{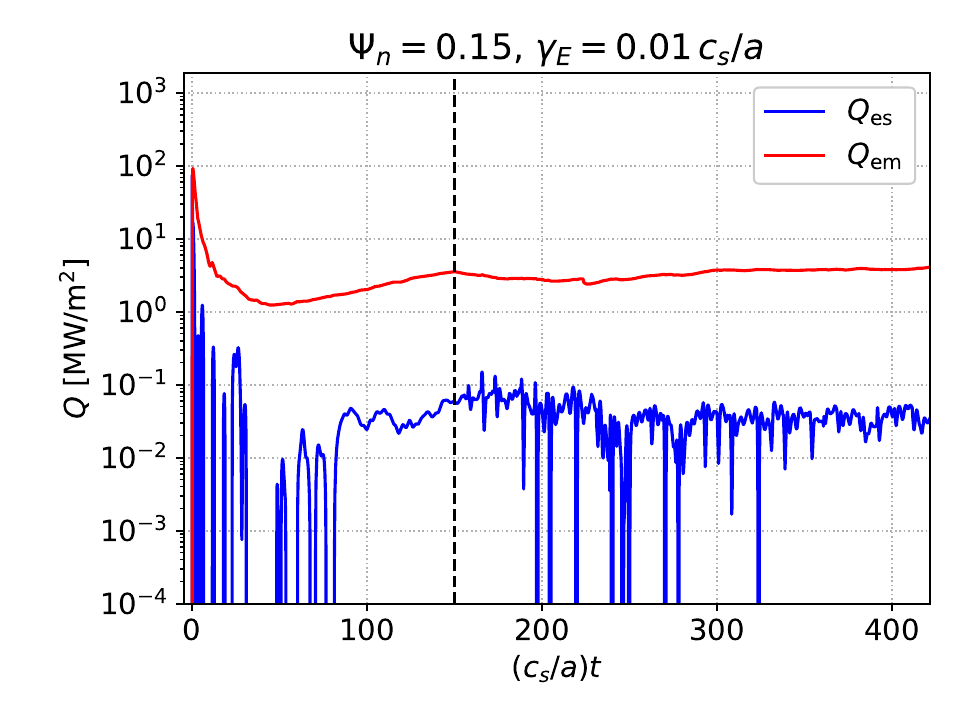}}\quad
    \subfloat[]{\includegraphics[width=0.48\textwidth]{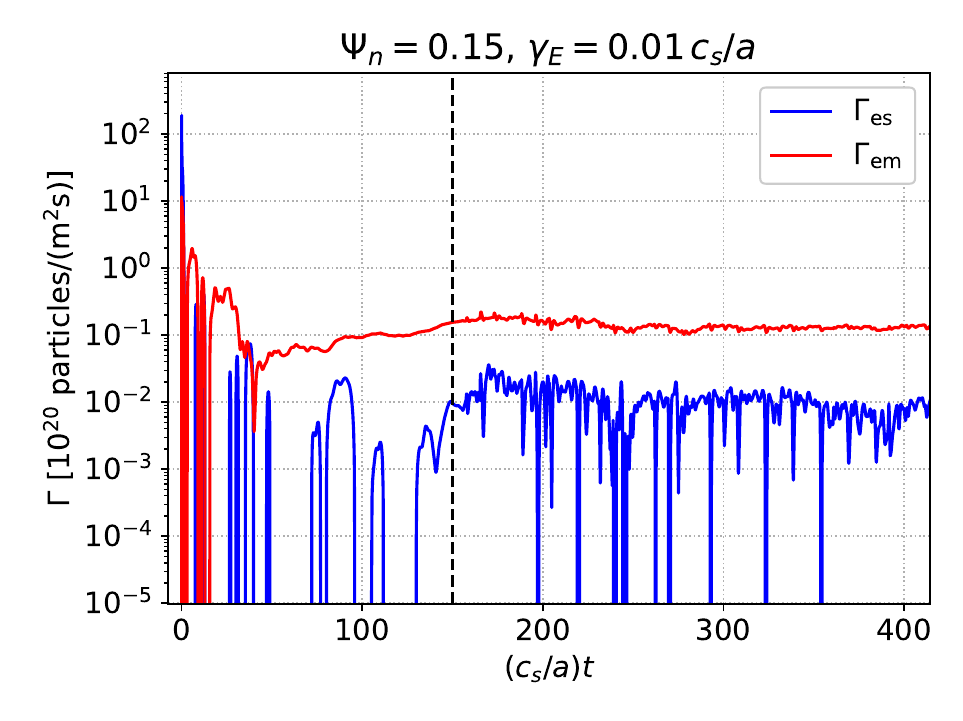}}
    \caption{Time trace of the electromagnetic and electrostatic heat (a) and particle (b) fluxes from the nonlinear simulation at $\Psi_n=0.15$ with $\gamma_E=0.01\,c_s/a$. The equilibrium flow shear is active from $t=150\,a/c_s$, as indicated by the dashed vertical line.}
    \label{fig:psin015_exb}
\end{figure}

\begin{figure}
    \centering
    \subfloat[]{\includegraphics[width=0.48\textwidth]{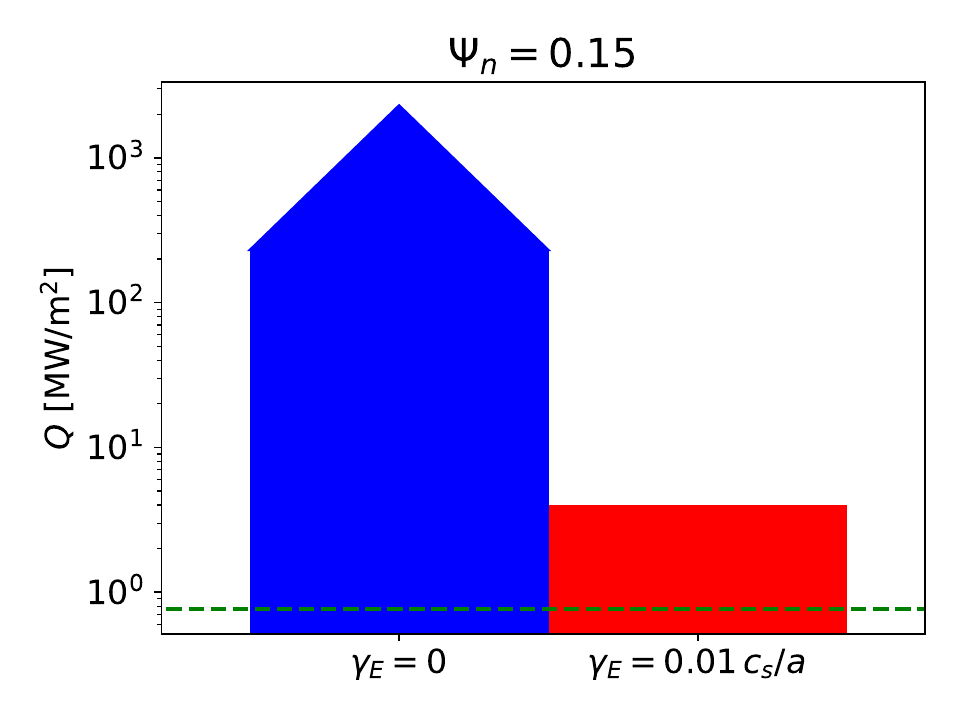}}\quad
    \subfloat[]{\includegraphics[width=0.48\textwidth]{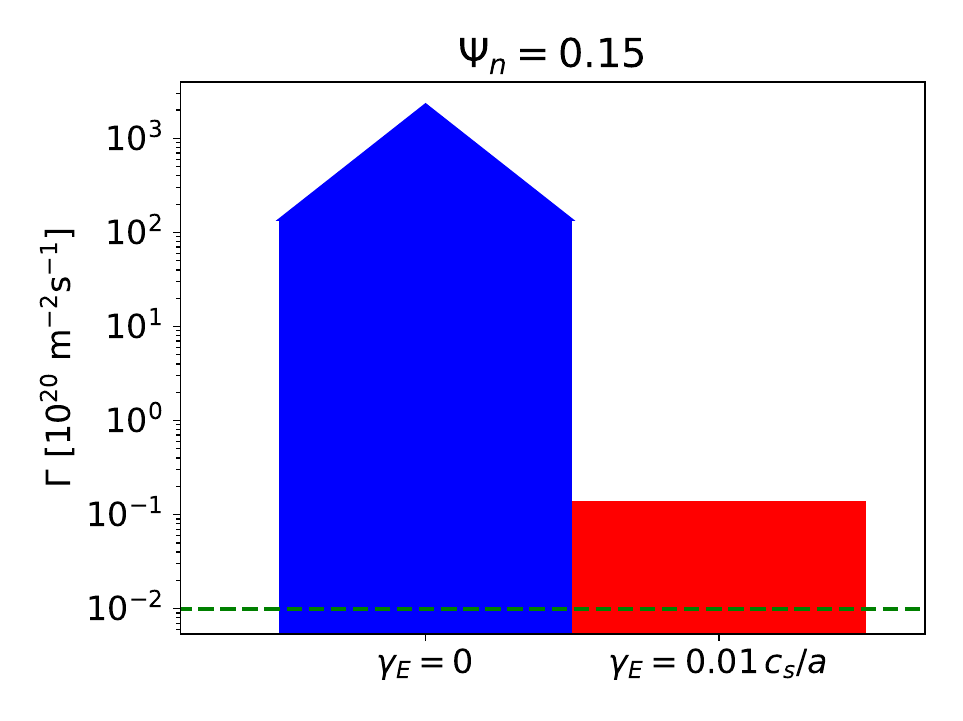}}
    \caption{Total heat (a) and particle (b) fluxes from nonlinear simulations at $\Psi_n=0.15$ with (red) and without (blue) equilibrium flow shear. The heat and particle fluxes in the simulation without equilibrium flow shear exceed the value of $10^2$~MW/m$^2$ and $10^{22}$~particles/(m$^2$s) with no saturation within the simulation time considered here, as indicated by the triangle on the top of the bar. The dashed horizontal line denotes the target heat and particle flux values at $\Psi_n=0.15$ computed from the JETTO heat and particle sources.}
    \label{fig:psin015_nl_bar}
\end{figure}

The sensitivity of $\gamma^{\rm KBM}(k_y \rho_s=0.05)$ to $\theta_0$ shown in Fig.~\ref{fig:theta0}~(a), together with the low hybrid-KBM growth rate, implies that equilibrium flow shear may be important here even at small shearing rates near the diamagnetic level, $\gamma_E^{\rm dia}$, where: 
\begin{equation}
\label{eqn:flow_shear}
    \gamma_E^\mathrm{dia} = \frac{\rho}{q}\frac{\mathrm{d}}{\mathrm{\rho}}\Bigl(\frac{E_r}{R B_\theta}\Bigr) = \frac{\rho}{q}\frac{\mathrm{d}}{\mathrm{\rho}}\biggl[\frac{1}{R B_\theta}\biggl(\frac{1}{n_i e}\frac{\mathrm{d}p_i}{\mathrm{d}r}+v_\phi B_\theta - v_\theta B_\phi\biggr)\biggr]\,,
\end{equation}
with $E_r$ the neoclassical radial electric field, $n_i$ and $p_i$ the thermal ion density and pressure, $v_\theta$ and $v_\phi$ the poloidal and toroidal velocities, and $B_\theta$ and $B_\phi$ the poloidal and toroidal magnetic field components.
In the JINTRAC flat-top operating point considered here, $v_\theta$ is evaluated using NCLASS \citep{nclass} while $v_\phi=0$.   
Fig.~\ref{fig:psin015_exb} shows the time trace of the electrostatic and electromagnetic heat and particle fluxes from a nonlinear simulation at $\Psi_n=0.15$ with $\gamma_E = \gamma_E^\mathrm{dia} \simeq  0.01\, c_s/a$. When the equilibrium flow shear is included at $t=150\,a/c_s$, the heat flux reduces slightly and saturates at approximately 5 MW/m$^2$. The runaway fluxes are completely avoided in the simulation with $\gamma_E =0.01\, c_s/a$.
Fig.~\ref{fig:psin015_nl_bar} shows, however, that, even with this level of flow shear, the saturated heat and particle fluxes nevertheless remain above the values required to achieve a transport steady state on this surface in STEP-EC-HD with the available heat and particle sources.  

\begin{figure}
    \centering
    \subfloat[]{\includegraphics[width=0.48\textwidth]{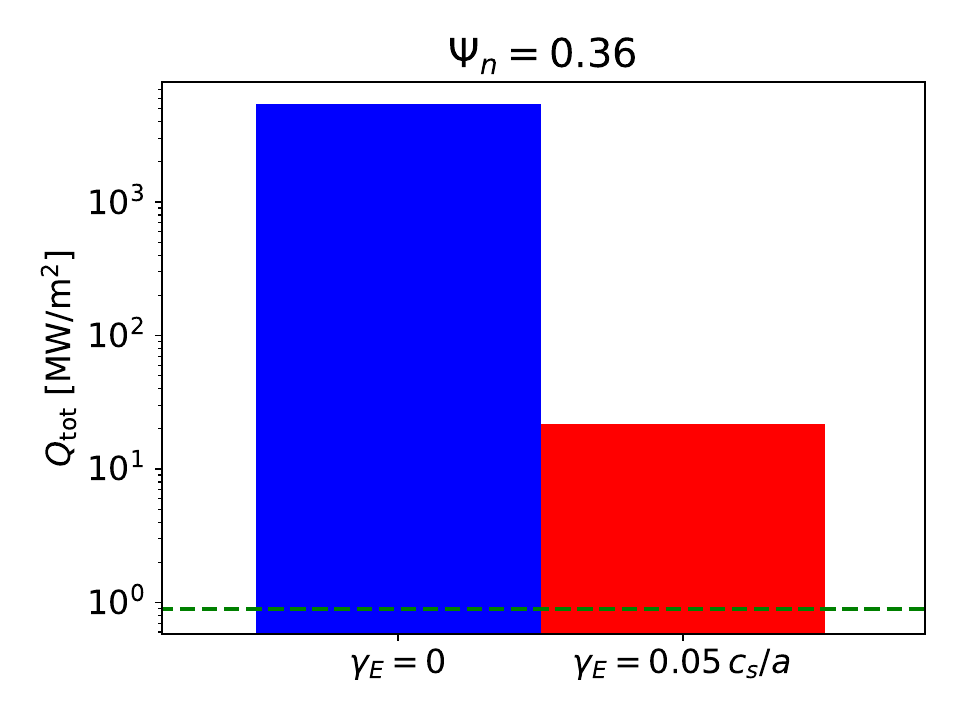}}\quad
    \subfloat[]{\includegraphics[width=0.48\textwidth]{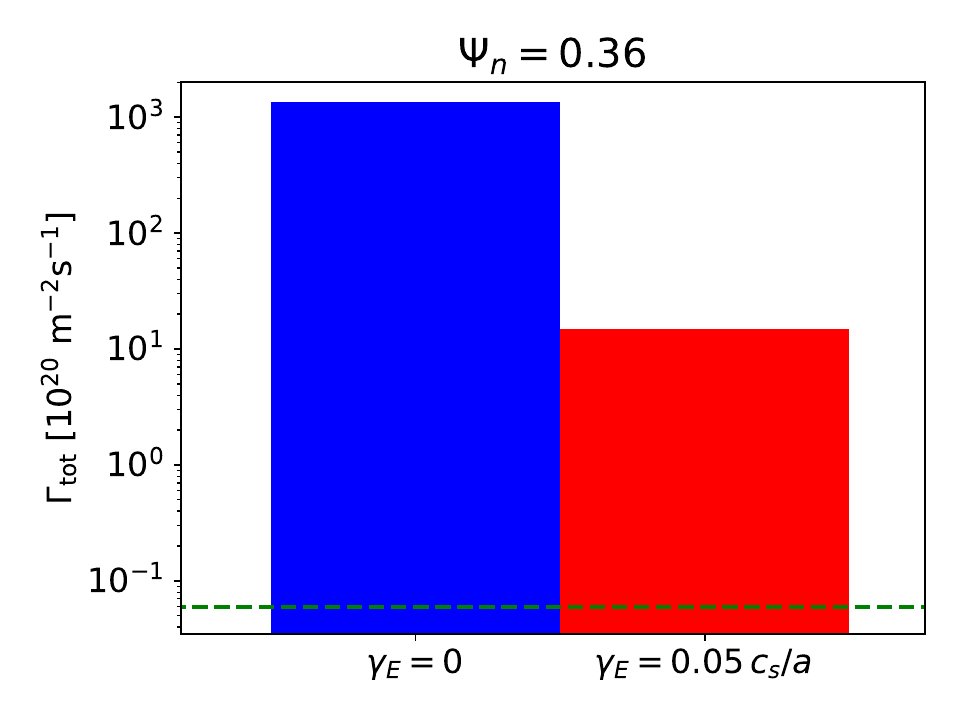}}
    \caption{Total heat (a) and particle (b) fluxes from nonlinear simulations at $\Psi_n=0.36$ with (red) and without (blue) equilibrium flow shear. The dashed horizontal line denotes the target flux value at $\Psi_n=0.36$ computed from JETTO heat and particle sources.}
    \label{fig:psin036_nl}
\end{figure}

\vspace*{2mm}
\noindent
{\em Further out radially where hybrid-KBMs dominate ($\Psi_n=0.36$):}\\ 
Fig.~\ref{fig:psin036_nl} compares the heat and particle fluxes averaged over time in the saturated phase from nonlinear simulations without and with equilibrium flow shear of the order of the diamagnetic level, $\gamma_E^\mathrm{dia}\approx 0.05\, c_s/a$. 
Although this nonlinear simulation at $\gamma_E=0$ appears to saturate within the time simulated, the corresponding heat and particle fluxes exceed STEP-EC-HD transport steady state values for this surface by more than three orders of magnitude. 
When flow shear is included, both heat and particle fluxes reduce significantly, but remain well above the transport steady state values, as shown in Fig.~\ref{fig:psin036_nl}.

\begin{figure}
    \centering
    \subfloat[]{\includegraphics[width=0.48\textwidth]{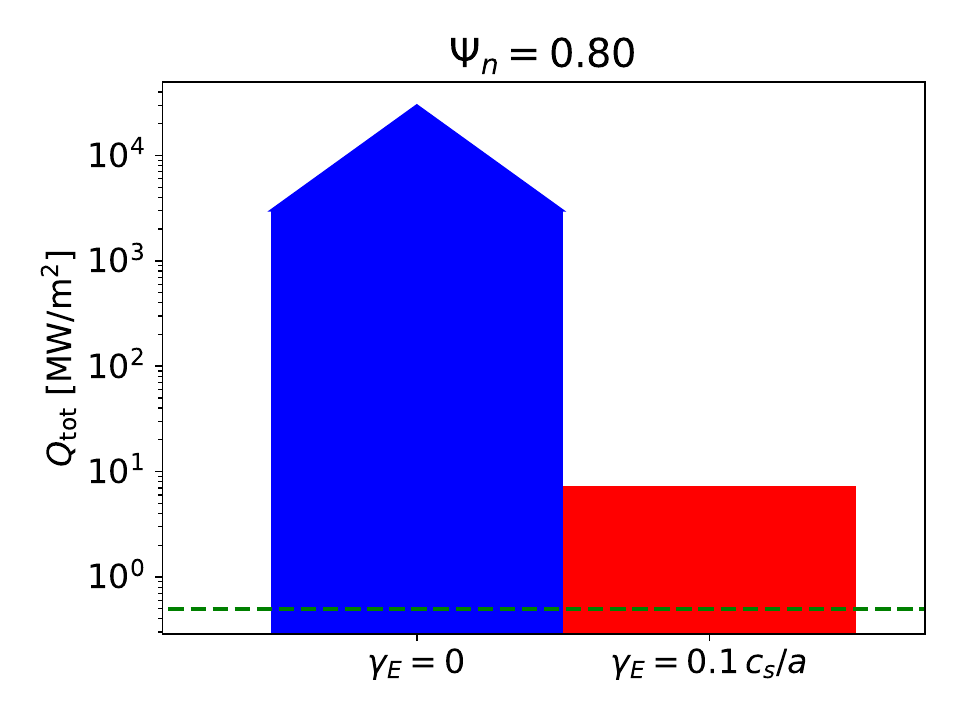}}\quad
    \subfloat[]{\includegraphics[width=0.48\textwidth]{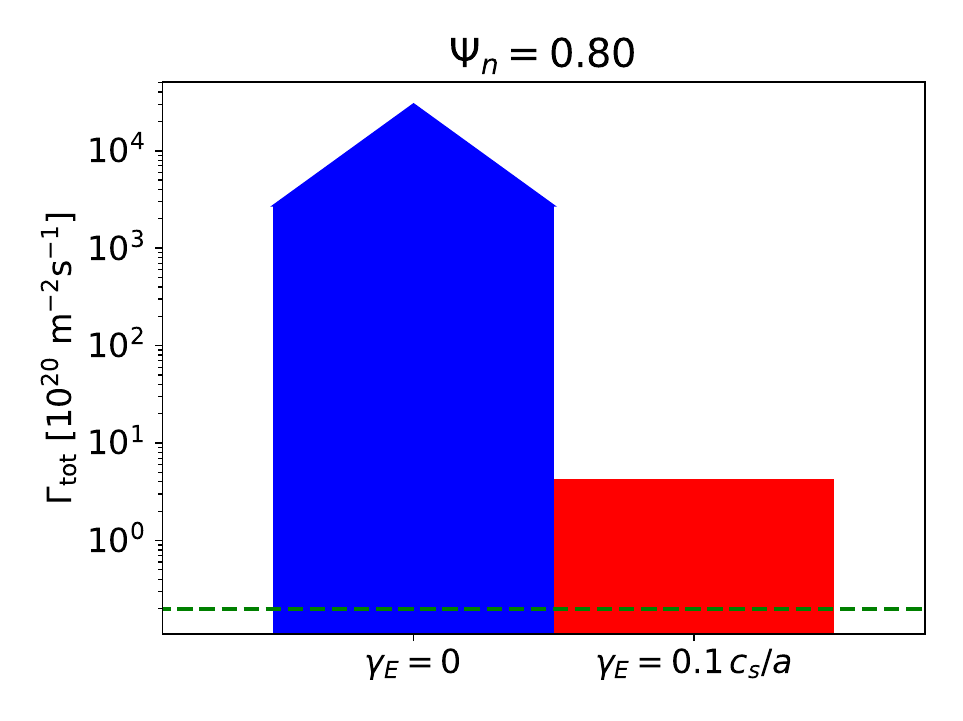}}
    \caption{Saturated value of the total heat (a) and particle (b) fluxes from nonlinear simulations at $\Psi_n=0.80$ with (red) and without (blue) equilibrium flow shear. The heat and particle fluxes in the simulation without equilibrium flow shear exceed $10^3$~MW/m$^2$ and $10^{23}$~particles/(m$^2$s) with no saturation within the simulation time considered here, as indicated by the triangular arrow at the top of the bar.
    The dashed horizontal line denotes the target flux value at $\Psi_n=0.80$ computed from the JETTO heat and particle sources.}
    \label{fig:psin080_nl}
\end{figure}

\vspace*{2mm}
\noindent
{\em The outermost flux surface, where hybrid-KBMs also dominate ($\Psi_n=0.8$):}\\ 
Nonlinear simulations deliver similar findings at $\Psi_n=0.8$. 
Fig.~\ref{fig:psin080_nl} shows the heat and particle fluxes from nonlinear simulations without and with equilibrium flow shear ($\gamma_E=0.1\,c_s/a$). The simulation without equilibrium flow shear does not saturate within the considered simulation time interval, reaching values larger than $10^3$~MW/m$^2$. The simulation with equilibrium flow shear saturates at much lower values, but turbulent fluxes remain an order of magnitude larger than the transport steady state values for STEP-EC-HD at $\Psi_n=0.8$.

\vspace*{2mm}
\noindent
{\em Summary of Section 2:}\\
This brief nonlinear survey suggests that anomalous transport would be dominated by hybrid-KBM turbulence across a wide radial region in STEP-EC-HD. In the absence of equilibrium flow shear, the associated turbulent fluxes exceed values required to reach a transport steady state in STEP-EC-HD by several orders of magnitude.  The fluxes are very sensitive to flow shear, and including a diamagnetic level of $\gamma_E$ reduces the fluxes significantly.  Nevertheless this is insufficient to reduce fluxes to the values required in a transport steady state; i.e. STEP-EC-HD is not a transport steady state. To reach such a kinetic equilibrium, the temperature and density profiles need to be evolved self-consistently with turbulent transport from hybrid-KBMs.  This clearly motivates the need for a computationally affordable reduced model to describe turbulent transport in STEP-relevant regimes.  The development of such a model is detailed in the next section.

\section{A reduced transport model for STEP}
\label{sec:ql_metric}

Transport will play an important role in governing all aspects of the plasma scenarios that are viable in STEP, including the fusion power that can be achieved in the flat-top. One of the most urgent priorities for the STEP programme is to develop robust physics-based models to predict plasma confinement.
The highest fidelity approach available would use nonlinear gyrokinetic simulations to compute the turbulent transport, but in STEP-like regimes this is prohibitively expensive computationally even using local gyrokinetics codes. A reduced core transport model is urgently required to describe turbulent transport in high-$\beta$ STEP regimes faithfully and in a computationally affordable way. 

Here we describe in detail one approach to obtain such a reduced turbulent transport model, retaining first-principles linear physics from gyrokinetic calculations, and exploiting a set of nonlinear gyrokinetic simulations in STEP relevant regimes to obtain an empirical functional relationship between the quasi-linear and the nonlinear fluxes. We stress that this model is not intended for application beyond STEP plasma conditions, although the general approach may have value for reduced model development in broader plasma regimes. Our approach has some parallel with earlier efforts to improve reduced model descriptions of magnetic flutter transport, where the goal was to better describe transport from KBMs in JET discharges (see \citep{kumar2021}).

In local $\delta f$ gyrokinetics, where the distribution function $f$ is the sum of the equilibrium distribution function $f_0$ and a perturbation $\delta f$,  the nonlinear electrostatic $\mathbf{E}\times \mathbf{B}$ drift term competes with the linear drive, i.e. $\gamma \delta f \sim (k_y \delta \phi/B) k_x \delta f$, at sufficient perturbation amplitude when  $\delta \phi/B \sim \gamma/(k_x k_y)$.
Quasi-linear models assume saturation occurs when $\partial_x p \sim k_x \delta p$, allowing the total electrostatic heat flux to be estimated as $Q^{\rm ES} \sim \delta p k_y\delta \phi/B \sim (\gamma/k_x^2) \partial_x p$.  If the turbulence is isotropic, with $k_x \sim k_y \sim k_{\perp}$, this can be written as $Q^{\rm ES} \sim (\gamma/k_\perp^2) \partial_x p$, corresponding to a turbulent thermal diffusivity of $\gamma/k_\perp^2$. 

\subsection{Quasi-linear reduced transport model} \label{sec:metric}

In the quasi-linear inspired reduced model developed here, the saturation rule describes the total heat flux\footnote{Imposing a saturation rule for the total heat flux implies saturation levels for all the fields, because these are related through the quasi-linear weights; this differs from quasi-linear models where the saturation rule is applied directly to the electrostatic potential, but is similar to the concept of turbulence intensity used to define saturation in some models \citep{kinsey2008}.}, which is modelled as\footnote{Note that the relationship between the total heat flux and $\gamma/k_\perp^2$ depends on the turbulent regime and might not be linear (see e.g. \cite{ricci2006};\cite{giacomin2020};\cite{dudding2022}). For example, if $\delta p \sim (q \delta \phi/T) p$, the electrostatic heat flux becomes $Q^{ES} \sim (\gamma^2/k_\perp^4) (qB/T) k_{\perp} p$.  Including the parameter $\alpha$ in Eq.~\eqref{eqn:redmod} allows the model to describe such a different turbulence regime.}
\begin{equation}
\label{eqn:redmod}
    Q_\mathrm{ql} = Q_0 \Lambda^\alpha\,,
\end{equation}
where the quasi-linear metric $\Lambda$ (involving $\gamma/k_\perp^2$, all fields, and describing flow shear) is at the heart of the model, and $Q_0$ and $\alpha$ are constants chosen to fit the total turbulent heat transport in the regime of interest. 
In Sec.~\ref{sec:application}, $Q_0$ and $\alpha$ will be optimised to describe the total heat flux from hybrid-KBM turbulence in STEP.  Firstly we  define $\Lambda$ and then we specify the species decomposition of the model transport fluxes.

In ballooning space local perturbed modes are labelled by bi-normal perpendicular wavenumber and ballooning parameter, $k_y$ and $\theta_0$, respectively.  For each mode (i.e. $k_y$ and $\theta_0$) and field $\chi_i \in \{e\delta\phi/T_e, \delta A_\parallel/(\rho_s B_0), \delta B_\parallel/B_0\}$, we define a perpendicular wavenumber averaged in poloidal angle $\theta$, following \cite{jenko2005},  as:
\begin{equation}
    \label{eqn:kperp}
    \langle k_\perp^2\rangle_{k_y, \theta_0, i} = \frac{\int \mathrm{d}\theta\; k_\perp^2(\theta, k_y, \theta_0) \,J(\theta) \,|\chi_i(\theta)|^2}{\int \mathrm{d}\theta\; J(\theta)|\chi_i(\theta)|^2}\,,
\end{equation}
where $k_\perp^2(\theta, k_y, \theta_0) = g^{yy}(\theta)k_y^2 + 2g^{xy}(\theta) k_x k_y + g^{xx}(\theta)k_x^2$, $k_x = \hat{s} \theta_0 k_y$, $g^{yy}$, $g^{xx}$ and $g^{xy}$ are metric tensor elements, and $J(\theta)$ is the Jacobian (these geometric quantities are defined for GS2 in \cite{kotschenreuther1995} and \cite{dorland2000}). 

By using $\langle k_\perp^2\rangle_{k_y, \theta_0, i}$ from Eq.~\eqref{eqn:kperp} and summing over all three fields, we define
\begin{equation}
\label{eqn:ql_core}
    \hat{\Lambda}(k_y, \theta_0) = \sum_{i=1}^3 \frac{\max_\theta|\chi_i(k_y,\theta_0, \theta)|}{\max_\theta| \chi_1(k_y,\theta_0, \theta) |} \;\frac{\gamma(k_y, \theta_0)}{\langle k_\perp^2(k_y, \theta_0, \theta)\rangle_{\theta,i}}\,.   
\end{equation}
When modes are strongly electromagnetic, $\hat{\Lambda}(k_y, \theta_0)$ has substantial contributions from magnetic perturbations\footnote{The quantity $\hat{\Lambda}(k_y, \theta_0)$ defined in Eq.~\eqref{eqn:ql_core} is implemented in GS2 and available as an output from commit \texttt{415983d}.}. On the other hand, when $\delta A_\parallel$ and $\delta B_\parallel$ vanish (e.g. at low $\beta$), Eq.~\eqref{eqn:ql_core} reduces to the electrostatic model of \cite{jenko2005}. 

The dependence of $\hat{\Lambda}$ on $\theta_0$ in Eq.~\eqref{eqn:ql_core} can be exploited to capture the effect of equilibrium flow shear\footnote{A similar novel approach to flow shear has recently been included in a new quasi-linear model for momentum transport  \citep{sun2024}.} on the quasi-linear turbulent transport. Equilibrium flow shear advects local modes in the ballooning parameter at a rate $\mathrm{d}\theta_0/\mathrm{d}t=\gamma_E/\hat{s}$, which is accounted for by averaging $\hat{\Lambda}(k_y, \theta_0)$ over a suitable $\theta_0$ interval:
\begin{equation}
\label{eqn:lambdaky}
    \bar{\Lambda}(k_y) = \frac{1}{\theta_{0,\mathrm{max}}(k_y, \gamma_E)}\int_0^{\theta_{0,\mathrm{max}}(k_y,\gamma_E)}\hat{\Lambda}(k_y, \theta_0)\, \mathrm{d}\theta_0 \,,
\end{equation}
where
\begin{equation}
\label{eqn:theta0}
    \theta_{0,\mathrm{max}} = \min\Bigl(\frac{\gamma_E}{\hat{s}\gamma},\ \pi\Bigr)\,.
\end{equation}
The $\theta_0$ interval is the advection range over one growth time, from the typically most unstable mode at $\theta_0=0$ to $\theta_{0,\mathrm{max}}=\gamma_E/(\hat{s} \gamma)$, with $\gamma$ the growth rate at $\theta_0=0$. 
In the low flow shear limit, $\gamma_E \ll \hat{s} \gamma$, $\bar{\Lambda}(k_y)$ reduces to $\hat{\Lambda}(k_y,0)$; in the high flow shear limit where $\gamma_E > \hat{s} \gamma \pi$, the average is calculated over a reduced $\theta_0$ range $\left[ 0, \pi \right]$, justified by periodicity.
The $\theta_0$ average in Eq.~\eqref{eqn:lambdaky} allows the quasi-linear model to capture flow shear suppression of turbulence at finite $\gamma_E$. We note that if $\gamma(\theta_0)$ is narrowly peaked close to $\theta_0=0$, $\bar{\Lambda}(k_y)$ can be sharply reduced even at small values of $\gamma_E$. As shown in Fig.~\ref{fig:theta0} and in \cite{kennedy2023a}, the growth rate of hybrid-KBMs can peak strongly at $\theta_0=0$ in STEP. 

The quasi-linear metric $\Lambda$ in Eq.~\eqref{eqn:redmod} is then obtained by integrating $\bar{\Lambda}(k_y)$ in Eq.~\eqref{eqn:lambdaky} over $k_y$, and normalised to $\rho_* c_s$,
\begin{align}
\label{eqn:ql}
    \Lambda & = \frac{1}{\rho_* c_s}\int \mathrm{d}k_y \bar{\Lambda}(k_y) \nonumber \\
    &=\frac{1}{\rho_* c_s}\int \mathrm{d}k_y\frac{1}{\theta_{0,\mathrm{max}}}\int_0^{\theta_{0,\mathrm{max}}}\mathrm{d}\theta_0\,\sum_{i=1}^3 \frac{\max_\theta|\chi_i(k_y,\theta_0, \theta)|}{\max_\theta|\chi_1(k_y,\theta_0, \theta)|} \frac{\gamma(k_y, \theta_0)}{\langle k_\perp^2(k_y, \theta_0, \theta)\rangle_{\theta,i}} \,.
\end{align}
This quantity depends only on linear physics, so its evaluation requires only linear gyrokinetic simulations over the unstable spectrum in $(k_y, \theta_0)$.  We highlight that $\Lambda$ has been derived rather generally without any specific assumption on the turbulence regime. The parameters $Q_0$ and $\alpha$, which appear in Eq.~\eqref{eqn:redmod}, are the only free parameters to influence the turbulence saturation level in this reduced model.  

Species heat and particle fluxes are obtained 
\textcolor{myred}{(from Eqs.~\eqref{eqn:heat_species}~and~\eqref{eqn:particle_species} below)} using the total model heat flux, $Q_\mathrm{ql}=Q_0\Lambda^\alpha$ defined in Eq.~\eqref{eqn:redmod} \textcolor{myred}{that sets the turbulence saturation amplitude}, together with the quasi-linear weights $Q_{\mathrm{l},s}/Q_{\mathrm{l}}$ and $\Gamma_{\mathrm{l},s}/Q_{\mathrm{l}}$, where $Q_{\mathrm{l},s}$ and $\Gamma_{\mathrm{l},s}$ are the linear heat and particle fluxes for species $s$, and $Q_\mathrm{l} = \sum_s Q_{\mathrm{l},s}$ is the total linear heat flux;
\textcolor{myred}{all linear fluxes are taken from the final time point in a converged linear simulation for a given mode (i.e. $k_y, \theta_0$). The quasi-linear weights apportion contributions to transport fluxes for each species for a given mode (depending on $k_y$ and  $\theta_0$) must be included in the integrands of Eqs \eqref{eqn:heat_species} and \eqref{eqn:particle_species} to obtain the total} species heat and particle fluxes:  
\begin{align}
\label{eqn:heat_species}
   Q_{\mathrm{ql},s} & = Q_0 \Lambda^{\alpha-1}\biggl(\frac{1}{\rho_* c_s}\biggr) \int\mathrm{d}k_y\;\frac{1}{\theta_{0,\mathrm{max}}}\int_0^{\theta_{0,\mathrm{max}}} \mathrm{d}\theta_0 \frac{Q_{\mathrm{l},s}(k_y,\theta_0)}{Q_{\mathrm{l}}(k_y,\theta_0)} \hat{\Lambda}(k_y, \theta_0)\\
\label{eqn:particle_species}
   \Gamma_{\mathrm{ql},s} & = Q_0 \Lambda^{\alpha-1}\biggl(\frac{1}{\rho_* c_s}\biggr) \int\mathrm{d}k_y\;\frac{1}{\theta_{0,\mathrm{max}}}\int_0^{\theta_{0,\mathrm{max}}} \mathrm{d}\theta_0 \frac{\Gamma_{\mathrm{l},s}(k_y,\theta_0)}{Q_{\mathrm{l}}(k_y,\theta_0)}  \hat{\Lambda}(k_y, \theta_0)\,,   
\end{align}
where we note that $\sum_s  Q_{\mathrm{ql},s} = Q_\mathrm{ql}=Q_0\Lambda^\alpha$.

\subsection{Application to STEP relevant regimes}
\label{sec:application}
The parameters $Q_0$ and $\alpha$, which link $\Lambda$ to the quasi-linear heat flux $Q_\mathrm{ql}$ via Eq.~\eqref{eqn:redmod}, are tuned here to optimise the description of the total heat transport from hybrid-KBM turbulence in STEP. 
Suitable values are determined by fitting the nonlinear saturated heat fluxes from a set of  nonlinear gyrokinetic simulations for STEP relevant local equilibria.
This database contains nonlinear simulations performed using local parameters from various reference surfaces in STEP-EC-HD. \textcolor{myred}{The parameter scans reported in \citet{kennedy2023a} and \citet{giacomin2024} already cover many of the most important local parameters, including pressure gradient, $\beta$, flow shear, $q$ and $\hat{s}$, to which hybrid-KBM turbulence is sensitive. Those scans provide the bulk of the database used to develop the quasi-linear model.}
In detail, we include several parameter scans at $\Psi_n=0.49$ with $\gamma_E\in \{0.05\,, 0.1\,, 0.2\}\,c_s/a$, $q\in \{3\,, 3.5\,, 4.0\,, 4.5\}$, $\beta\in\{0.005\,, 0.02\,, 0.09\}$, and $L_{p,\mathrm{ref}}/L_p\in \{0.8\,, 1.0\,, 1.2\}$ from \cite{giacomin2024} and an additional scan with $\hat{s} \in \{0.3\,, 0.6\,, 1.2\}$ for $\gamma_E=0$. The database also includes the simulations at $\Psi_n=0.36$ ($\gamma_E=0$ and $\gamma_E=0.05\,c_s/a$) and at $\Psi_n=0.8$ ($\gamma_E=0.1\,c_s/a$)  described in Sec.~\ref{sec:nonlinear}, as well as an additional scan with $q\in\{3.0\,, 3.5\,, 4.0\,, 4.5\}$ for $\gamma_E=0$ at $\Psi_n=0.36$.

\begin{figure}
    \centering
    \includegraphics[scale=0.7]{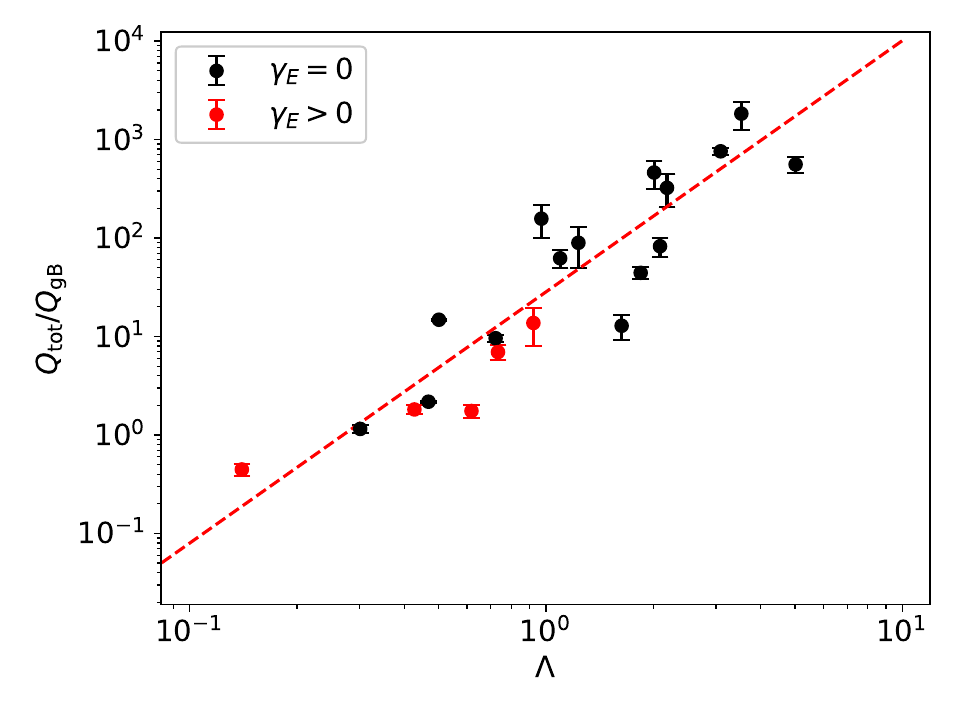}
    \caption{Normalised total heat flux $Q_\mathrm{tot}/Q_\mathrm{gB}$ from nonlinear gyrokinetic simulations as a function of $\Lambda$. The dashed line represents the best weighted fit, $Q_\mathrm{tot}=Q_0 \Lambda^\alpha$, with $Q_0$ and $\alpha$ the fitting coefficients. The error bar in the heat flux is taken as the standard deviation of the heat flux time trace in the saturated phase. Black and red markers correspond to simulations without and with equilibrium flow shear, respectively.}
    \label{fig:ql_fit}
\end{figure}

Fig.~\ref{fig:ql_fit} compares the total normalised heat flux, $Q/Q_\mathrm{gB}$, from nonlinear simulations to that from the best-fit quasi-linear model defined in Eq.~\eqref{eqn:redmod}, as functions of $\Lambda$, where $\Lambda$ is defined in Eq.~\eqref{eqn:ql} and obtained from linear gyrokinetic simulations. 
Each point in Fig.~\ref{fig:ql_fit} corresponds to a nonlinear simulation and the color distinguishes simulations with and without equilibrium flow shear.
The uncertainty in the average nonlinear heat flux from each simulation is taken as the standard deviation in the saturated phase\footnote{The stationary phases are identified by applying the augmented Dickey-Fuller test to the flux time traces, as described in \cite{giacomin2024}.}.
These uncertainties are accounted for by using a weighted fit to obtain the best values for the model parameters $Q_0$ and $\alpha$.  This gives $Q_0/Q_\mathrm{gB}\simeq 25$ and $\alpha\simeq 2.5$, with a fitting quality of $R^2\simeq 0.8$.
The unweighted fit returns values of $\alpha$ and $Q_0$  that differ approximately by 20\%, which can be considered as an estimate of the uncertainty in the fit values. 
The quasi-linear model reproduces acceptably well the nonlinear heat flux across a range of more than three orders of magnitude, although we highlight significant scatter.  The discrepancy between the reduced model and simulations approaches an order of magnitude in a few cases, illustrating a limitation in the reduced transport model. 
A comparison of the nonlinear species heat and particle fluxes to the quasi-linear predictions given by Eqs.~\eqref{eqn:heat_species}~and~\eqref{eqn:particle_species} is reported in Appendix~\ref{sec:appendix}.

Turbulent transport in the nonlinear simulations of Fig.~\ref{fig:ql_fit} is predominantly dominated by hybrid-KBM turbulence, but in two low $\beta$ cases turbulence is from electrostatic ITG and TEM. 
In principle MTMs may generate magnetic stochastic transport if the separation between the magnetic islands from the tearing perturbation is smaller than the island width (see, e.g., \cite{guttenfelder2011}, \cite{doerk2011}, \cite{nevins2011}, \cite{giacomin2023a}),  but this physics is \emph{not} captured here or in most other quasi-linear models since it would require predictions of the saturated magnetic island width generated by the underlying MTM instability. On the other hand, the nonlinear gyrokinetic simulations of \cite{giacomin2024} suggest that magnetic stochastic transport should be modest on most STEP-EC-HD surfaces that have been analysed.  Other simulations, however, suggest that stochastic transport may be significant in some spherical tokamak plasmas from NSTX~\citep{guttenfelder2011} and MAST~\citep{giacomin2023a}.  It is entirely possible that MTM induced magnetic stochastic transport will be substantial under some plasma conditions in STEP, and investigating this is an important area for future work.  It is therefore likely that the reduced turbulent transport model developed here will need to be extended to account for magnetic stochastic turbulent transport from MTMs. We note that the development of such models is an active research area \cite{rafiq2016,curie2022,hamed2023,hornsby2024}.  

\begin{figure}
    \centering
    \subfloat[]{\includegraphics[width=0.47\textwidth]{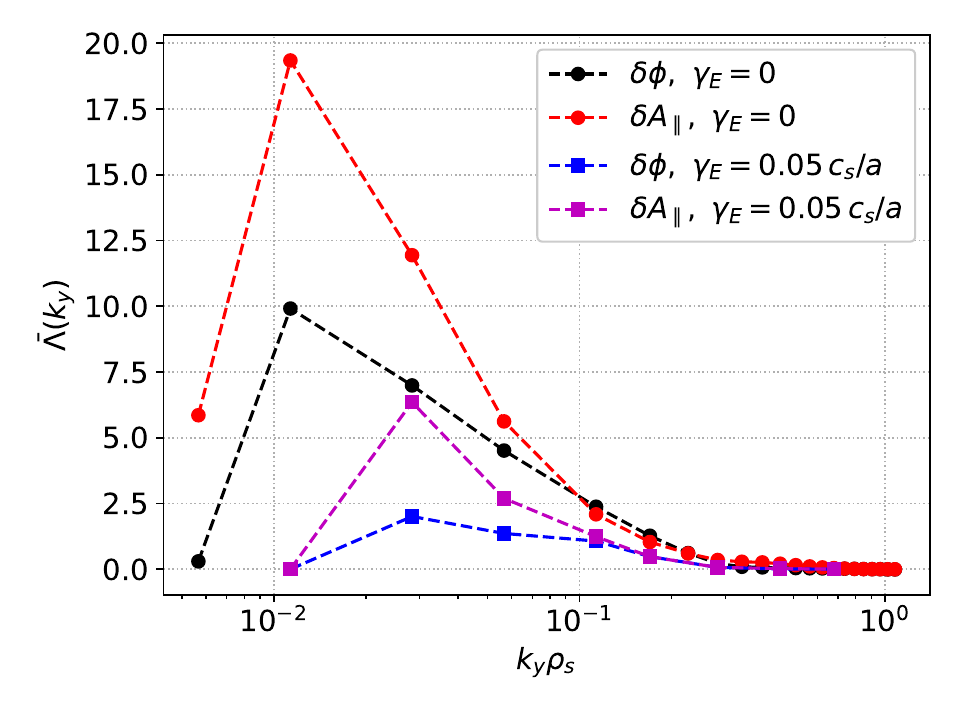}}\quad
    \subfloat[]{\includegraphics[width=0.47\textwidth]{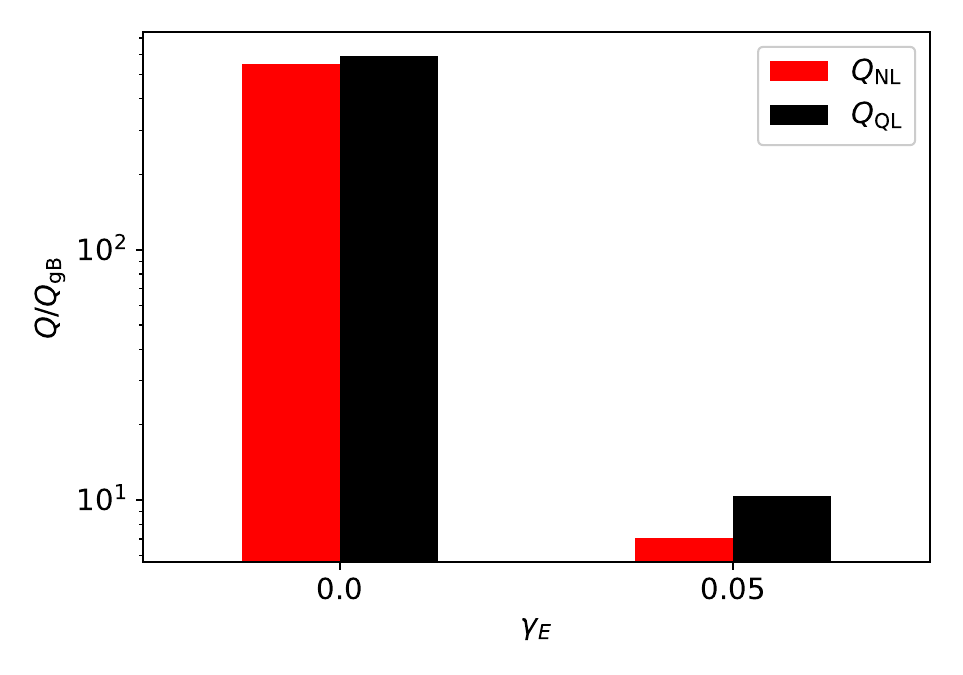}}
    \caption{(a) Contributions of the $\delta \phi$ and $\delta A_\parallel$ terms in Eq.~\eqref{eqn:lambdaky} to $\bar{\Lambda}(k_y)$ without (round markers) and with (squared markers) equilibrium flow shear. The contribution from $\delta B_\parallel$ is negligible and not shown. (b) Comparison between the nonlinear and the quasi-linear total heat flux for the two cases considered in (a). These simulations are performed at $\Psi_n=0.36$ of STEP-EC-HD.}
    \label{fig:comp_ql}
\end{figure}

Fig.~\ref{fig:comp_ql}~(a) shows examples of the $\delta \phi$ and $\delta A_{\parallel}$ contributions to the $k_y$ spectrum of $\bar{\Lambda}(k_y)$ for the surface at $\Psi_n=0.36$ of STEP-EC-HD without and with flow shear ($\gamma_E=0$ and $\gamma_E = 0.05\, c_s/a$).
Both contributions to $\bar{\Lambda}(k_y)$ peak at low $k_y$ with and without flow shear, but at $\gamma_E=0$ the peaks are significantly larger and at longer wavelength. We note that in this local equilibrium the $\delta A_\parallel$ contribution to $\bar{\Lambda}(k_y)$ dominates both at $\gamma_E=0$ and $\gamma_E = 0.05\, c_s/a$, highlighting the need to include $\delta A_\parallel$ in Eq.~\eqref{eqn:ql}.

Fig.~\ref{fig:comp_ql}~(b) compares the nonlinear and quasi-linear model heat fluxes, where reduced model used $Q_0$ and $\alpha$ in Eq.~\eqref{eqn:redmod} from the fit in Fig.~\ref{fig:comp_ql}. The quasi-linear and nonlinear heat fluxes agree within 30\%, though it is pertinent to note that both simulations were included in the database used to obtain $Q_0$ and $\alpha$. The reduction in the heat flux attributable to flow shear approaches two orders of magnitude, and this is largely captured by the quasi-linear model.  The reduction in the quasi-linear heat flux is slightly weaker than found in the nonlinear simulations, where the saturated total heat flux is 30\% lower at $\gamma_E=0.05\,c_s/a$.

We underline that the nonlinear heat flux calculations in the two cases of Fig.~\ref{fig:comp_ql} requires more than $5 \times 10^5$ core-hours for each simulation, corresponding to a week running on 32 nodes of the ARCHER2 high performance computing facility (Edinburgh, UK); in contrast computing the quasi-linear metric takes approximately 100 core-hours, corresponding to 10 minutes running on 4 nodes of ARCHER2. 
Computing the quasi-linear metric is at least a factor of $10^3$ less expensive than computing the nonlinear heat flux.   

\begin{figure}
    \centering
    \subfloat[]{\includegraphics[width=0.47\textwidth]{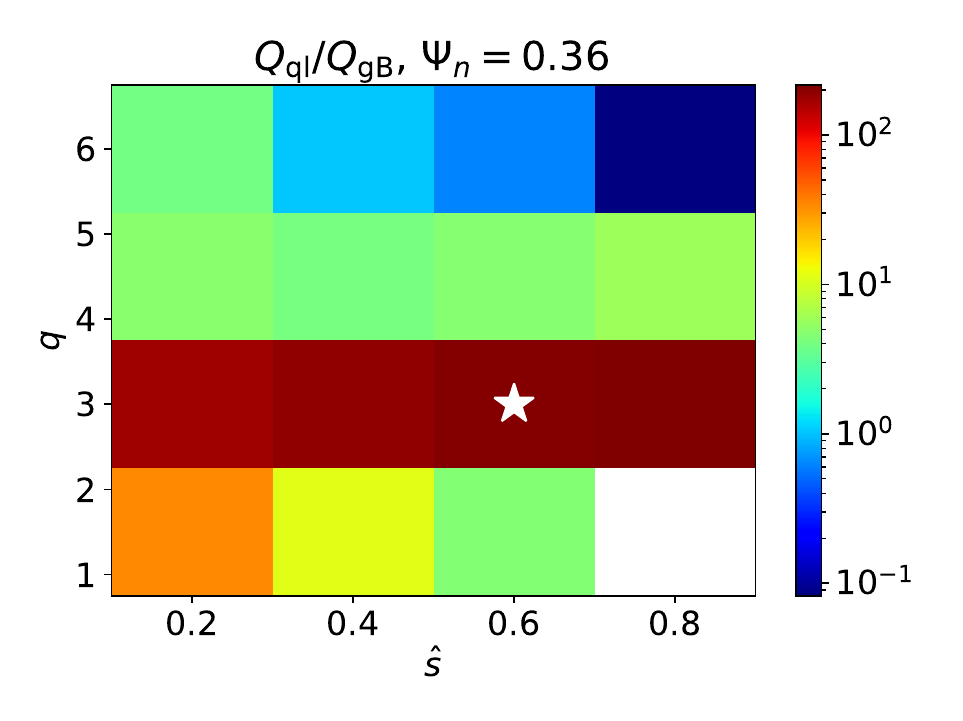}}
    \subfloat[]{\includegraphics[width=0.47\textwidth]{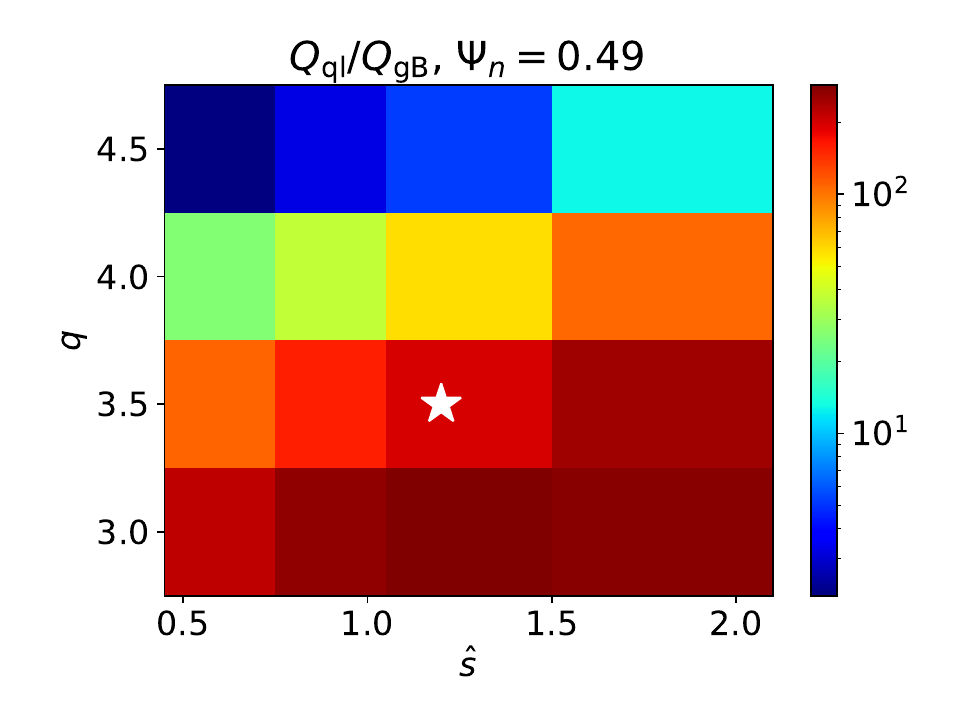}}\\
    \caption{Two-dimensional $(q, \hat{s})$ scan of the quasi-linear heat flux (normalized to $Q_\mathrm{gB}$) at $\Psi_n=0.36$ (a) and $\Psi_n=0.49$ (b). The white star indicates the nominal value of $q$ and $\hat{s}$ of the chosen STEP flat-top operating point. The simulation at $\hat{s}=0.8$, $q=1.5$ and $\Psi_n=0.36$ is stable. In this figure, the quasi-linear heat flux is computed without equilibrium flow shear.  }
    \label{fig:sqscan}
\end{figure}

Motivated by the analysis of \cite{giacomin2024} that finds a significant influence of $q$ on hybrid-KBM turbulence at mid-radius in STEP-EC-HD, Fig.~\ref{fig:sqscan} explores the sensitivity of the quasi-linear heat flux to $q$ and $\hat{s}$ at $\Psi_n=0.36$ and $\Psi_n=0.49$.
Equilibrium flow shear is not included in this two-dimensional parameter scan. At $\Psi_n=0.49$, $Q_\mathrm{ql}/Q_\mathrm{gB}$ increases monotonically as $\hat{s}$ increases or $q$ decreases, which is consistent with the nonlinear $q$ scan presented in \cite{giacomin2024}). At $\Psi_n=0.36$, the dependence of $Q_\mathrm{ql}/Q_\mathrm{gB}$ on $q$ and $\hat{s}$ is non-monotonic, with a maximum at $q=3.0$; at $q$ values higher or smaller than the nominal value, the quasi-linear heat flux decreases as $\hat{s}$ increases.
In particular, Fig.~\ref{fig:sqscan} shows the presence of a low turbulent transport regime that could be achieved by varying $\hat{s}$ and $q$ from their nominal values, which are denoted by the white star in Fig.~\ref{fig:sqscan}; determining whether such local equilibrium conditions can be accessed across a significant radial region, while still satisfying operational constraints, clearly requires integrated scenario modelling  \citep{tholerus2024}.

\section{Flux-driven STEP simulations}
\label{sec:t3d}

Sec.~\ref{sec:nonlinear} demonstrates that a diamagnetic level of equilibrium flow shear is insufficient to reduce the hybrid-KBM-driven heat flux below the level required to achieve a transport steady state in STEP-EC-HD with the assumed sources.  
The imbalance between the turbulent fluxes crossing a surface and the available sources will result in the density and temperature profiles undergoing transport evolution from this state.
Other local equilibrium parameters, such as $\beta$, $\beta'$, collisionality and eventually $q$, $\hat{s}$ and $\Delta'$, are also affected by transport evolution, with the self-consistent Grad-Shafranov equilibrium and current evolving on a longer resistive timescale: local transport is affected by these local equilibrium quantities as well as the temperature and density gradients.

In this section, we report on the implementation of the quasi-linear reduced model of Sec.~\ref{sec:ql_metric} in the T3D code, and its exploitation to evolve the temperature and density profiles of electron and main ion plasma species in the STEP flat-top operating point introduced in Sec.~\ref{sec:overview}.   

\subsection{Implementation of the reduced model in T3D}

T3D~\citep{t3d} is a transport code that has been recently developed from TRINITY~\citep{barnes2010}. It is a very flexible and modular tool that facilitates the implementation of new transport models.    
The transport equations solved by T3D are (see~\citep{barnes2010}):
\begin{align}
\label{eqn:density}
    \frac{\partial n_s}{\partial t} + \frac{1}{V'}\frac{\partial}{\partial \Psi}(V'\overline{\Gamma_s}) &= \overline{S_{n, s}}\,,\\
    \label{eqn:pressure}
    \frac{3}{2}\frac{\partial p_s}{\partial t} + \frac{1}{V'}\frac{\partial}{\partial \Psi}(V'\overline{Q_s}) &= \frac{3}{2}n_s\sum_u \nu_{su}(T_u - T_s) + \overline{S_{p,s}}\,,
\end{align}
where $n_s$ and $p_s$ are the density and pressure of the species $s$, $\Psi$ is the flux label, $V$ is the volume enclosed by the surface $\Psi$, $V'$ is the derivative of $V$ with respect to $\Psi$, $\Gamma_s$ and $Q_s$ are the total particle and heat fluxes of a given species $s$ including the classical, neoclassical and turbulent contributions, and $S_{n,s}$ and $S_{p,s}$ are the particle and the heat sources. The overline denotes a flux surface average. The term proportional to $\nu_{su}$ is the collisional energy exchange term, with $\nu_{su}$ taken from \cite{nrl}.
Density and temperature profiles are discretised on a uniformly spaced radial grid, and gradients are computed using a sixth order finite difference scheme on interior radial grid points and a second order finite difference scheme at the boundary. A non-zero Dirichlet boundary condition is applied at the outermost radial grid point, while the physical zero flux condition is applied at the magnetic axis\footnote{The first radial grid point is located at $r=\Delta r/2$ where $\Delta r$ is the radial grid spacing. \textcolor{myred}{Fluxes and gradients are evaluated on interior radial grid points (apart from the magnetic axis where fluxes vanish) up to the outermost radial grid point $r_{N_r}-\Delta r/2$. The value of density and temperature at the magnetic axis can be evaluated using the zero flux on-axis boundary condition.}}. 
Eqs.~\eqref{eqn:density}~and~\eqref{eqn:pressure} are evolved using a semi-implicit method. 
Further details on the transport equations and on their numerical implementation are reported in \cite{barnes2010}.

Evaluating the fluxes $\Gamma_s$ and $Q_s$ in Eqs.~\eqref{eqn:density}~and~\eqref{eqn:pressure} requires a transport model. 
For instance, the coupling between T3D and GX~\citep{mandell2022}, a local gyrokinetic code developed to run natively on Graphics Processing Units (GPUs), allows T3D to submit nonlinear gyrokinetic simulations at each radial location to compute  $\Gamma_s$ and $Q_s$ from first-principles simulations. A total number of $(N_r-1)(N_p+1)$ simulations per iteration is required, where $N_r$ is the number of radial grid points and $N_p$ is the number of evolved profiles.  \textcolor{myred}{We note that $N_p+1$ evaluations are required at each radial location to compute the local flux derivatives with respect to each evolved profile gradient (see \citet{barnes2010} for details)}. The quasi-neutrality condition is applied to impose the density value of the last evolved species.  Our T3D simulations for STEP use the quasi-linear reduced transport flux equations~\eqref{eqn:heat_species} and \eqref{eqn:particle_species}, which have been implemented in T3D, and exploit a coupling to GS2 for the linear calculations required to compute the quasi-linear fluxes.

\begin{figure}
    \centering
    \includegraphics[scale=0.42]{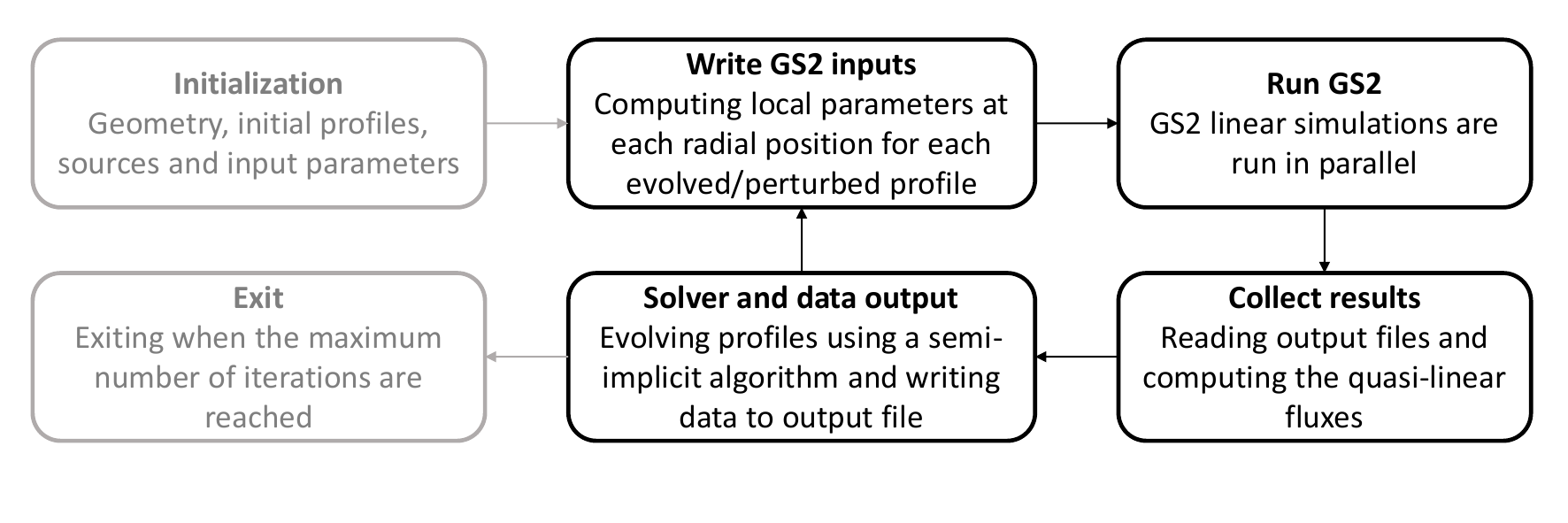}
    \caption{Workflow of T3D-GS2. The main cycle is decomposed in four main blocks (in black): writing GS2 input files, running GS2, computing the quasi-linear heat flux and evolving the density and temperature profiles. The blocks in grey are executed only once per simulation.}
    \label{fig:workflow}
\end{figure}

The workflow of T3D-GS2 is shown in Fig.~\ref{fig:workflow}. After initialisation, where the input file (containing run configuration data, such as the radial grid resolution, the maximum number of iteration, and the transport model selection), geometry\footnote{Local equilibrium shaping parameters, $q$ and $\hat{s}$ are computed from the initial state and kept constant during the T3D simulation, while local $\beta'$ is updated every time step consistently with the pressure profile. There is a plan in the future to allow the magnetic equilibrium in T3D to be recomputed self-consistently from the pressure profile every $N$ time steps.}, initial profiles and sources are loaded, T3D-GS2 enters into the main loop, which is divided in four substeps. In the first substep, T3D writes the GS2 input files for each radial grid point and $(k_y, \theta_0)$ mode, according to the specified numerical resolutions. Several GS2 calculations are then distributed over the available MPI tasks. Depending on the numerical resolutions and on the number of linear gyrokinetic simulations to be performed, T3D-GS2 can efficiently scale to thousands of cores. 
Once all the GS2 linear simulations are completed, results are gathered and the quasi-linear fluxes are computed using Eqs.~\eqref{eqn:heat_species}~and~\eqref{eqn:particle_species}. In the last sub-step, density and pressure profiles are evolved using Eqs.~\eqref{eqn:density}~and~\eqref{eqn:pressure}.
The transport calculation terminates when the maximum number of iterations or the maximum simulation time is reached. It is possible to restart from the final state of a previous T3D run, which is invaluable when a transport steady state solution (characterized by power and particle losses that match sources and profiles that are constant in time) is not reached within a single run.
 
\subsection{Evolution of density and temperature profiles in a STEP case}
\label{sec:fluxdriven}

Here we present results from the first flux-driven transport simulations to account for hybrid-KBM turbulence in a STEP flat-top operating point.

\subsubsection{Transport simulation set-up and assumptions}

In this first analysis, we have made a number of simplifying approximations that  are detailed below.
First, only density and temperature profiles of electrons and a single thermal ion species are evolved. The quasi-neutrality is enforced on the ion species, such that $n_i=n_D+n_T=n_e$ where $n_D=n_T=n_i/2$ with $n_D$ and $n_T$ being the deuterium and tritium density, respectively. The deuterium and tritium temperature are set to be equal to each other, i.e., only one ion temperature profile if evolved, with $T_i=T_D=T_T$.
We note that the nonlinear gyrokinetic simulations used to develop the reduced model parameters, $Q_0$ and $\alpha$ in Eq.~\eqref{eqn:redmod}, excluded impurities, helium ash and fast $\alpha$ particles.  It will be an important future research direction to include these additional ion species in nonlinear gyrokinetic simulations, and to extend the reduced model to accommodate their impacts on the turbulent fluxes. 
Second, the geometrical shaping parameters, $q$ and $\hat{s}$ are held constant while evolving the pressure profile. If the initial and final pressure profiles are similar, the magnetic equilibrium should change only slightly, although even small variation of safety factor or magnetic shear may impact turbulence (see, e.g., the dependence on the safety factor in Fig.~\ref{fig:sqscan}). 
On the other hand, the value of $\beta'$ is evolved self-consistently with the pressure profile in our T3D simulations, and this is known to have a strong impact on turbulence in STEP-EC-HD (see \citet{giacomin2024}).
Finally, the equilibrium flow shear has been computed from Eq.~\eqref{eqn:flow_shear}, where $E_r$ is taken from the JINTRAC STEP flat-top operating point, and has been kept constant as the kinetic profiles evolve.
There is no external torque in STEP, so equilibrium flow shear is relatively small ($\gamma_E\lesssim 0.05\,c_s/a$).  Sensitivity to $\gamma_E$ is addressed specifically through T3D simulations with $\gamma_E(r)$ varied by 40\% around $\gamma_E^\mathrm{dia}$.  
In addition, the model parameters $\alpha$ and $Q_0$ are varied within 20\% (which is the uncertainty given by the weighted/unweighted fit in Fig.~\ref{fig:ql_fit}) to investigate the impacts of uncertainties on these parameters. 
The reference case is performed using the values of $Q_0$ and $\alpha$ from the fit in Sec.~\ref{sec:ql_metric}, and setting $\gamma_E = [0.01, 0.01, 0.02, 0.03, 0.05]\,c_s/a$ at $r/a =[0.16, 0.33, 0.49, 0.65, 0.82]$, respectively.

Our T3D simulations use six radial grid points uniformly distributed over the radial domain, extending from the magnetic axis to $r/a=0.9$ close to the pedestal top.  At each radial position and for each profile, a total of $n_{k_y}\times n_{\theta_0} = 12 \times  6 = 72$ linear modes are evolved, using a non-uniform grid spacing in $k_y$ and $\theta_0$ to improve resolution at low $k_y$ and low $\theta_0$, which is the region contributing most strongly to the quasi-linear metric. A minimum $k_y\rho_s$ is imposed to resolve only toroidal mode numbers $n>1$, and its value depends on radial location and plasma profiles. Each mode is evolved by an independent GS2 simulation in ballooning space. 
GS2 simulations are performed using 33 grid points along $\theta$, 24 pitch-angles and 10 energy grid points. 
The total number of GS2 simulations in ballooning space carried out at each step is $N=(N_r-1)(N_p+1)n_{k_y}n_{\theta_0} = 1440$, \textcolor{myred}{where $N_p=3$ as $n$, $T_e$ and $T_i$ are evolved here}.  The GS2 simulations are performed in parallel depending on the number of available MPI tasks. 
The auxiliary heating power density profile from electron cyclotron resonant heating ($P_\mathrm{EC} =150$~MW in total), radiated power density profile ($P_\mathrm{rad}=340$~MW in total), and particle source from pellet injection (corresponding to a fueling rate $\sim 10^{22}$ particles/s) are taken from the JINTRAC simulation of STEP-EC-HD (see \cite{tholerus2024} for details), and kept constant throughout the T3D transport calculation.  Fusion $\alpha$-particle heating and collisional energy exchange contributions to Eq.~\eqref{eqn:pressure} evolve with the density and temperature profiles and are calculated self-consistently. The ion density is multiplied by a dilution factor, $f_\text{He} = 0.9$, when computing the fusion power in order to account for the dilution effect from helium ash. 
Motivated by the presence of subdominant MTM turbulence in the STEP initial condition \citep{giacomin2024}, a small additional Rechester-Rosenbluth-like contribution is included in the electron heat flux proportional to $(a/L_{T_e})^3$ (see \cite{doerk2011}). This helps convergence, while it has a negligible impact in the final T3D steady state.
Neoclassical transport is included in the following T3D simulations and is computed using the neoclassical code NEO~\citep{belli2008}.

\subsubsection{Transport simulation results}
\begin{figure}
    \centering
    \subfloat[]{\includegraphics[height=0.21\textheight]{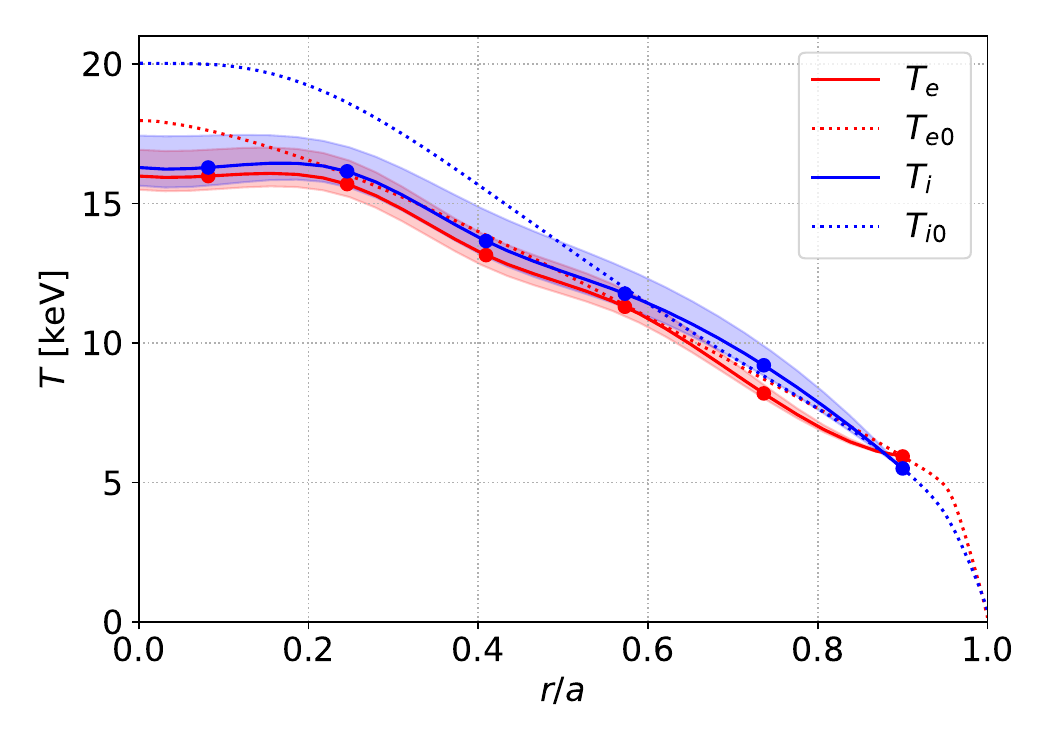}}\qquad
    \subfloat[]{\includegraphics[height=0.21\textheight]{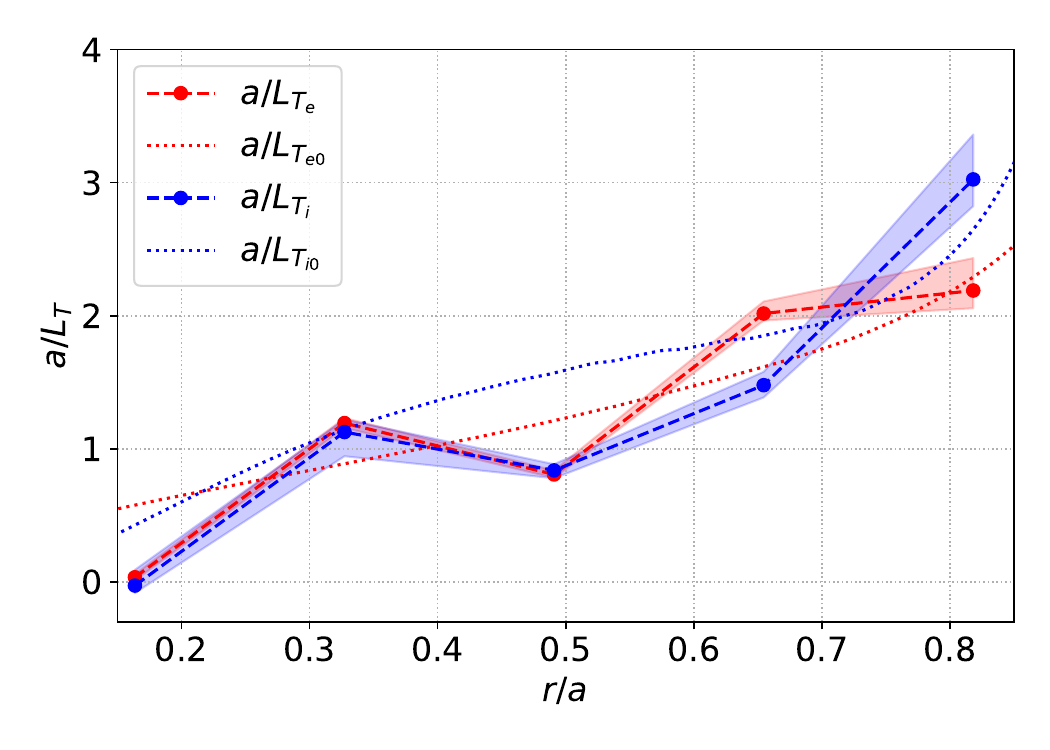}}\\
    \subfloat[]{\includegraphics[height=0.21\textheight]{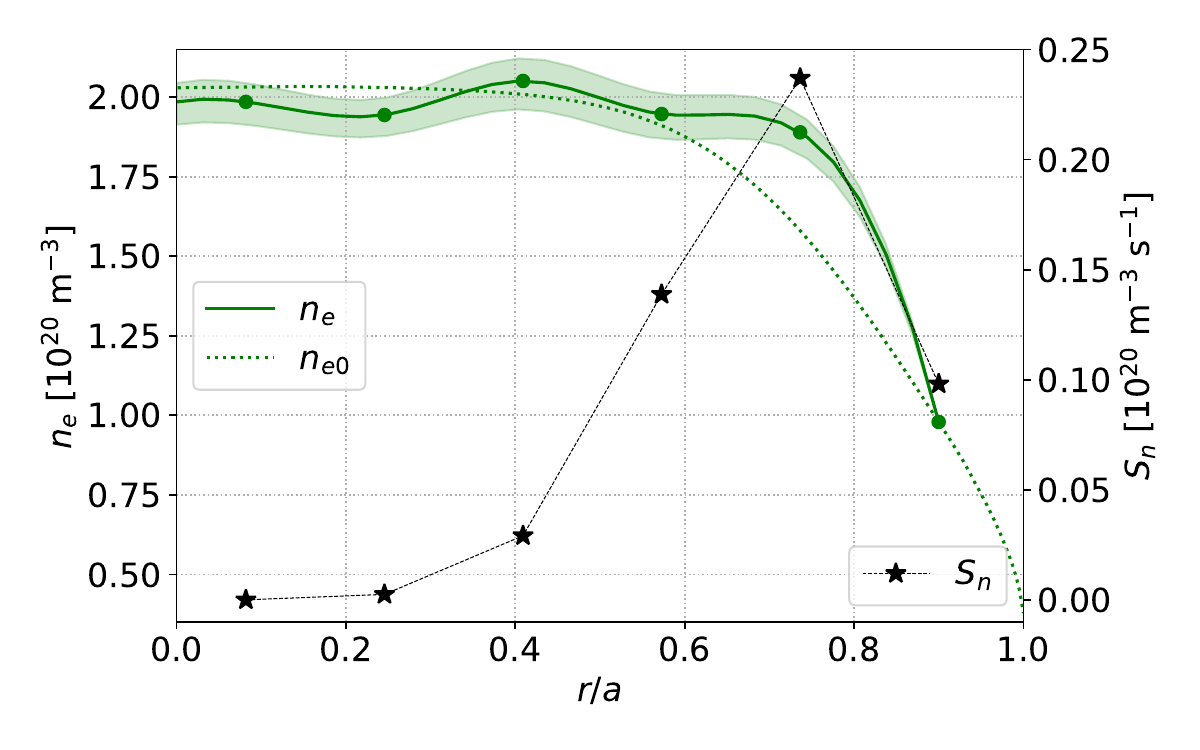}}    
    \subfloat[]{\includegraphics[height=0.21\textheight]{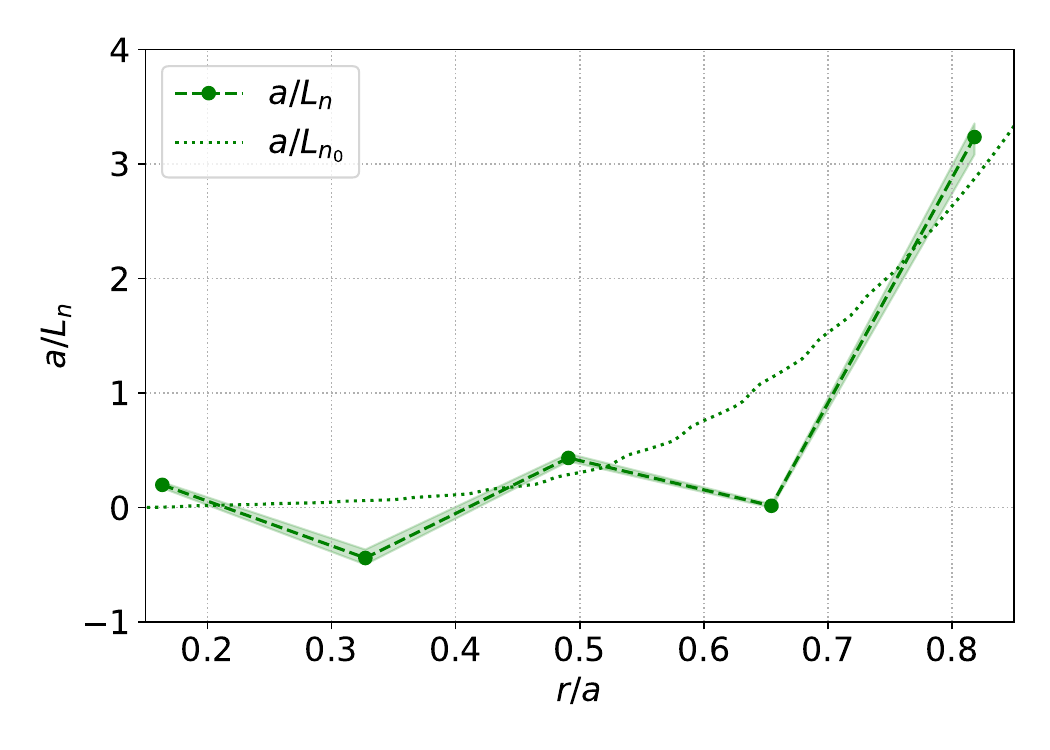}}\\
    \subfloat[]{\includegraphics[height=0.21\textheight]{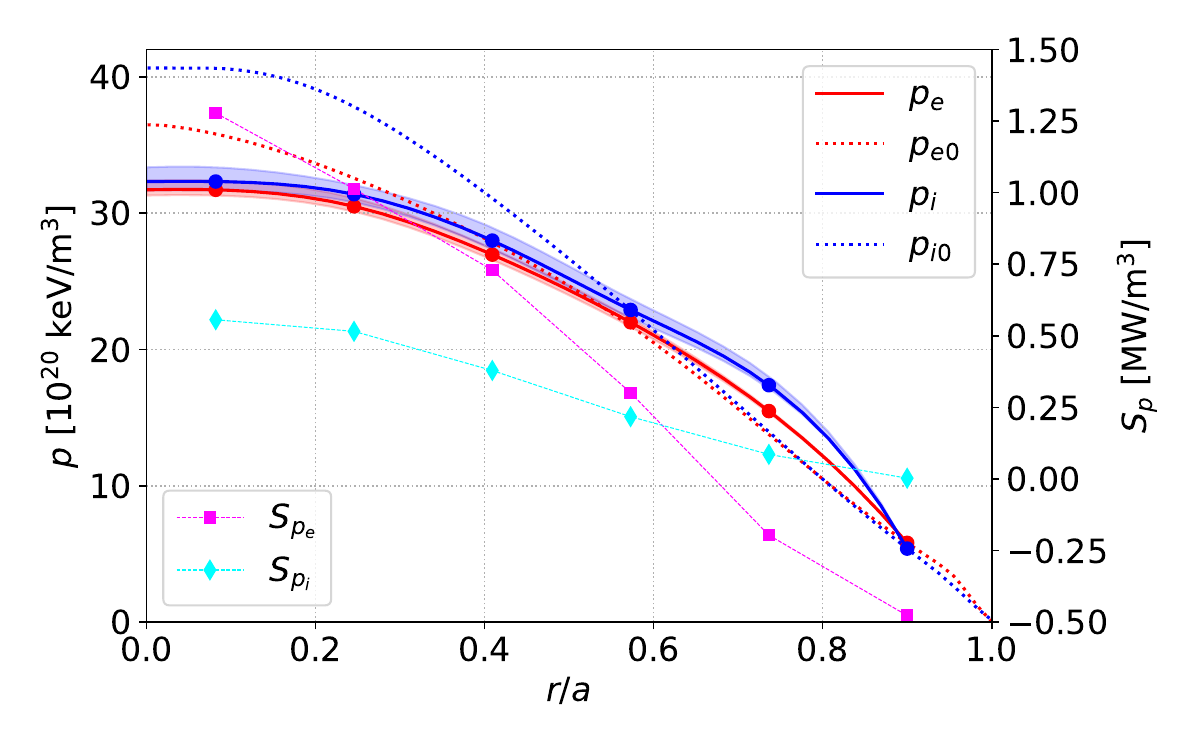}}
    \subfloat[]{\includegraphics[height=0.21\textheight]{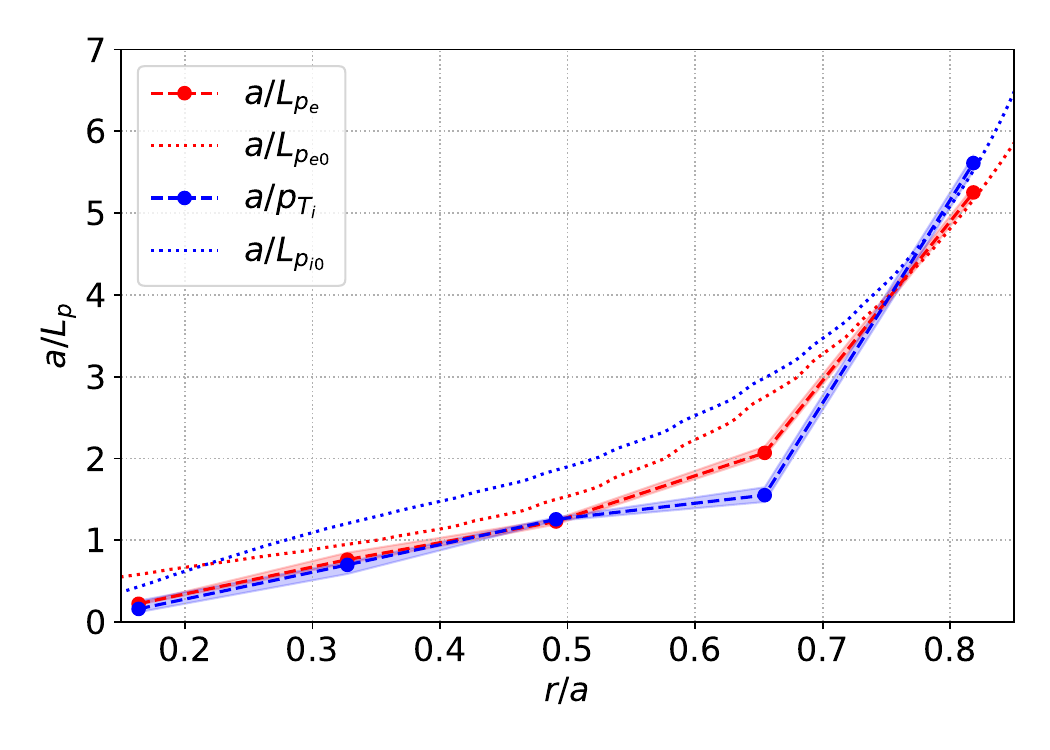}}\\
    \caption{Comparison of initial (dotted lines) and final (solid lines) temperature (a), density (c) and pressure (e) profiles as well as their inverse gradient scale length [(b), (d) and (f)]. The initial profiles are taken from the JINTRAC analysis \citep{tholerus2024}. The round markers denotes the position of the radial grid points in the T3D grid. The value of the density and pressure at the outermost radial grid point is imposed by the finite Dirichlet boundary condition. The shaded area represents the profile variation corresponding to a $\pm$40\% variation in $\gamma_E$ and a $\pm$20\% variation in $Q_0$ and $\alpha$. The solid line represents an interpolation through the T3D radial grid points. \textcolor{myred}{The particle source (star markers) as well as the electron (square markers) and ion (diamond markers) total power sources are shown in (c) and (e).}}
    \label{fig:profiles}
\end{figure}

Fig.~\ref{fig:profiles} compares the initial temperature, density and pressure profiles and their gradients taken from the JINTRAC STEP flat-top operating point with the corresponding profiles from the final state of the reference T3D simulation. Fluxes and gradients in T3D are computed at the mid-point between each radial grid point.
The final electron temperature profile values are lower than in the initial condition, with the largest reduction occurring near the magnetic axis where $T_e$ decreases by approximately 10\%.
The final normalised logarithmic electron temperature gradient, $a/L_{T_e}$, oscillates radially around its initial profile, and the largest difference is in the inner core where the final electron temperature profile is flatter than the initial one.

A significant reduction is observed in the ion temperature profile for $r/a<0.4$, where there is a 20\% reduction in $T_i$ near the magnetic axis.
On the other hand, the initial and final ion temperature profiles are very similar in the outer core at $r/a>0.5$. 
The ion temperature gradient is smaller than the initial temperature gradient over the entire radial domain, except near $r/a\simeq 0.8$ where it is slightly larger. 
We note that the initial ion temperature profile is significantly higher than the electron temperature despite this flat-top operating point being dominated by electron heating: this arises because the transport model used to design STEP-EC-HD assumed negligible turbulent ion heat transport.  This feature is not recovered in the T3D calculation using our reduced model for hybrid-KBM turbulence, where the ion turbulent transport is non negligible and the final $T_i$ and $T_e$ profiles are very similar, with $a/L_{T_i}$ reducing significantly compared to its initial value between $r/a\simeq0.4$ and $r/a\simeq 0.65$.

\begin{figure}
    \centering
    \subfloat[]{\includegraphics[height=0.23\textheight]{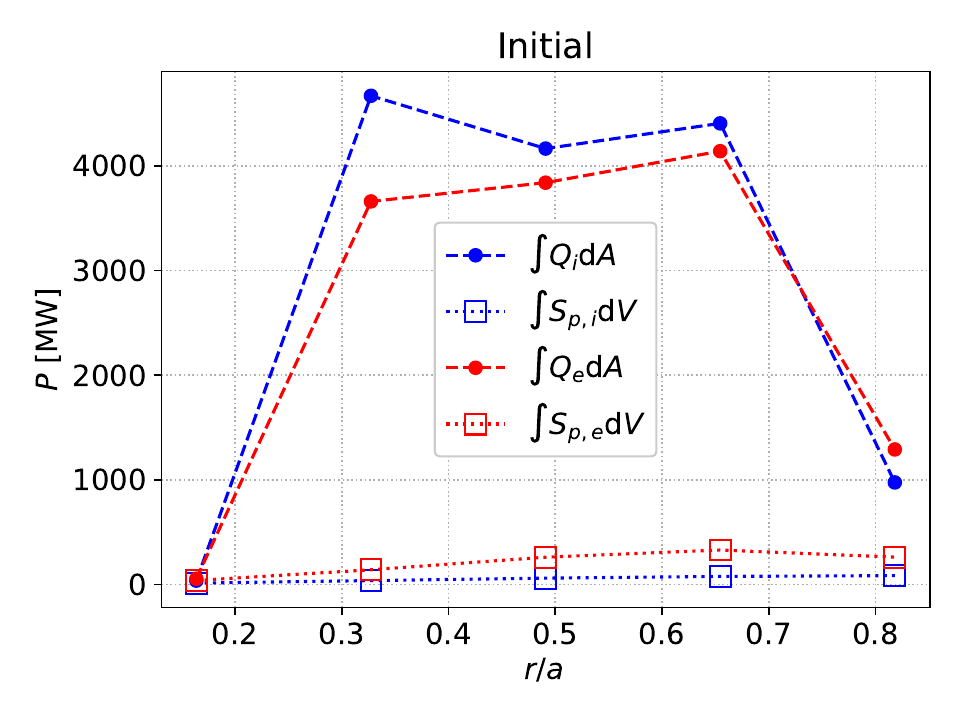}}
    \subfloat[]{\includegraphics[height=0.23\textheight]{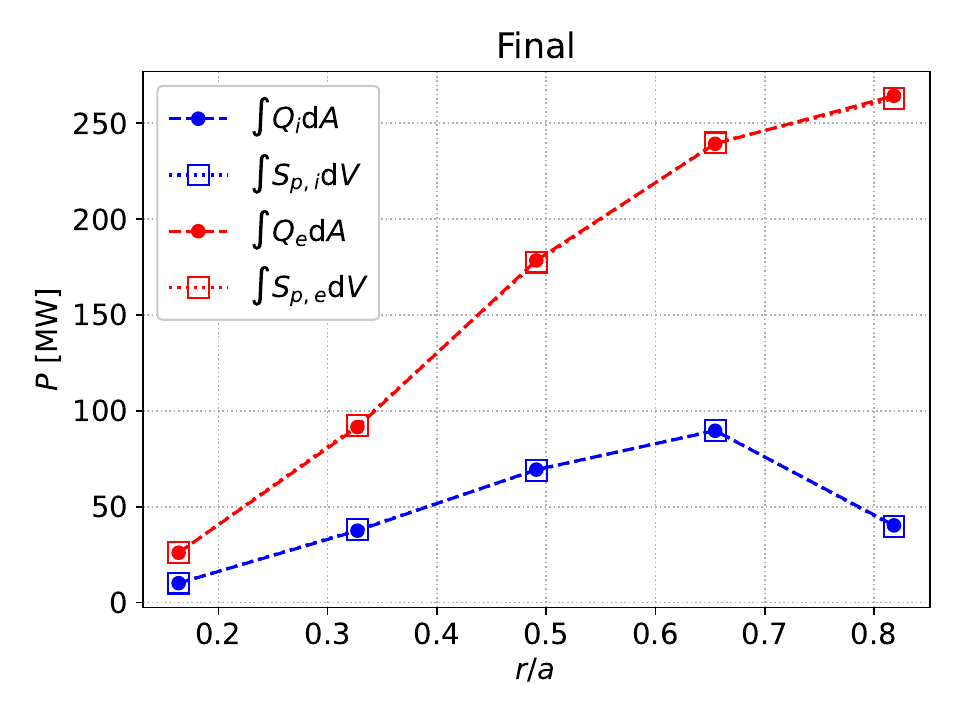}}\\
    \subfloat[]{\includegraphics[height=0.23\textheight]{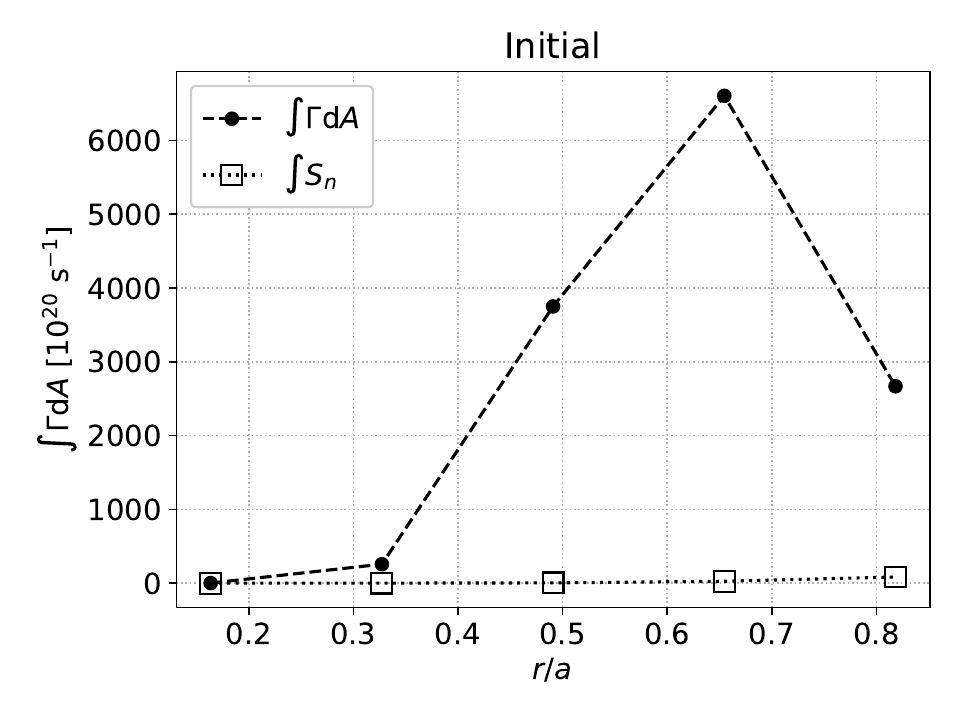}}\quad
    \subfloat[]{\includegraphics[height=0.23\textheight]{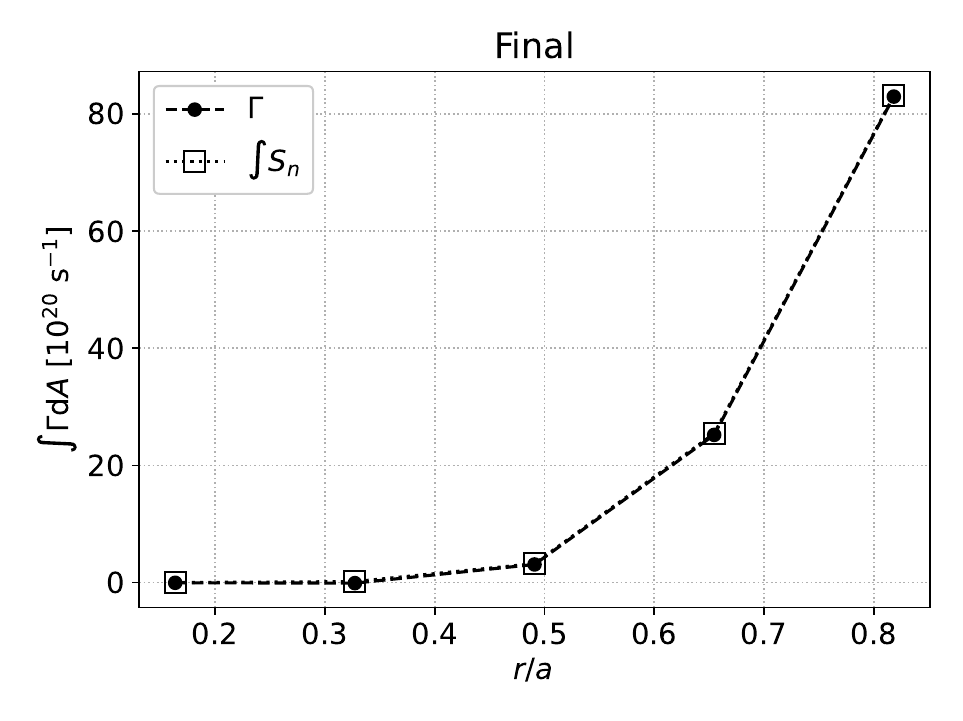}}
    \caption{Power [(a) and (b)] and particle [(c) and (d)] balance in the initial and final state of the reference T3D simulation. Heat and particle sources are denoted with open markers, while heat and particle losses with solid markers. The initial state is taken from JINTRAC-JETTO STEP flat-top.}
    \label{fig:balance}
\end{figure}

The final T3D electron density is close to the initial profile for $r/a \leq 0.6$, but higher for $r/a>0.6$: i.e. there is a small increase in the edge density gradient for $r/a\simeq 0.8$ and a relaxation of the gradient around $r/a\simeq 0.65$. \textcolor{myred}{We note that the density increase occurs in proximity of the maximum density source due to pellet injection, shown in Fig.~\ref{fig:profiles}~(c).} 
The particle confinement time evaluated from the T3D steady state solution is $\tau_p = \int n_e \mathrm{d} V / \int S_n\mathrm{d} n = 12.5$~s, which is close to that assumed in STEP-EC-HD \citep{tholerus2024}.
We highlight that the T3D steady state ion temperature and density gradients at $r/a\simeq 0.8$ are larger than in the initial condition from STEP-EC-HD. This may appear counter-intuitive, especially given that the reduced model heat and particle fluxes are much larger than the available sources at the STEP-EC-HD initial condition, as illustrated in Fig.\ref{fig:balance}. On the other hand, the dependence of the turbulent fluxes on the temperature and density gradients is non-monotonic in STEP-EC-HD because of the stabilizing $\beta'$ effect, as pointed out in \cite{giacomin2024}. The profile evolution in the initial phase leads to a steepening of temperature and density profile in the outer core region, which causes a local increase of the $\beta'$ value and a subsequent reduction of turbulent fluxes, thus sustaining larger density and temperature gradients in the outer core region.  We note, however, that the variation of other parameters, such as $\beta$ and collisionality, the evolution of the fusion power and the contribution from the neoclassical fluxes make the dynamics regulating the profile evolution strongly nonlinear: this effect cannot be solely explained by the dependence of turbulent fluxes on gradients.       

Fig.~\ref{fig:profiles} shows also a comparison between the initial and final pressure profile: the reduced final pressure for $r/a<0.6$ is consistent with lower $T_e$ and $T_i$ in this region of the plasma; for $r/a>0.6$ both the electron and ion final pressure profiles are larger than  the initial profiles, mostly because of the increased density.
If we do not separate transport and radiation losses, the energy confinement time using the final steady state pressure profile gives $\tau_E = \frac{3}{2}\int p \mathrm{d} V / P_\mathrm{net} = 1.5$~s, with $P_\mathrm{net} = P_\mathrm{\alpha} + P_\mathrm{EC}$.  If we exclude radiation losses from the net loss power using $P^{*}_\mathrm{net} = P_\mathrm{\alpha} + P_\mathrm{EC} - P_\mathrm{rad}$, this gives $\tau^*_E = \frac{3}{2}\int p \mathrm{d} V / P^{*}_\mathrm{net} = 4.6$~s. These energy confinement times are similar to those assumed in the original STEP flat-top operating point\footnote{The T3D steady state energy confinement time is slightly higher than in the original STEP-EC-HD ($\tau_E = 1.24$~s and $\tau_E^* = 4.20$~s) because of the larger density in the outer core, which leads to a slightly larger $\int p \mathrm{d} V$.} \citep{tholerus2024}. 
\textcolor{myred}{We highlight, however, that the radial domain evolved by these T3D simulations does not include the pedestal region. The outermost radial grid point boundary condition of the density and temperature is taken from near the pedestal top of the JINTRAC initial profiles (see Fig.~\ref{fig:profiles}); the pedestal model used in the JINTRAC modelling of this STEP flat-top operating point is described in \citet{tholerus2024}. Any variation of the pedestal top, and therefore of the outer boundary condition, is likely to impact core transport and the energy confinement time. Assessing the sensitivity on the outer boundary condition, i.e. on the density and temperature values at the pedestal top, is clearly a very important topic for future investigation.}

The shaded area in Fig.~\ref{fig:profiles} indicates the range of profiles variation obtained from T3D simulations with local values of $\gamma_E$ varied by $\pm$40\% from the nominal value, and also includes additional T3D simulations where $Q_0$ and $\alpha$ are varied by $\pm$20\% to assess the impacts of uncertainties in the reduced model parameters.
The variation in the density and temperature profiles is comparatively small, which may seem surprising given the large sensitivity of fluxes to $\gamma_E$ in STEP-EC-HD. This relatively weak sensitivity of the kinetic profiles on the shearing rate can be explained (i) \textcolor{myred}{by T3D converging to a regime where the hybrid-KBM turbulence is closer to marginal, especially at low $k_y$, in the transport steady state, owing reductions of the density gradient in the core and increases in $\beta'$ at the edge;} and (ii) by large transport stiffness, where small variations in the temperature and density gradients lead to large changes in the turbulent flux. High transport stiffness improves the robustness of the T3D steady state.

\begin{figure}
    \centering
    \subfloat[]{\includegraphics[height=0.23\textheight]{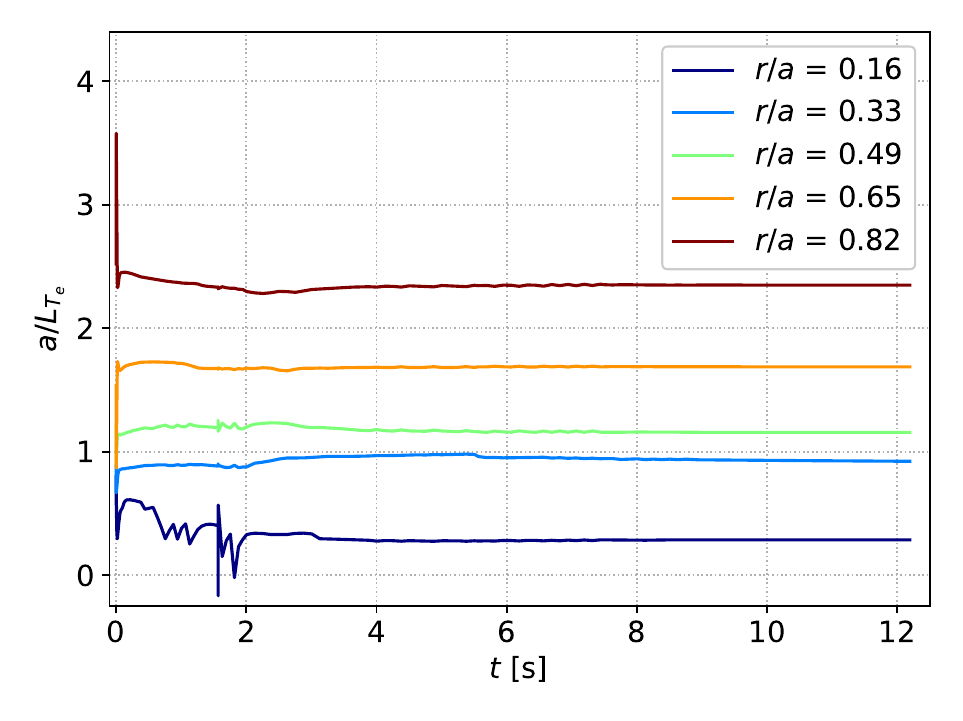}}\quad
    \subfloat[]{\includegraphics[height=0.23\textheight]{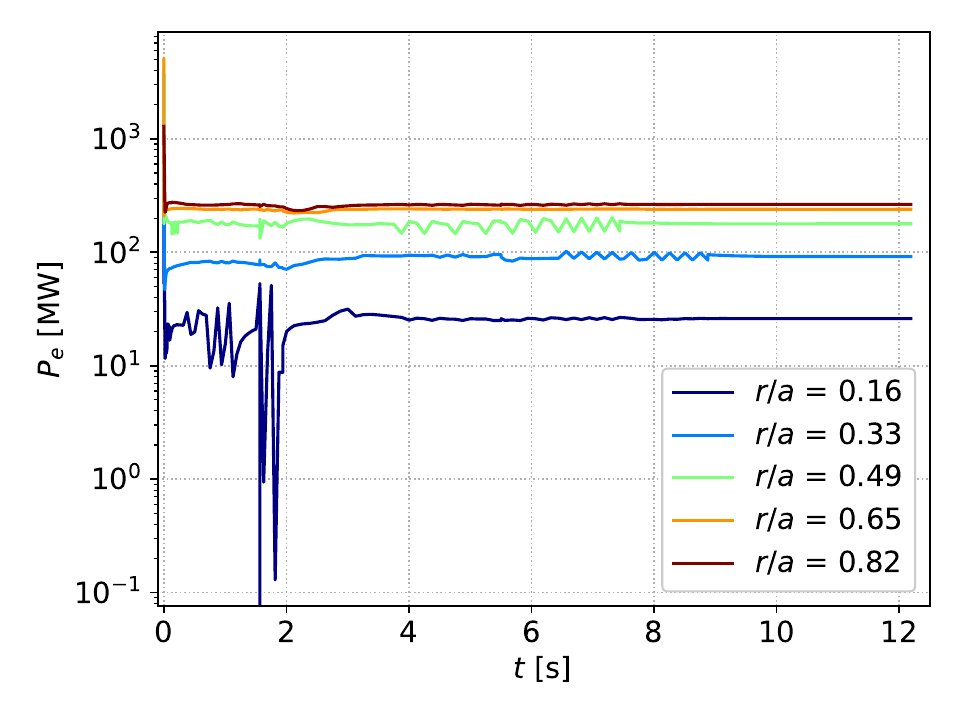}}\\
    \subfloat[]{\includegraphics[height=0.23\textheight]{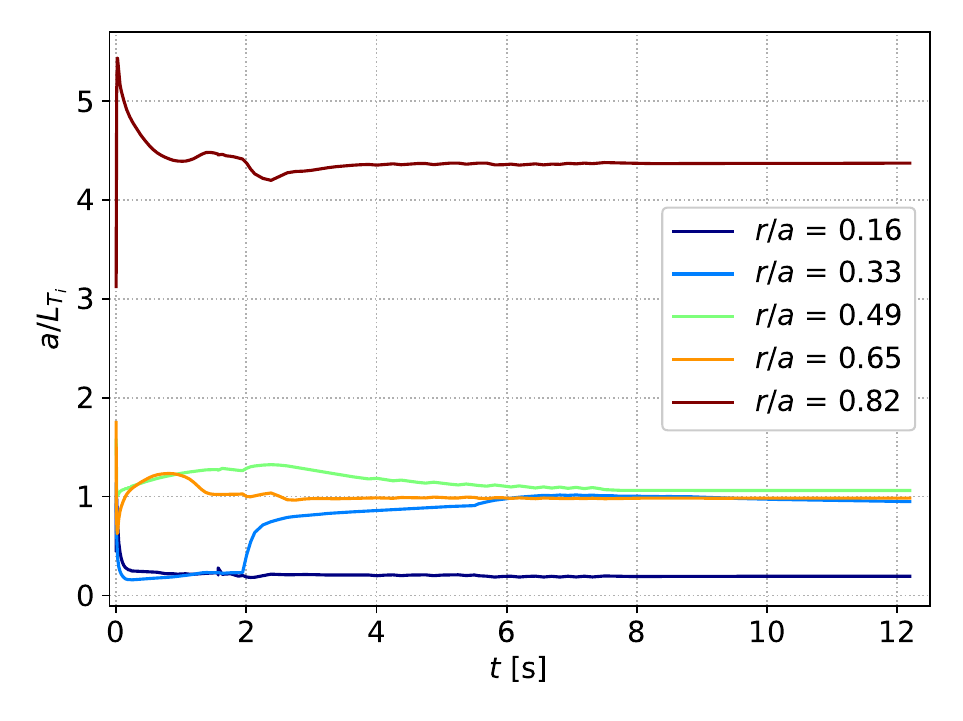}}\quad
    \subfloat[]{\includegraphics[height=0.23\textheight]{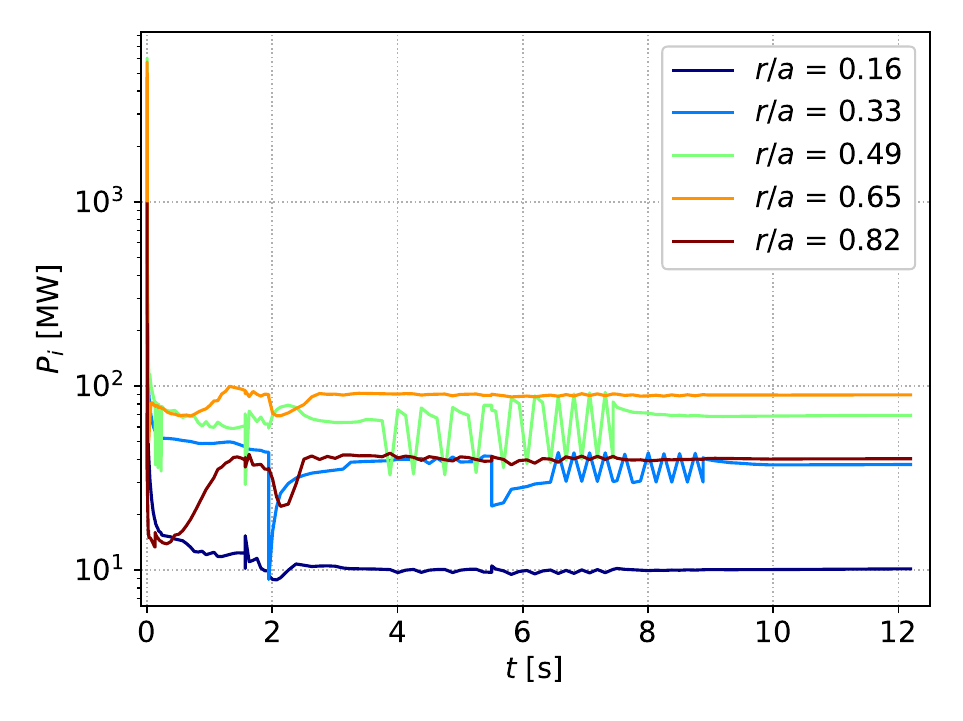}}\\
    \subfloat[]{\includegraphics[height=0.23\textheight]{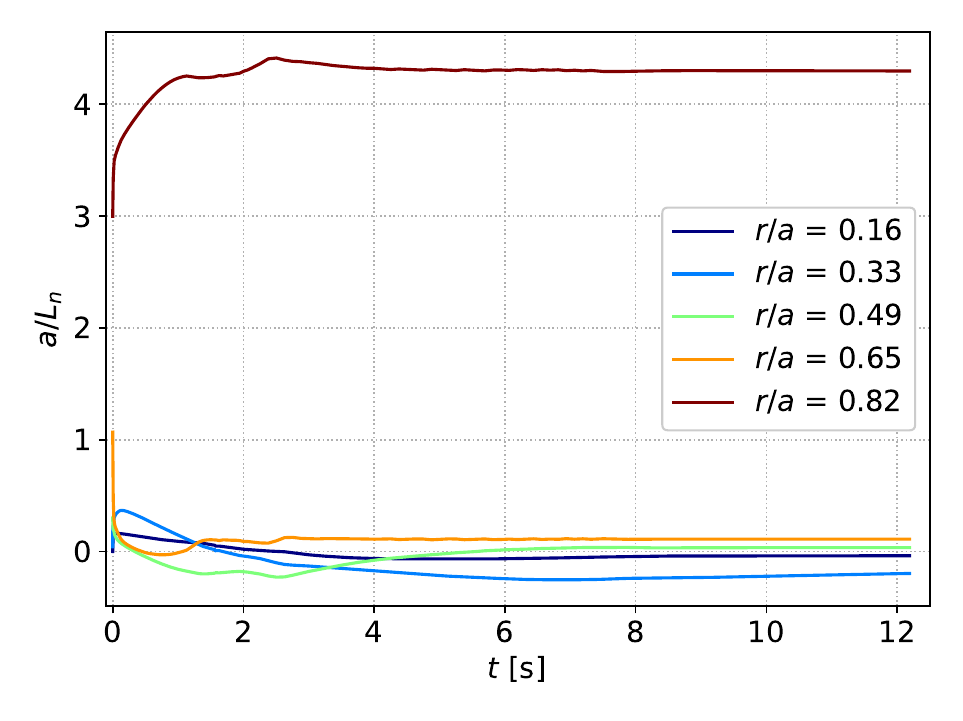}}\quad
    \subfloat[]{\includegraphics[height=0.23\textheight]{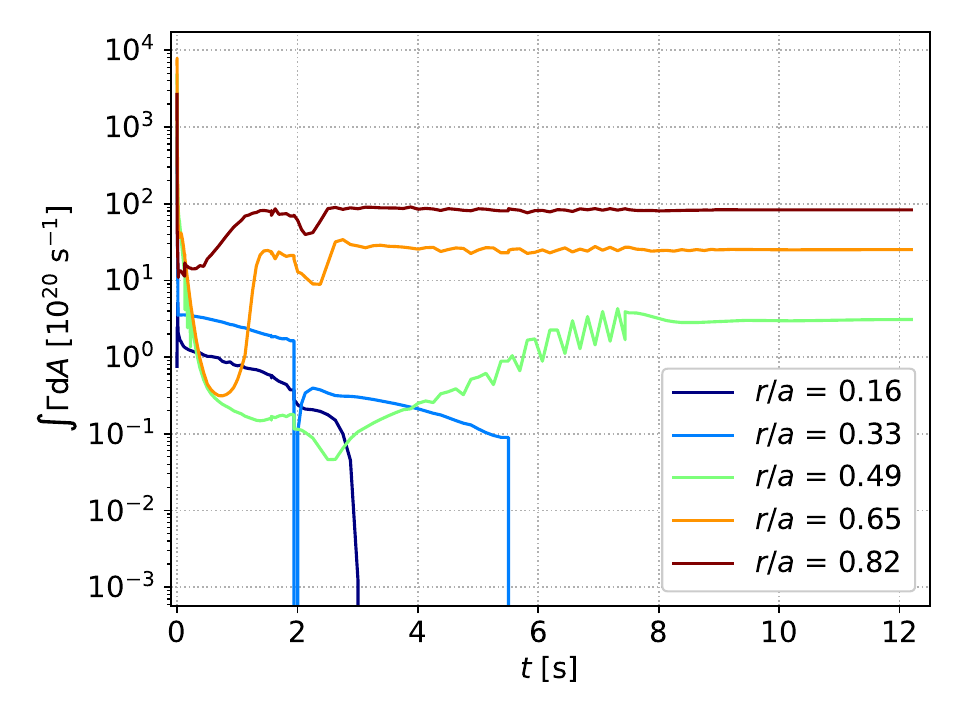}}\\
    \caption{Time evolution of the density and temperature gradients as well as of the power and particle losses in the reference T3D simulation. The initial profiles are taken from the JETTO STEP-EC-HD flat-top operating point. Different colors represent different radial surfaces (fluxes and gradients in T3D are computed at the mid-point between each radial grid point).}
    \label{fig:evolution}
\end{figure}

Fig.~\ref{fig:evolution} shows the time evolution of the density and temperature gradients as well as of the power and particle losses over the duration of the T3D reference simulation. 
The time traces can be divided in three main phases. The first phase is very short ($0<t\lesssim 0.1$~s, 50 solver iterations) and is characterized by a fast evolution of the profiles that leads to a substantial reduction of power and particle losses. This fast initial phase is caused by the strong initial particle and power imbalance,  as shown in Fig.~\ref{fig:balance}~(a)~and~(c).
The second phase ($0.1\lesssim t\lesssim 9$~s, 100 solver iterations) is characterized by a slow evolution of gradients and fluxes towards a steady state solution. 
In this phase, the particle fluxes at $r/a\simeq 0.16$ and $r/a\simeq 0.33$ drop to negligible values, which is consistent with an approach to steady state with no particle source in the core. There is a transitory ripple in the time evolution of the power and particle losses at $r/a\simeq 0.33$ and $r/a\simeq 0.49$, while the density and temperature gradients vary smoothly; this is caused by low $k_y$ hybrid-KBMs being stabilized or destabilized while T3D profiles are slowly converging to a steady state solution, and it is amplified by stiff transport.
Apart from the ripple and slow increase of particle flux at $r/a\simeq 0.5$, power and particle losses do not vary significantly in the second phase. 
The final phase (40 solver iterations), reaches the transport steady state, where the kinetic gradients, power and particle losses are constant in time.
Fig.~\ref{fig:balance} shows excellent agreement between transport losses and sources in the T3D transport steady state.

\begin{figure}
    \centering
    \subfloat[]{\includegraphics[height=0.23\textheight]{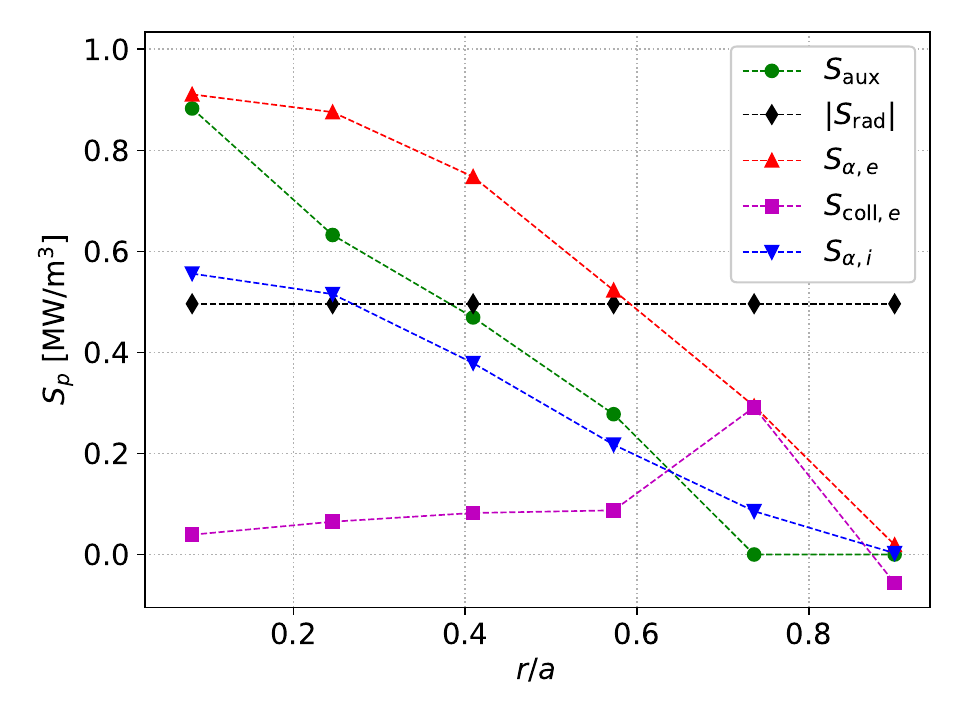}}\quad
    \subfloat[]{\includegraphics[height=0.23\textheight]{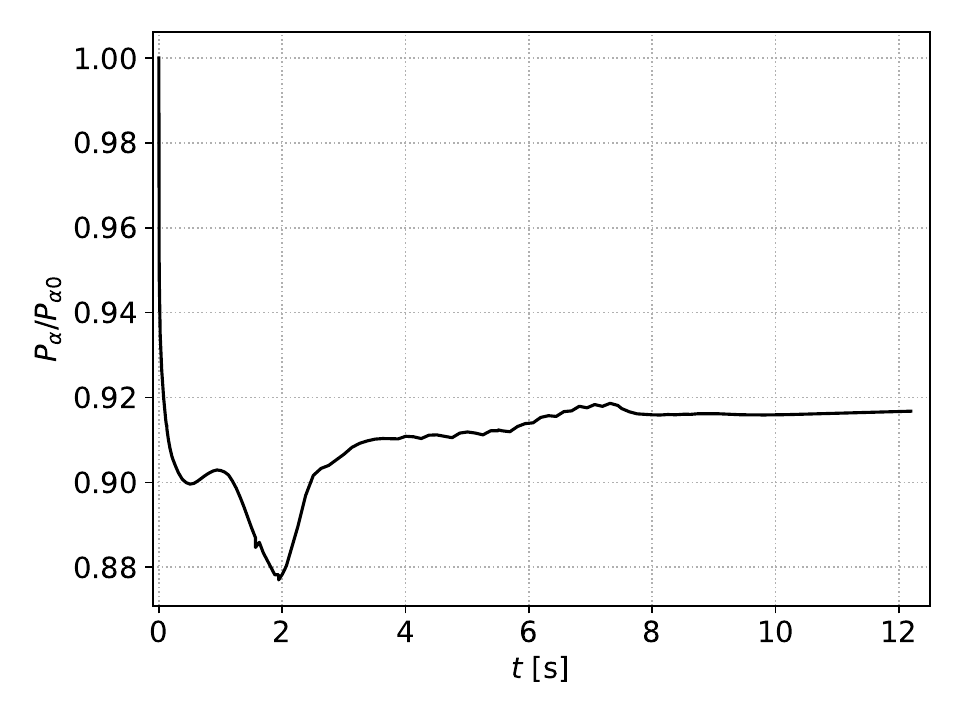}}
    \caption{\textcolor{myred}{(a) Power source radial profile in the final T3D steady state of the auxiliary heating ($S_\mathrm{aux}$), $\alpha$-particle heating ($S_{\alpha, e}$ and $S_{\alpha, i}$), collisional exchange ($S_{\mathrm{coll}, e}=-S_{\mathrm{coll}, i}$) and radiation ($S_\mathrm{rad}$).} (b) Fusion power evolution in the reference T3D simulation. The fusion power is normalized to its initial value of 1.7 GW.}
    \label{fig:fusion_evol}
\end{figure}

\textcolor{myred}{The heating power density radial profiles from auxiliary and $\alpha$-particle heating, collisional exchange, and radiation are shown in Fig.~\ref{fig:fusion_evol}~(a) for the final T3D steady state, while the} evolution of the fusion power is shown in Fig.~\ref{fig:fusion_evol}~(b). Consistently with the pressure profile evolution, the fusion power rapidly decreases in the initial phase, which is followed by a slower evolution. The final fusion power is approximately 10\% smaller than its initial value.

\section{Gyrokinetic analysis of the T3D transport steady state}
\label{sec:analysis}

The aim of this section is to verify whether fluxes from nonlinear gyrokinetic simulations remain consistent with our reduced model in the transport steady state from T3D.  This verifies whether the quasi-linear reduced transport model of Eqs~.\eqref{eqn:heat_species}~and~\eqref{eqn:particle_species} faithfully describes turbulent transport in an equilibrium that was not in the database used to tune the model parameters in Sec.~\ref{sec:ql_metric}.  This test is performed at the three outermost T3D radial grid points, i.e. $r/a \in \{0.5,0.65,0.8\}$, which are the most critical surfaces for fusion performance\footnote{The pressure profile in the core for $r/a<0.4$ is rather flat and any failure of the reduced transport model in this region should only weakly influence fusion performance.}.

\subsection{Linear analysis }

\begin{figure}
    \centering
    \subfloat[]{\includegraphics[height=0.22\textheight]{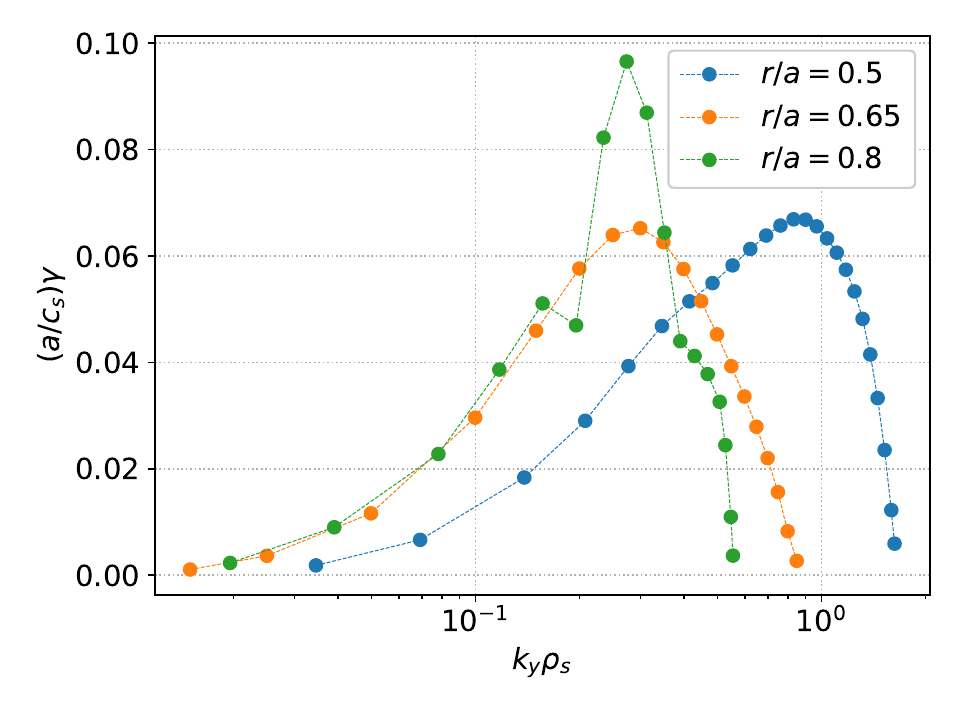}}\quad
    \subfloat[]{\includegraphics[height=0.22\textheight]{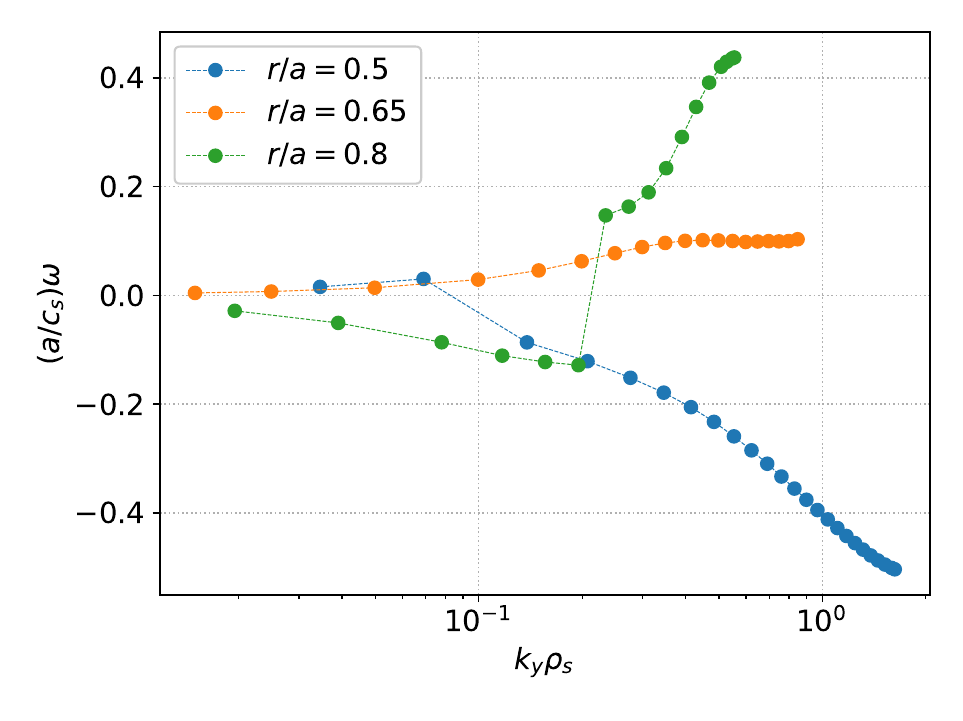}}
    \caption{Growth rate (a) and mode frequency (b) as functions of $k_y$ at $r/a=0.5$ (blue dots), $r/a=0.65$ (orange dots) and $r/a=0.8$ (green dots) from linear simulations of the new STEP density and temperature profiles.}
    \label{fig:linear_newstate}
\end{figure}

\begin{figure}
    \centering
    \subfloat[]{\includegraphics[height=0.22\textheight]{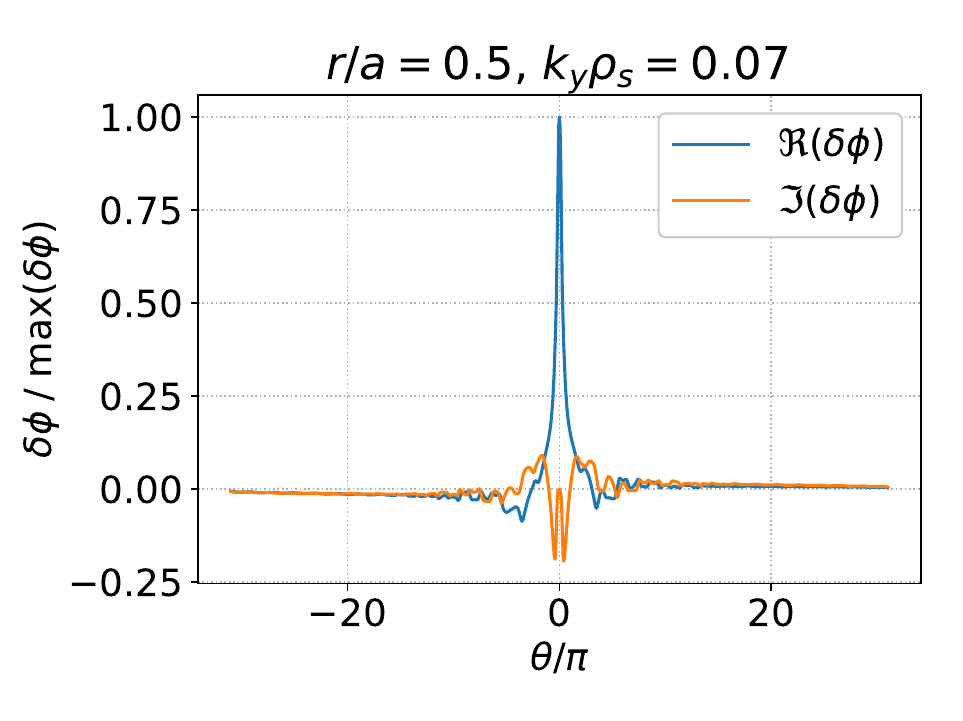}}
    \subfloat[]{\includegraphics[height=0.22\textheight]{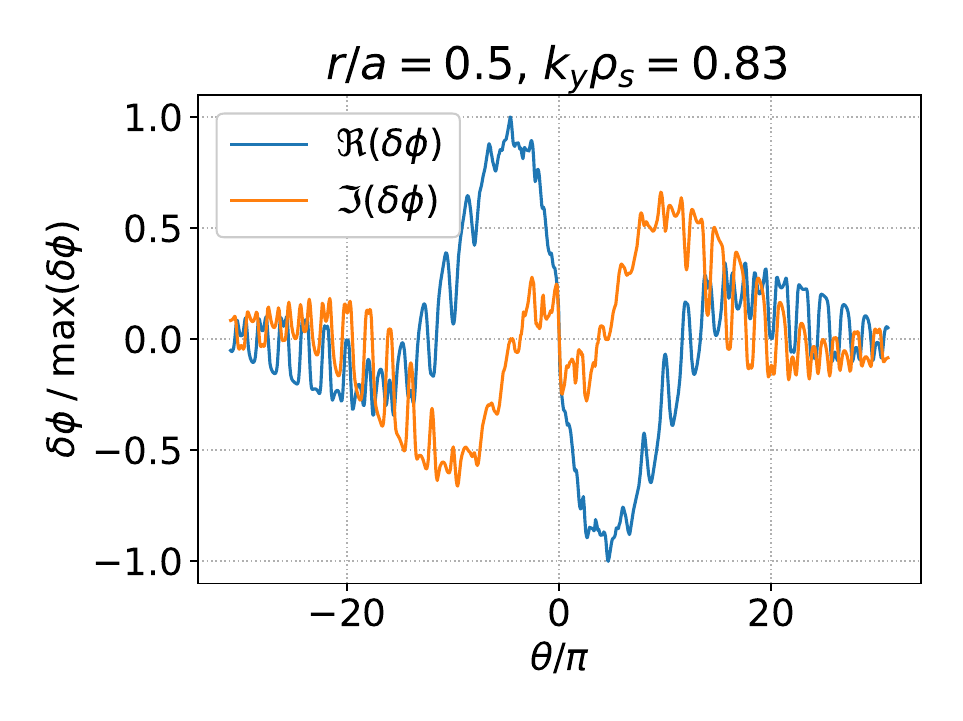}}
    \caption{Parallel mode structure of $\delta \phi$ at $k_y\rho_s\simeq 0.07$ (a) and $k_y\rho_s\simeq 0.83$ corresponding to the maximum growth rate of the hybrid-KBM and MTM instabilities, respectively, at $r/a=0.5$ of the new STEP profiles obtained from the T3D reference simulation.}
    \label{fig:linear_eig_newstate}
\end{figure}

Linear simulations are performed on these three surfaces with GS2, using numerical resolutions similar to those used in Sec.~\ref{sec:linear}, except setting $n_\text{period} = 32$ at $r/a=0.5$.
The dominant mode growth rates and mode frequencies are shown as functions of $k_y$ in Fig.~\ref{fig:linear_newstate}.  
The mode frequency at $r/a=0.5$ shows a discontinuity between $k_y\rho_s=0.15$ and $k_y\rho_s=0.2$, which corresponds to a transition of the dominant mode from hybrid-KBMs to MTMs.
The $\delta \phi$ parallel mode structures at $k_y\rho_s \simeq 0.07$ and $k_y\rho_s \simeq 0.83$, corresponding to the fastest growing hybrid-KBM and MTM respectively, are shown in Fig.~\ref{fig:linear_eig_newstate}. The hybrid-KBM eigenfunction is narrow in $\theta$ and has even parity, while the MTM eigenfunction has odd parity and is much more extended in $\theta$. 
A sign change on the mode frequency is also observed at $r/a=0.8$ around $k_y\rho_s \simeq 0.2$. In this case, however, the parallel structures of modes with positive and negative frequency are similar. This is consistent with the complex nature of the hybrid-KBM, involving both trapped electron and ion dynamics, discussed in \cite{kennedy2023a}.

\begin{figure}
    \centering
    \subfloat[]
    {\includegraphics[height=0.15\textheight]{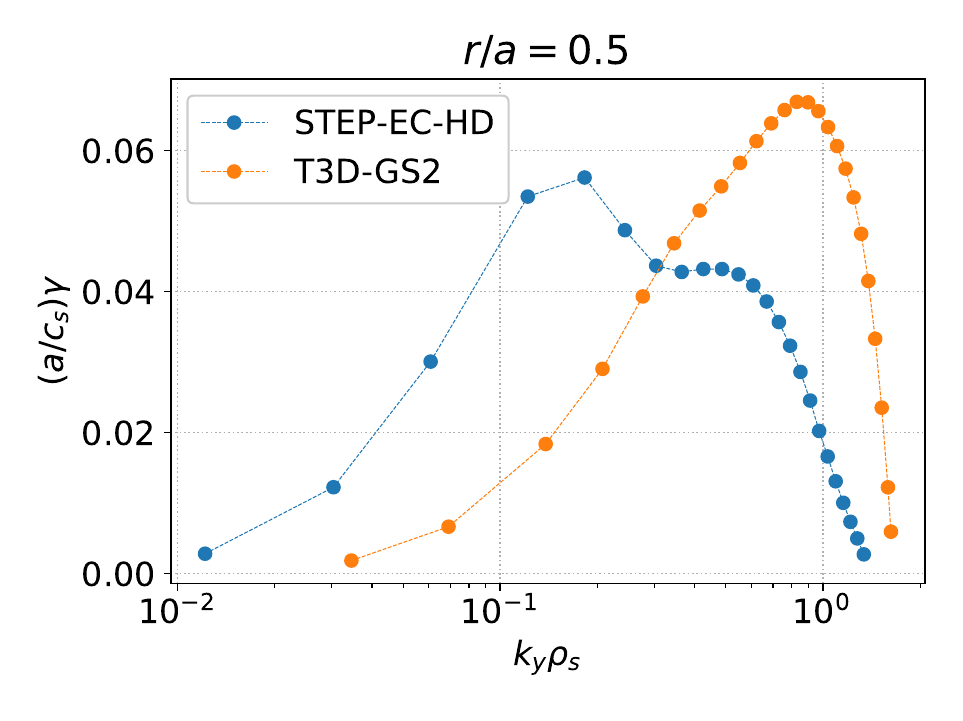}}\quad
    \subfloat[]{\includegraphics[height=0.15\textheight]{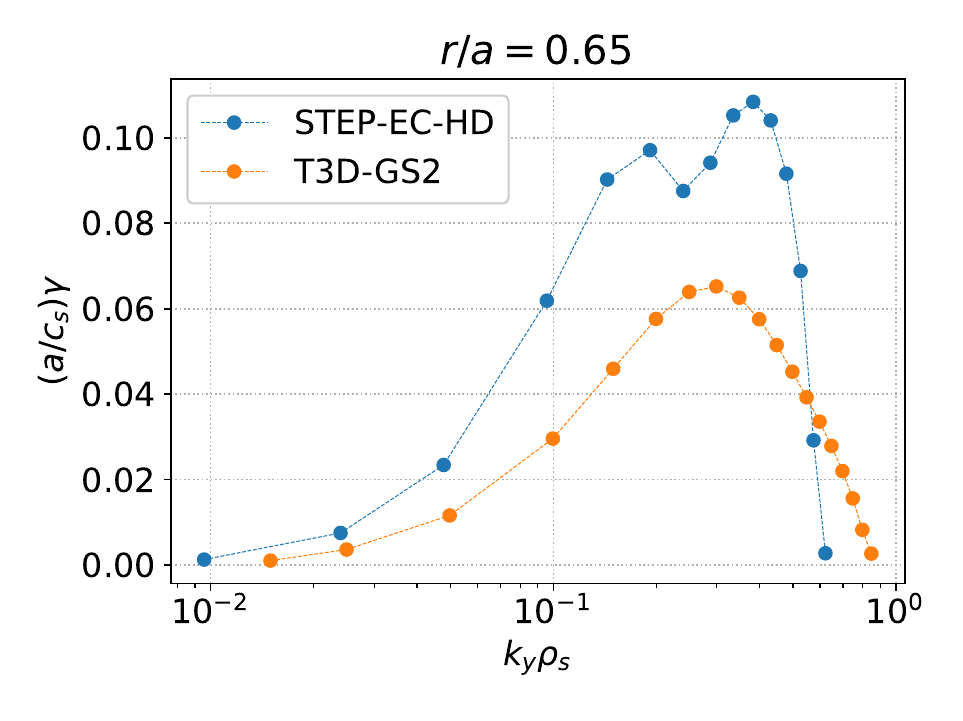}}\quad
    \subfloat[]{\includegraphics[height=0.15\textheight]{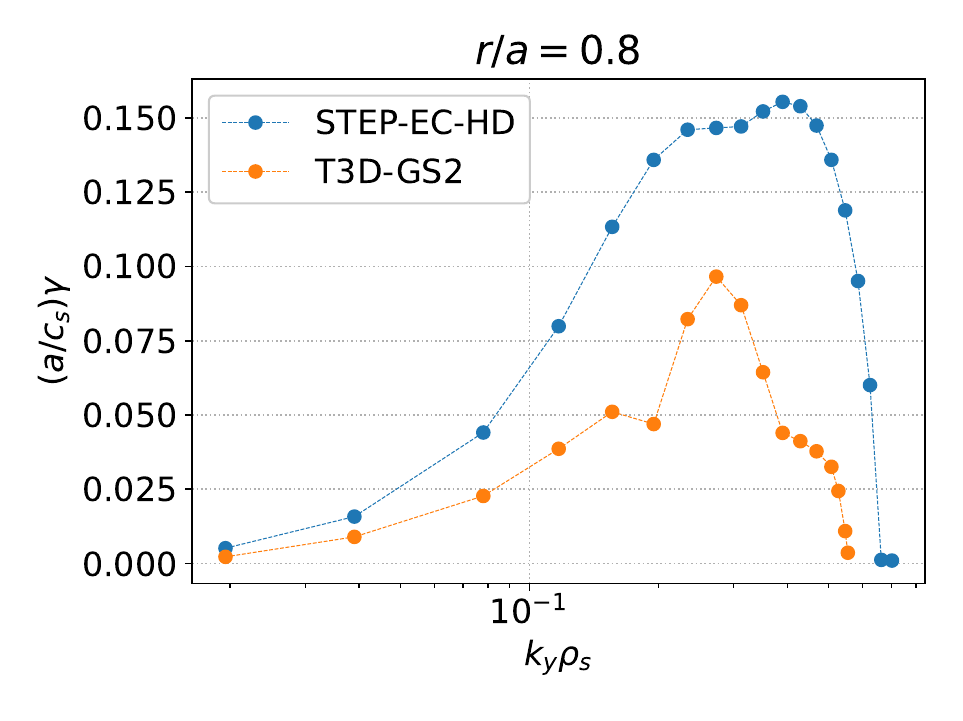}}
    \caption{Growth rate comparison between the initial (STEP-EC-HD) and final (T3D-GS2) profiles at $r/a=0.5$ (a),  $r/a=0.65$ (b) and $r/a=0.8$ (c).}
    \label{fig:linear_comparison}
\end{figure}

Fig.~\ref{fig:linear_comparison} shows a comparison of the growth rate values between the initial and final STEP plasma profiles at $r/a=0.5$, $r/a=0.65$ and $r/a=0.8$.
Beyond the difference on the dominant instability at $r/a=0.5$ (the hybrid-KBM and MTM instabilities dominate in the original STEP case and in the new T3D steady state, respectively), a strong stabilization at low $k_y$ and a substantial growth rate maximum shift at higher $k_y$ are observed in the new T3D steady state. 
At both $r/a=0.65$ and $r/a=0.8$ cases, where the dominant instability remains the hybrid-KBM, there is a clear reduction of the growth rates across most of the spectrum, including at low $k_y$ which typically dominates the turbulent fluxes.
This linear comparison  underlies the substantial reduction of turbulent flux predictions by the reduced transport model between the initial condition and the transport steady state.  This will now be compared with nonlinear gyrokinetic simulations.

\subsection{Nonlinear analysis}
\label{sec:NL_T3D}

\begin{figure}
    \centering
    \subfloat[]{\includegraphics[height=0.22\textheight]{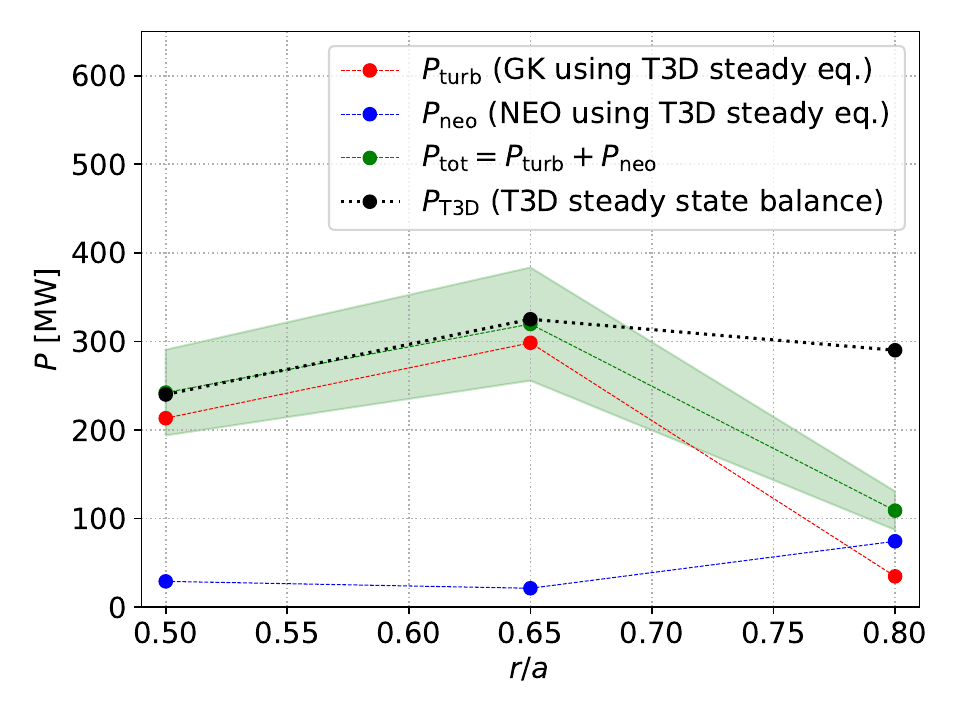}}\quad
    \subfloat[]{\includegraphics[height=0.22\textheight]{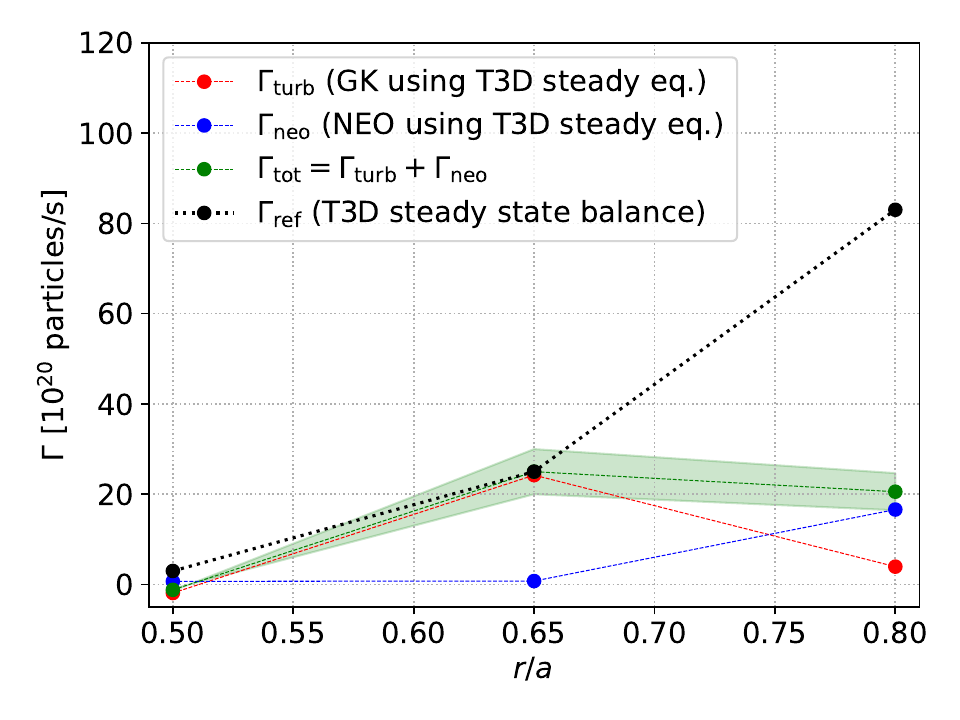}}
    \caption{Turbulent (red dots), neoclassical (blue dots) and total (green dots) power (a) and particle (b) losses at $r/a=0.5$, $r/a=0.65$ and $r/a=0.8$ of the new STEP profiles, compared to the expected heating power and particle fueling from the original STEP case (black dots). The green shaded area corresponds to a $\pm$20\% of the total heat (a) and particle (b) transport. The turbulent fluxes are evaluated from GENE nonlinear gyrokinetic simulations and the neoclassical fluxes are evaluated from NEO simulations, both considering the T3D steady state profiles.}
    \label{fig:nonlinear_newstate_overview}
\end{figure}

Here we briefly show results from nonlinear gyrokinetic simulations performed using GENE at $r/a = [0.5, 0.65, 0.8]$ in the T3D steady state solution: these simulations used the numerical resolutions listed in table~\ref{tab:res_newstate}. These simulations include equilibrium flow shear with $\gamma_E = [0.02,0.03,0.05 ]\,c_s/a$ at $r/a = [0.5,0.65,0.8]$, respectively.  Fig.~\ref{fig:nonlinear_newstate_overview} plots the total heat and particle transport losses across these three surfaces in the transport steady state, comparing values from the quasi-linear model in T3D with neoclassical and turbulent fluxes computed using NEO and GENE.
  
\begin{table}
    \centering
    \begin{tabular}{cccccccc}
    \toprule
    $\mathbf{r/a}$ & $\boldsymbol{\Delta k_y\rho_s}$ & $\boldsymbol{k_{y,\mathrm{max}}\rho_s}$ & $\boldsymbol{\Delta k_x\rho_s}$ & $\boldsymbol{k_{x,\mathrm{max}}\rho_s}$ &  $\boldsymbol{n_\theta}$ & $\boldsymbol{n_\mu}$ & $\boldsymbol{n_v}$ \\
    \midrule
    0.50 & 0.035 & 1.65 & 0.030  & 2.0 & 32 & 16 & 28\\ 
    0.65 & 0.030 & 1.10 & 0.026 & 1.70 & 32 & 16 & 28\\
    0.80 & 0.019 & 0.90 & 0.034 & 2.21 & 64 & 16 & 28\\
    \bottomrule
    \end{tabular}
    \caption{Numerical resolutions used in nonlinear GENE simulations on three surfaces from the T3D transport steady state, $r/a=[0.5, 0.65,0.8]$.}
    \label{tab:res_newstate}
\end{table}

At $r/a=0.5$ and $r/a=0.65$, Fig.\ref{fig:nonlinear_newstate_overview} shows relatively good agreement between the nonlinear gyrokinetic heat flux and the reduced transport model in T3D.  The neoclassical heat flux is negligible at these radial locations.
There is also good agreement between the gyrokinetic and reduced model particle losses here, and we note that both neoclassical and turbulent particle fluxes are negligible at $r/a=0.5$.
 
Turbulent transport in the GENE nonlinear simulation at $r/a=0.5$ is mainly driven by hybrid-KBMs, in spite of linearly dominant MTMs (see Fig.\ref{fig:linear_newstate}): the nonlinearly dominant mode fields $\delta \phi(\theta)$ and $\delta A_\parallel(\theta)$ have even and odd parity, respectively. Good agreement in Fig.~\ref{fig:nonlinear_newstate_overview} at $r/a=0.5$ suggests that  transport from these hybrid-KBMs is captured faithfully by the quasi-linear reduced transport model.

At $r/a=0.8$, the quasi-linear model overestimates the heat and particle fluxes from the nonlinear simulation, despite turbulent transport being solely driven by hybrid-KBMs. Turbulent fluxes should therefore  be described well by the quasi-linear model. 
The disagreement might be due to some inaccuracy affecting the quasi-linear model, as visible in Fig.~\ref{fig:ql_fit} where a few points exhibit order of magnitude differences between the quasi-linear and the nonlinear heat flux predictions. \textcolor{myred}{In addition, the local equilibrium at $r/a=0.8$ might be near a marginal state where a small increase in drive can boost the turbulent fluxes significantly: this marginal condition, where simulations are more difficult and nonlinear effects intrinsically important, might not be described well by the quasi-linear model, although further investigations, which are outside the aim of this work, would be required before drawing firm conclusions}. We note that transport stiffness should allow particle and power balance to be restored with a modest increase in the drive and minimal impact on the overall density and temperature profiles. In fact, transport stiffness makes flux-driven simulations less sensitive to input parameters and improves the robustness of the steady state profile predictions. 

In summary, Fig.~\ref{fig:nonlinear_newstate_overview} shows that gyrokinetic transport fluxes for the T3D transport steady state in STEP  are broadly consistent with the quasi-linear reduced transport model of Eqs.~\eqref{eqn:heat_species} and \eqref{eqn:particle_species}. This gives confidence in the validity of the reduced model within the parameter space explored in this T3D transport simulation. 
It is important to stress, however, that this is a first iteration to develop a suitable reduced model for STEP: further guidance from higher-fidelity nonlinear gyrokinetic simulations will be required to overcome important limitations of the reduced transport model presented here, including the absence of fast $\alpha$ effects. Nevertheless the reduced model and T3D-GS2 are powerful tools that could be further developed and used more extensively, together with existing integrated modelling codes, to assist in the design of new STEP plasmas and flat-top operating points.

\section{Conclusions}
\label{sec:conclusion}

Nonlinear gyrokinetic calculations have demonstrated the need to account for transport from hybrid-KBM turbulence in the design of STEP plasma scenarios \citep{giacomin2024}.  Scenario design therefore needs to model the transport from such turbulence at modest computational cost, which motivates the development of a new reduced transport model to describe hybrid-KBM turbulence in STEP-like regimes.
The reduced transport model developed here is inspired by quasi-linear theory, and gives the transport fluxes of Eqs.\eqref{eqn:heat_species} and \eqref{eqn:particle_species}.  The quasi-linear model is applied to the total heat flux through a quasi-linear metric, $\Lambda$,  that lies at the heart of the model. The quantity $\Lambda$ is computed directly from linear gyrokinetic calculations: it is fully electromagnetic with contributions from all three fields and it exploits the dependence of linear modes on the ballooning parameter, $\theta_0$, to include the impact of flow shear on turbulent saturation. To this point the model is rather general, but it contains two free parameters, $Q_0$ and $\alpha$, that link $\Lambda$ to the total heat flux. 
Here, the reduced model total heat flux is optimised for hybrid-KBM turbulence by tuning these parameters to best match results from a set of nonlinear gyrokinetic simulations for STEP-relevant parameters, as shown in Fig.~\ref{fig:ql_fit}.  The saturation rule in Eq.~\eqref{eqn:redmod} is in itself rather general, and with suitable different choices of coefficients $Q_0$ and $\alpha$ it may also be applicable to other turbulence regimes.

The new quasi-linear model is first applied to explore the dependence of STEP transport fluxes  on safety factor and magnetic shear.  This recovers the strong non-monotonic dependence of transport fluxes on $q$ that was previously revealed on a mid-radius surface using nonlinear gyrokinetic simulations \cite{giacomin2024}.
Evaluating the quasi-linear fluxes requires only linear gyrokinetic calculations, which are considerably cheaper computationally than nonlinear gyrokinetics. A new reduced model with sufficient fidelity to replace nonlinear gyrokinetics offers enormous potential benefits, including making the exploration of larger regions of STEP parameter space more tractable.

This reduced transport model is implemented in T3D \citep{t3d}, which includes a flexible transport solver based on the algorithm developed for the Trinity code \citep{barnes2010}.  Implementation of the model is achieved by coupling T3D to the local gyrokinetic code GS2, which is run linearly to determine the quasi-linear fluxes (see Fig.~\ref{fig:workflow} for a summary of the code workflow).
T3D-GS2 calculations are performed  from an initial condition set to the STEP-EC-HD flat-top operating point \cite{tholerus2024}.  To test the sensitivity of the resulting transport steady state to uncertainties, calculations are performed with reasonable variations in the reduced model parameters $Q_0$ and $\alpha$, and in flow shear.  
These initial T3D-GS2 simulations converge to a transport steady state, with  density and temperature profiles that are only weakly affected by the aforementioned variations (see shaded area in Fig.~\ref{fig:profiles}).  The robustness of the solution is likely due to profile relaxation in the presence of highly stiff transport, which is not captured in gradient driven simulations \cite{giacomin2024}.
The transport steady state pressure profile from T3D is 20\% lower on-axis but broader than the initial STEP-EC-HD pressure profile, and the energy confinement times are similar ($\tau_E^* = 4.6$~s in the T3D steady state, while $\tau_E^*=4.2$~s in the STEP-EC-HD).  The total fusion power in the T3D steady state is lower than in STEP-EC-HD by 10\%, though we note that fusion power is a highly sensitive parameter that may change  significantly when assumptions from this first physics-based transport calculation are relaxed. 

In order to verify whether turbulent transport remains consistent with the reduced model fluxes in the T3D solution and with the modelled sources and sinks, nonlinear gyrokinetic calculations were performed at the three outermost radial grid points of the T3D transport steady state, at $r/a = [0.5, 0.65, 0.8]$.  These surfaces are the most critical for fusion performance. The dominant microinstabilities are similar to those in STEP-EC-HD, with hybrid-KBMs dominant at $r/a>0.6$ and MTMs dominant at $r/a=0.5$. Hybrid-KBM linear growth rates, including at low $k_y$, are lower in the transport steady state than in the STEP-EC-HD initial condition, which is consistent with the substantial reduction in turbulent transport found nonlinearly. It is important to point out that local gyrokinetics appears to describe turbulence well in the new T3D steady state solution, where radial turbulent structures are much smaller and transport is much closer to threshold than found in STEP-EC-HD. This nonlinear gyrokinetic analysis supports that (i) the reduced model is a reasonable description of nonlinear fluxes found on these surfaces
and (ii) the T3D steady state is compatible with the modelled sources.  The reduced transport model developed here could be used more routinely to account for hybrid-KBM turbulent transport in STEP scenario development, which is one of STEP most pressing needs. 

First flux-driven calculations using this physics-based reduced model suggest that a transport steady state exists for STEP, which is broadly compatible with assumed sources in STEP-EC-HD and delivers close to the target fusion power. The $\beta^{\prime}$ stabilisation at the edge appears to play a key role in sustaining the stored energy. We highlight, however, that these initial calculations have made some simplifying assumptions that could impact on the transport steady state: (i) most magnetic equilibrium parameters are held constant: shaping, safety factor (and $\hat{s}$), but not $\beta'$ which evolves with the pressure profile, (ii) impurities (e.g. from thermalised helium) and fast-$\alpha$ particles are neglected, (iii) the radial electric field is held constant, and (iv) external heat and particle sources and radiation are constant while fusion $\alpha$-heating and electron-ion exchange evolve with the kinetic profiles.
We also highlight that a route to access this transport steady state has not been demonstrated.

\begin{figure}
    \centering
    \includegraphics[scale=0.4]{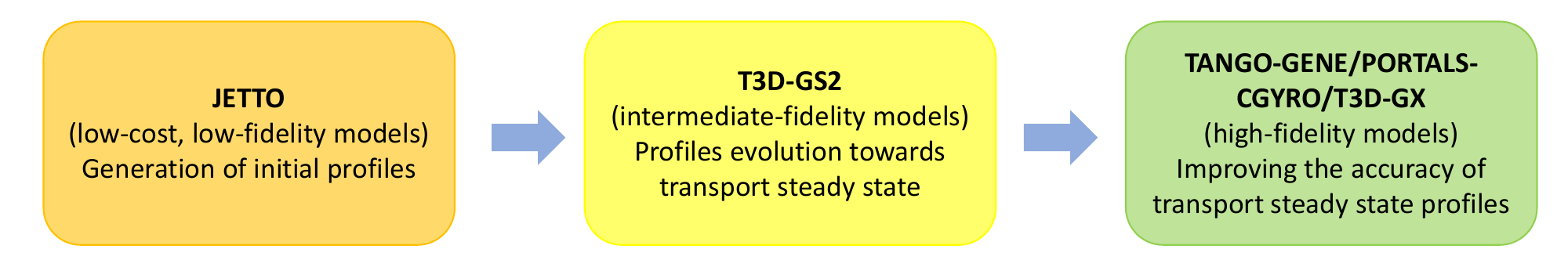}
    \caption{Hierarchy of tools that can be used in tandem to design plasma scenarios, involving lower cost and fidelity (left) and higher cost and fidelity (right) transport simulations. T3D-GS2 with the quasi-linear model provides an intermediate step between computationally cheap low-fidelity  and more demanding high-fidelity transport simulations using nonlinear gyrokinetics. }
    \label{fig:hierarchy}
\end{figure}

In the reference T3D simulation, quasi-linear heat fluxes are computed approximately 300 times at each radial location before the transport steady state is reached, at a total computational cost of approximately $10^5$~core-hours. If fluxes were computed using nonlinear local gyrokinetics, this computational cost would have exceeded $10^8$~core-hours as a single calculation requires more than $ 10^5$~cores-hours in STEP-like regimes.
Starting from the reduced model transport steady state from T3D as an initial condition, however, should significantly reduce the computational cost of the nonlinear gyrokinetic approach to such a flux-driven calculation. 
It can be envisaged that scenario design for future machines, including STEP, will benefit from exploiting a hierarchy of tools similar to that outlined in Fig.~\ref{fig:hierarchy}: integrated modelling codes like JETTO, with lower computational cost and fidelity transport models, providing initial profiles; refinement of transport predictions with intermediate fidelity physics-based models like the quasi-linear reduced model in T3D-GS2; and further honing with higher fidelity flux-driven codes like \textcolor{myred}{TANGO-GENE}, T3D-GX or \textcolor{myred}{PORTALS}-CGYRO, based on nonlinear gyrokinetic fluxes. 
This approach may also be used to improve the fidelity of models used in integrated modelling, e.g. by implementing the reduced transport model of the intermediate approach directly into JETTO (which would require either using a fast linear eigensolver or building a surrogate model).

This work motivates many exciting future research directions. The reduced model can be exploited to seek improved STEP operating points, e.g. through optimising the $q$ profile or other equilibrium properties. More sophisticated and demanding higher fidelity flux-driven calculations are needed where auxiliary heating, current drive, magnetic equilibrium, and flow shear are recomputed self-consistently with the kinetic profiles. The reduced model should be extended to include impurities and fast $\alpha$-particles, and it may also be necessary to include magnetic stochastic turbulent transport from MTMs if these become dominant when hybrid-KBMs are suppressed.
A further key application for T3D-GS2 and the new model will be to explore the STEP $I_p$ ramp-up phase, and to help find a route to access the high performance flat-top steady state. Last, but certainly not least, it will be important to seek suitable experimental regimes where this new quasi-linear model can be validated.

\section*{Acknowledgements}

We would like to thank T. Adkins, E. Belli, J. Candy, B. Chapman-Oplopoiou, A. Di Siena, T. G{\"o}rler, P. Ivanov, H. Meyer, and F. Sheffield for useful discussions. 

\section*{Funding}
This work has been funded by the Engineering and Physical Sciences Research Council (grant numbers EP/R034737/1 and EP/W006839/1) and by STEP, a UKAEA programme to design and build a prototype fusion energy plant and a path to commercial fusion.
Simulations have been performed on the ARCHER2 UK National Supercomputing Service under the project e607, on the Cambridge Service for Data Driven Discovery (CSD3) facility operated by the University of Cambridge Research Computing Service that is partially funded by the Engineering and Physical Sciences Research Council (grant EP/T022159/1), and on the Viking cluster that is a high performance compute facility provided by the University of York.

\section*{Declaration of interests}
The authors report no conflict of interest.

\section*{Data availability}
To obtain further information on the data and models underlying this paper please contact PublicationsManager@ukaea.uk.

\appendix

\section{Quasi-linear species flux contributions}
\label{sec:appendix}

\begin{figure}
    \centering
    \subfloat[Electron heat flux]{\includegraphics[height=0.16\textheight]{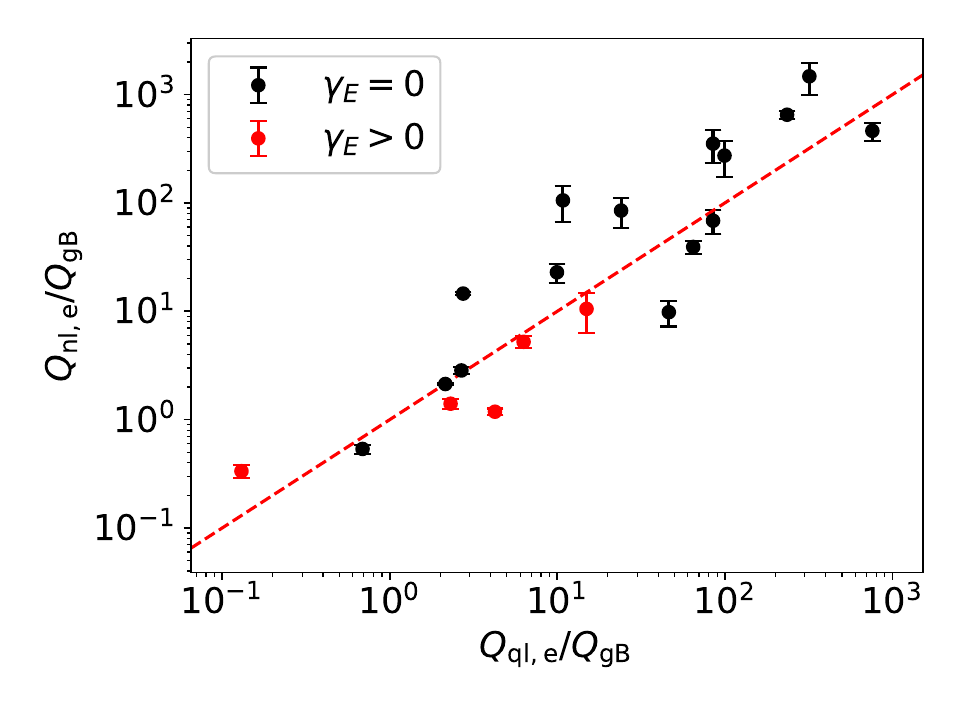}}
    \subfloat[Ion heat flux]{\includegraphics[height=0.16\textheight]{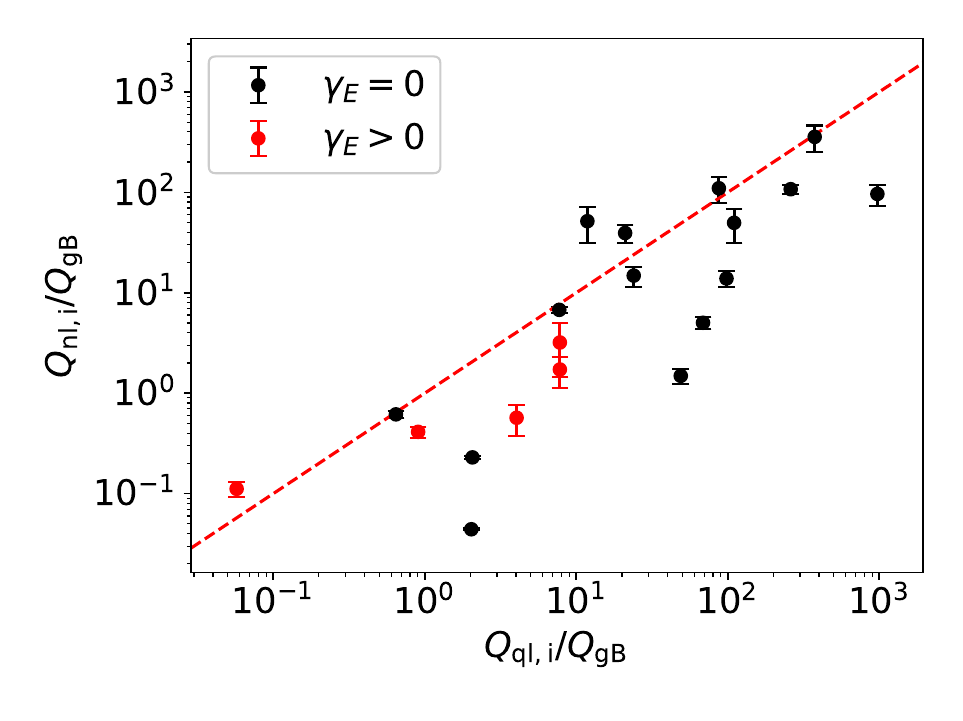}}
    \subfloat[Particle flux]{\includegraphics[height=0.16\textheight]{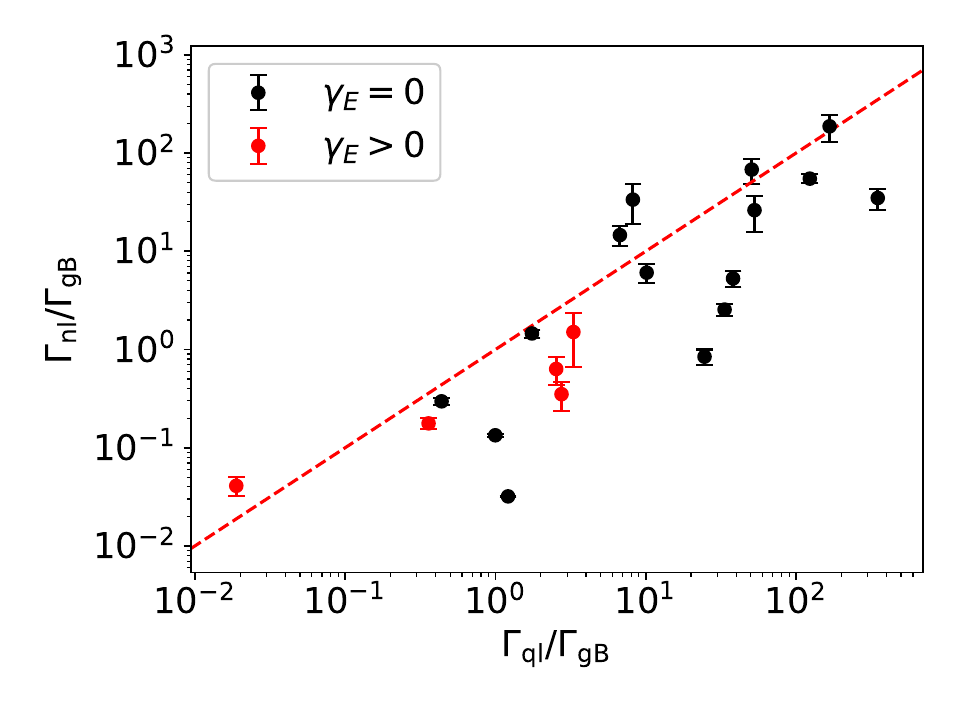}}
    \caption{Comparison of the electron (a) and main ion (b) heat flux and total particle flux (c) computed from the nonlinear gyrokinetic database of Sec.~\ref{sec:ql_metric} to the quasi-linear heat and particle flux predictions given by Eqs.~\eqref{eqn:heat_species}~and~\eqref{eqn:particle_species}.}
    \label{fig:species_breakdow}
\end{figure}

The quasi-linear reduced transport model is developed starting from the total heat flux, as detailed in Sec.~\ref{sec:ql_metric} (see Eq.~\eqref{eqn:redmod}). Each species particle and heat flux contribution is then determined from the quasi-linear weights $Q_{\mathrm{l},s}(k_y, \theta_0)/Q_l(k_y, \theta_0)$ and $\Gamma_{l,s}(k_y, \theta_0)/Q_\mathrm{l}(k_y, \theta_0)$, which multiply the quasi-linear metric $\hat{\Lambda}(k_y, \theta_0)$ before integrating over $\theta_0$ and $k_y$ (see Eqs.~\eqref{eqn:heat_species}~and~\eqref{eqn:particle_species}).
Fig.~\ref{fig:species_breakdow} compares the nonlinear electron and main ion heat and particle fluxes computed from the nonlinear gyrokinetic database described in Sec.~\ref{sec:ql_metric} to the quasi-linear predictions given by Eqs.~\eqref{eqn:heat_species}~and~\eqref{eqn:particle_species}. Parameters $\alpha$ and $Q_0$ have the values determined in the fit of the saturation model in Eq.~\eqref{eqn:redmod} to the total heat flux from STEP nonlinear gyrokinetic simulations shown in Fig.~\ref{fig:ql_fit}.

An overall good agreement is observed between nonlinear and quasi-linear electron heat flux in Fig.~\ref{fig:species_breakdow}~(a), while the quasi-linear ion heat flux as well as the quasi-linear particle flux tend to overestimate the corresponding nonlinear quantities, as shown in Figs.~\ref{fig:species_breakdow}~(b)~and~(c). This is due to a slight overestimation of the quasi-linear weights $Q_{\mathrm{l},i}(k_y, \theta_0)/Q_\mathrm{l}(k_y, \theta_0)$ and $\Gamma_\mathrm{l}(k_y, \theta_0)/Q_\mathrm{l}(k_y, \theta_0)$ with respect to nonlinear flux ratios, $Q_{\mathrm{nl},i}(k_y, \theta_0)/Q_\mathrm{nl}(k_y, \theta_0)$ and $\Gamma_{\mathrm{nl}}(k_y, \theta_0)/Q_{\mathrm{nl}}(k_y, \theta_0)$. 
The agreement between the quasi-linear and the nonlinear ion heat flux (and the quasi-linear and the nonlinear particle flux) could be improved by introducing species-dependent coefficients, but this would further constrain the reduced model to the specific STEP regime of STEP-EC-HD.

\bibliographystyle{jpp}
\bibliography{bibliography}

\end{document}